\documentclass[
	final,
	paper=a4,
	pagesize=auto,
	fontsize=10pt,
	version=last,
]{scrartcl}

\usepackage[utf8]{inputenc}
\usepackage[
	english,
]{babel}

\makeatletter
\newcommand{\LoadPackagesNow}{}
\newcommand{\LoadPackageLater}[2][]{%
   \g@addto@macro{\LoadPackagesNow}{%
      \usepackage[#1]{#2}%
   }%
}
\makeatother

\usepackage{xspace}
\usepackage{xargs}
\usepackage{ifthen}
\usepackage{etoolbox}

\KOMAoptions{%
	numbers=noenddot,
	twoside=semi, 
}

\setkomafont{section}{\Large\rmfamily\bfseries}
\setkomafont{subsection}{\large\rmfamily\bfseries}
\setkomafont{subsubsection}{\rmfamily\bfseries}
\setkomafont{paragraph}{\rmfamily\bfseries}
\setkomafont{pageheadfoot}{\smaller\scshape\rmfamily}

\usepackage[headsepline,automark]{scrlayer-scrpage}
\clearmainofpairofpagestyles
\clearplainofpairofpagestyles
\rehead{\pagemark}
\rohead{\pagemark}
\lehead{\headmark}
\lohead{\shortauthor}
\automark[section]{section}

\usepackage{calc}
\usepackage[margin=3cm,a4paper]{geometry}
\usepackage{framed}
\usepackage{mdframed}
\usepackage{enumitem}
\usepackage[toc]{appendix}

\usepackage[
	bottom,
	stable,      
	multiple,    
	flushmargin,
	hang,
]{footmisc}
\usepackage{mathpazo} 

\usepackage{microtype}

\usepackage{relsize}
\usepackage[normalem]{ulem}      
\usepackage{soul}		           
\usepackage{url}
\usepackage{lipsum}
\usepackage{comment}

\usepackage{array}
\usepackage{multirow}
\usepackage{longtable}
\usepackage{hhline}
\usepackage{tabularx}

\usepackage[table]{xcolor}
\usepackage{pgfplots}
\pgfplotsset{compat=1.16}
\usepackage{tikz}
\usepackage{tkz-euclide}
\usetikzlibrary{arrows,arrows.meta,shapes,calc,decorations.pathreplacing,decorations.pathmorphing,
	intersections,quotes,angles,positioning,fit,petri,through,backgrounds}
\tikzstyle{blackdot}=[shape=circle,fill=black,minimum size=1mm,inner sep=0pt,outer sep=0pt]

\usepackage{graphicx}

\usepackage{caption}
\usepackage{subcaption}
\usepackage[
   tbtags,    
   sumlimits,  
   nointlimits, 
   namelimits, 
   reqno,     
]{amsmath}

\usepackage{amsfonts}
\usepackage{mathrsfs}
\usepackage{dsfont}
\usepackage{amssymb}
\usepackage{units}
\LoadPackageLater{amsthm}
\LoadPackageLater{thmtools}
\usepackage[fixamsmath,disallowspaces]{mathtools}
\mathtoolsset{showonlyrefs}
\mathtoolsset{centercolon=true}

\usepackage{bm} 
\makeatletter
\g@addto@macro\bfseries{\boldmath}
\makeatother

\allowdisplaybreaks[4] 
\numberwithin{equation}{section} 

\usepackage{csquotes}
\usepackage[
	style=alphabetic,
	sorting=ynt,
 	sortcites=true,
	giveninits=true,
	maxbibnames=99,
	backend=biber,
	maxalphanames=5,
	maxcitenames=2,
]{biblatex}
\renewbibmacro{in:}{}
\DeclareFieldFormat{pages}{#1}

\usepackage[textsize=tiny,english,colorinlistoftodos,
	disable,
	]{todonotes}

\definecolor{pdfurlcolor}{rgb}{0,0,0.6}
\definecolor{pdffilecolor}{rgb}{0.7,0,0}
\definecolor{pdflinkcolor}{rgb}{0,0,0.6}
\definecolor{pdfcitecolor}{rgb}{0,0,0.6}
\usepackage[
  colorlinks=true,
  urlcolor=pdfurlcolor,
  filecolor=pdffilecolor,
  linkcolor=pdflinkcolor,
  citecolor=pdfcitecolor,
  raiselinks=true,
  breaklinks,
  verbose,
  hyperindex=true,
  linktocpage=true,
  hyperfootnotes=false,
  bookmarks=true,
  bookmarksopenlevel=3,
  bookmarksopen=true,
  bookmarksnumbered=true,
  bookmarkstype=toc,
  bookmarksdepth=4,
  plainpages=false,
  pageanchor=true,
  pdfstartview=FitH,
  pdfpagemode=UseOutlines,
  pdfpagelabels=true,
  pdfpagelayout=OneColumn,
]{hyperref}

\LoadPackagesNow 

\newcommand{\ifargdef}[3][{}]{\ifthenelse{\equal{#2}{}}{#1}{#3}}

\hypersetup{
	pdfauthor={Martin Genzel and Alexander Stollenwerk},
	pdftitle={A Unified Approach to Uniform Signal Recovery From Non-Linear Observations},
	pdfsubject={Research Article},
	pdfcreator={PDF-LaTeX},
}

\newenvironment{tmpl}
{\begin{enumerate}[label={\textbf{Step~\tmplenum*.}},leftmargin=0em, itemindent=2.5em, align=left]}
	{\end{enumerate}}

\makeatletter
\newcommand*{\tmplenum}[1]{%
	\expandafter\@tmplenum\csname c@#1\endcsname%
}

\newcommand*{\@tmplenum}[1]{%
	$\ifcase#1\or\text{1a}\or\text{1b}\or\text{1c}\or\text{2}\or\text{3}\or\text{4}%
	\else\@ctrerr\fi$%
}
\AddEnumerateCounter{\tmplenum}{\@tmplenum}{1}
\makeatother

\newenvironment{highlight}{\begin{quote}\itshape}{\end{quote}}

\newenvironment{arabiclist}
{\begin{enumerate}[label={(\arabic*)}]}
	{\end{enumerate}}

\newenvironment{rmklist}
{\begin{enumerate}[label={(\arabic*)},itemindent=2em,leftmargin=0em]}
{\end{enumerate}}

\newenvironment{asslist}
{\begin{enumerate}[label={(\alph*)},itemindent=2em,leftmargin=0em]}
	{\end{enumerate}}

\renewcommand{\qedsymbol}{$_\blacksquare$}

\providecommand{\qedhere}{\hfill\qedsymbol}

\newtheoremstyle{claim}
	{\topsep}{\topsep}%
	{\itshape}
	{}
	{}
	{}
	{.5em}
	{{\bfseries\boldmath\thmname{#1} \thmnumber{#2}} \thmnote{(#3)}}

\newtheoremstyle{definition}
	{\topsep}{\topsep}%
	{}
	{}
	{}
	{}
	{.5em}
	{\textbf{\thmname{#1} \thmnumber{#2}} \thmnote{(#3)}}
	
\newtheoremstyle{algorithm}
	{\topsep}{\topsep}%
	{}
	{}
	{\bfseries\boldmath}
	{}
	{\newline}
	{\thmname{#1} \thmnumber{#2} \thmnote{(#3)}}

\mdfdefinestyle{boxed}{innertopmargin =6pt, splittopskip = \topskip, skipbelow=6pt, skipabove=12pt}
\mdfdefinestyle{emphframe}{linecolor=black, innertopmargin = 3pt, splittopskip = \topskip, skipbelow=6pt, skipabove=6pt}

\declaretheorem[style=claim,numberwithin=section]{theorem}

\declaretheorem[style=claim,sibling=theorem]{corollary}
\declaretheorem[style=claim,sibling=theorem]{proposition}
\declaretheorem[style=claim,sibling=theorem]{fact}
\declaretheorem[style=definition,sibling=theorem]{definition}
\declaretheorem[style=definition,sibling=theorem]{assumption}

\declaretheorem[style=definition,sibling=theorem]{problem}

\declaretheorem[style=definition,sibling=theorem,qed=$\Diamond$]{remark}
\declaretheorem[style=definition,sibling=theorem,qed=$\Diamond$]{example}
\declaretheorem[style=algorithm,sibling=theorem,%
	preheadhook={\begin{mdframed}[style=emphframe] \setcounter{mpfootnote}{\value{footnote}}},%
	postfoothook=\end{mdframed}]{experiment}
\declaretheorem[style=algorithm,sibling=theorem,%
	preheadhook={\begin{mdframed}[style=emphframe] \setcounter{mpfootnote}{\value{footnote}}},%
	postfoothook=\end{mdframed}]{algorithm}
\declaretheorem[style=definition,sibling=theorem,%
	preheadhook={\begin{mdframed}[style=boxed] \setcounter{mpfootnote}{\value{footnote}}},%
	postfoothook=\end{mdframed}]{recipe}

\newcommand{\opleft}[1]{\mathopen{}\left#1}
\newcommand{\opright}[1]{\right#1\mathclose{}}
\newcommandx{\braces}[4]{%
\ifstrequal{#3}{normal}{#1#4#2}{%
\ifstrequal{#3}{auto}{\left#1#4\right#2}{%
\ifstrequal{#3}{opauto}{\opleft#1#4\opright#2}{%
#3#1#4#3#2}}}%
}

\newcommand{\N}{\mathbb{N}} 
\newcommand{\Z}{\mathbb{Z}} 
\newcommand{\R}{\mathbb{R}} 
\newcommand{\eps}{\varepsilon} 

\renewcommand{\implies}{\Rightarrow} 
\newcommand{\suchthat}{\ \colon} 

\newcommand{\setcompl}[1]{#1^c} 
\newcommand{\cardinality}[1]{\abs{#1}} 
\newcommand{\union}{\cup} 
\newcommand{\bigunion}{\bigcup} 
\newcommand{\intersec}{\cap} 

\newcommandx{\intvcl}[3][1=normal]{\braces{[}{]}{#1}{#2, #3}} 
\newcommandx{\intvop}[3][1=normal]{\braces{(}{)}{#1}{#2, #3}} 
\newcommandx{\intvclop}[3][1=normal]{\braces{[}{)}{#1}{#2, #3}} 
\newcommandx{\intvopcl}[3][1=normal]{\braces{(}{]}{#1}{#2, #3}} 
\DeclareMathOperator*{\argmin}{argmin} 
\DeclareMathOperator{\sign}{sign}
\newcommandx{\abs}[2][1=normal]{\braces{\lvert}{\rvert}{#1}{#2}} 
\newcommandx{\ceil}[2][1=normal]{\braces{\lceil}{\rceil}{#1}{#2}} 
\newcommandx{\floor}[2][1=normal]{\braces{\lfloor}{\rfloor}{#1}{#2}} 
\newcommandx{\round}[2][1=normal]{\braces{[}{]}{#1}{#2}} 
\newcommandx{\der}[1]{D^{#1}} 
\newcommandx{\gradient}{\nabla} 
\newcommandx{\partder}[4][1={},4={}]{\frac{\partial^{#4} #2}{\partial #3^{#4}}\ifargdef{#1}{\Big|_{#1}}} 
\newcommandx{\integ}[4][1={},2={}]{\int_{#1}^{#2} #3 \, #4} 
\newcommandx{\asympffaster}[2][1=normal]{o\braces{(}{)}{#1}{#2}} 
\newcommandx{\asympfaster}[2][1=normal]{O\braces{(}{)}{#1}{#2}} 
\newcommandx{\asympeq}[2][1=normal]{\Theta\braces{(}{)}{#1}{#2}} 
\newcommandx{\asympsslower}[2][1=normal]{\omega\braces{(}{)}{#1}{#2}} 
\newcommandx{\asympslower}[2][1=normal]{\Omega\braces{(}{)}{#1}{#2}} 

\newcommand{\matr}[1]{\begin{bmatrix} #1 \end{bmatrix}} 
\newcommandx{\norm}[2][1=normal]{\braces{\|}{\|}{#1}{#2}} 
\renewcommandx{\sp}[3][1=normal]{\braces{\langle}{\rangle}{#1}{#2, #3}} 
\newcommandx{\End}[2][2={}]{\mathcal{L}\opleft( #1 \ifargdef{#2}{, #2} \opright)} 
\newcommand{\orthcompl}[1]{{#1}^\perp} 
\DeclareMathOperator{\spann}{\operatorname{span}} 
\newcommand{\T}{\mathsf{T}} 
\renewcommand{\vec}[1]{\boldsymbol{#1}} 
\newcommandx{\opnorm}[2][1=normal]{\norm[#1]{#2}_{2\to 2}} 
\newcommandx{\ball}[2][1={},2={}]{\mathbb{B}_{#1}^{#2}} 
\renewcommand{\S}{\mathbb{S}} 
\newcommand{\I}[1]{\vec{I}_{#1}}
\newcommand{\vnull}{\vec{0}}
\newcommand{\proj}{P}

\newcommandx{\measure}[2][1=normal]{\operatorname{vol}\braces{(}{)}{#1}{#2}} 
\newcommand{\indset}[1]{\chi_{#1}} 
\DeclareMathOperator{\supp}{supp} 
\newcommandx{\Leb}[3][1={},3=normal]{L^{#2}\ifargdef{#1}{\braces{(}{)}{#3}{#1}}{}} 
\newcommandx{\Lebnorm}[4][1=normal,3={2},4={}]{\norm[#1]{#2}_{#3}} 
\renewcommandx{\l}[3][1={},3=normal]{\ell^{#2}\ifargdef{#1}{\braces{(}{)}{#3}{#1}}} 
\newcommandx{\lnorm}[4][1=normal,3={2},4={}]{\norm[#1]{#2}_{#3}} 
\newcommandx{\Smooth}[4][1={},3={},4=normal]{C_{#3}^{#2}\ifargdef{#1}{\braces{(}{)}{#4}{#1}}} 
\newcommandx{\Schwartz}[2][1={},2=normal]{\mathscr{S}\ifargdef{#1}{\braces{(}{)}{#2}{#1}}} 
\newcommandx{\Schwartzpoly}[2][1=normal]{\braces{\langle}{\rangle}{#1}{\abs[#1]{#2}} } 
\newcommandx{\Tempdistr}[2][1={},2=normal]{\mathscr{S}'\ifargdef{#1}{\braces{(}{)}{#2}{#1}}} 
\newcommandx{\distrinp}[3][1=normal]{\braces{\langle}{\rangle}{#1}{#2, #3}} 
\newcommandx{\ft}[3][1=default,2=auto]{
\ifstrequal{#1}{default}{\widehat{#3}}{
\ifstrequal{#1}{long}{{\braces{(}{)}{#2}{#3}}^{\wedge}}{}}} 
\newcommandx{\ift}[3][1=default,2=auto]{
\ifstrequal{#1}{default}{\check{#3}}{
\ifstrequal{#1}{long}{{\braces{(}{)}{#2}{#3}}^{\vee}}{}}} 

\newcommand{\convhull}[1]{\operatorname{conv}(#1)} 
\newcommand{\cone}[1]{\operatorname{cone}(#1)} 
\newcommand{\meanwidth}[2][{}]{w_{#1}(#2)} 
\newcommand{\meanwidthalt}[2][{}]{\tilde{w}_{#1}(#2)} 
\newcommand{\effdim}[2][{}]{w_{#1}^2(#2)} 

\newcommandx{\prob}[2][1={},2=normal]{\mathbb{P}\ifargdef{#1}{\braces{[}{]}{#2}{#1}}}
\newcommandx{\mean}[2][1={},2=normal]{\mathbb{E}\ifargdef{#1}{\braces{[}{]}{#2}{#1}}}
\newcommandx{\var}[2][1={},2=normal]{\mathbb{V}\ifargdef{#1}{\braces{[}{]}{#2}{#1}}}

\newcommand{\distributed}{\sim}
\newcommandx{\Unif}[2][1=normal]{\mathsf{U}\braces{(}{)}{#1}{#2}} 
\newcommandx{\Normdistr}[3][1=normal]{\mathsf{N}\braces{(}{)}{#1}{#2, #3}} 
\newcommandx{\normsubg}[2][1=normal]{\norm[#1]{#2}_{\psi_2}} 

\newcommand{\Y}{\vec{y}} 
\newcommand{\ys}{\tilde{y}} 
\newcommand{\Ys}{\tilde{\vec{y}}} 
\newcommand{\fobs}{f} 
\newcommand{\scalfac}{\mu} 
\newcommand{\fout}{F} 
\newcommand{\noise}{\nu} 
\newcommand{\Noise}{\vec{\nu}} 
\renewcommand{\a}{\vec{a}} 
\newcommand{\A}{\vec{A}} 
\newcommand{\pv}{\vec{z}} 
\newcommand{\pvsolu}{\hat{\vec{z}}} 
\newcommand{\solu}[1]{\hat{#1}}
\newcommand{\grtr}{\vec{x}} 
\newcommand{\tfunc}{T} 
\newcommand{\sset}{K} 
\newcommand{\ssetgen}{H} 
\newcommand{\rset}{\mathcal{X}} 
\newcommand{\rsetgen}{\mathcal{H}} 
\newcommand{\clA}{\mathcal{A}}
\newcommand{\clB}{\mathcal{B}}
\newcommand{\mmcovar}[2][]{\rho_{#1}(#2)} 

\newcommand{\exloss}[1][{}]{\mathcal{E}_{#1}}

\newcommand{\lossemp}[1][{}]{\bar{\mathcal{L}}_{#1}}
\newcommand{\multiplterm}[1][{}]{\mathcal{M}_{#1}}
\newcommand{\multipl}{\xi}
\newcommand{\multiplDia}{r}
\newcommand{\multiplLip}[1][t]{L_{#1}}
\newcommand{\Lip}{\gamma}
\newcommand{\srradius}{\kappa}
\newcommand{\err}{\varepsilon}
\newcommand{\errDia}{\hat{r}}
\newcommand{\errLip}[1][t]{\hat{L}_{#1}}
\newcommand{\advterm}[1][{}]{\mathcal{N}_{#1}}
\newcommand{\quadrterm}[1][{}]{\mathcal{Q}_{#1}}

\newcommand{\vdir}{\vec{v}} 
\newcommand{\vdiralt}{\vec{w}} 
\newcommand{\probsuccess}{u} 
\newcommand{\probset}{\Omega}
\newcommand{\sigmaalg}{\mathsf{A}}

\newcommand{\subgparam}{L} 
\newcommand{\gaussian}{\vec{g}} 
\newcommand{\gaussianuniv}{g} 
\newcommand{\Covmatr}{\vec{\Sigma}} 
\newcommand{\talagfunc}{\gamma_2}
\newcommand{\rad}{R}

\newcommand{\suppset}{\mathcal{S}}
\newcommand{\ttau}{\vec{\tau}}
\newcommand{\covnumber}[2][t]{\mathcal{N}(#2, #1)}

\newcommand{\iset}{\mathcal{I}}
\newcommand{\nout}{m_0}
\newcommand{\tune}{\Delta}

\newcommand{\modulo}{\mathfrak{m}}
\newcommand{\sumf}{f}

\graphicspath{{images/}}

\begin{document}

\renewcommand*{\thefootnote}{\fnsymbol{footnote}}
\pagestyle{scrheadings}

\begin{center}
	\bfseries\larger[3]{A Unified Approach to Uniform Signal Recovery From Non-Linear Observations}
\end{center}

\vspace{1\baselineskip}
\begin{addmargin}[2em]{2em}
\begin{center}
\noindent{\normalsize\textbf{Martin Genzel}\footnote{Utrecht University, Mathematical Institute, Utrecht, Netherlands (e-mail: \href{mailto:m.genzel@uu.nl}{\texttt{m.genzel@uu.nl}}).} \qquad \textbf{Alexander Stollenwerk}\footnote{UCLouvain, ICTEAM Institute, ISPGroup, Louvain-la-Neuve, Belgium (e-mail: \href{mailto:alexander.stollenwerk@uclouvain.be}{\texttt{alexander.stollenwerk@uclouvain.be}}).}}
\end{center}

\vspace{1\baselineskip}
{\smaller
\noindent\textbf{Abstract.}
Recent advances in quantized compressed sensing and high-dimensional estimation have shown that signal recovery is even feasible under strong non-linear distortions in the observation process.
An important characteristic of associated guarantees is uniformity, i.e., recovery succeeds for an entire class of structured signals with a fixed measurement ensemble.
However, despite significant results in various special cases, a general understanding of uniform recovery from non-linear observations is still missing.
This paper develops a unified approach to this problem under the assumption of i.i.d.\ sub-Gaussian measurement vectors.
Our main result shows that a simple least-squares estimator with any convex constraint can serve as a universal recovery strategy, which is outlier robust and does not require explicit knowledge of the underlying non-linearity.
Based on empirical process theory, a key technical novelty is an approximative increment condition that can be implemented for all common types of non-linear models.
This flexibility allows us to apply our approach to a variety of problems in non-linear compressed sensing and high-dimensional statistics, leading to several new and improved guarantees.
Each of these applications is accompanied by a conceptually simple and systematic proof, which does not rely on any deeper properties of the observation model.
On the other hand, known local stability properties can be incorporated into our framework in a plug-and-play manner, thereby implying near-optimal error bounds.

\vspace{.5\baselineskip}

%

}
\end{addmargin}

\newcommand{\shortauthor}{Genzel and Stollenwerk: Uniform Signal Recovery From Non-Linear Observations}
%
\renewcommand*{\thefootnote}{\arabic{footnote}}
\setcounter{footnote}{0}


\thispagestyle{plain}

\section{Introduction}
\label{sec:intro}

This paper is concerned with the following fundamental reconstruction task:
\begin{problem}\label{prob:intro}
	Consider a set of signals $\rset \subset \R^p$ and let $\a_1, \dots, \a_m \in \R^p$ be a collection of measurement vectors.
	Moreover, let $\fout \colon \R^p \times \rset \to \R$ be a scalar output function.
	Under what conditions can the following recovery problem be solved \emph{uniformly} for all $\grtr \in \rset$:
	Assume that $\grtr$ is observed in the form
	\begin{equation}\label{eq:intro:meas}
		y_i \coloneqq \fout(\a_i, \grtr) + \noise_i, \quad i = 1, \dots, m,
	\end{equation}
	where $\noise_1,\dots, \noise_m \in \R$ is scalar noise.
	Given $\{(\a_i, y_i)\}_{i = 1}^m$, is it possible to recover the underlying signal $\grtr$ efficiently?
\end{problem}
The prototypical instance of Problem~\ref{prob:intro} is the approach of \emph{compressed sensing}.
Dating back to the seminal works of Cand\`{e}s, Romberg, Tao, and Donoho \cite{crt06a,crt06b,don06}, the traditional setup of compressed sensing focuses on the case of \emph{noisy linear observations}, i.e., we have $y_i = \sp{\a_i}{\grtr} + \noise_i$ for $i = 1, \dots, m$.
In this regime, Problem~\ref{prob:intro} is nowadays fairly well understood, underpinned by various real-world applications, efficient algorithmic methods, and a rich theoretical foundation; see \cite{fh13} for a comprehensive overview.
In a nutshell, compressed sensing has proven that signal recovery is still feasible when $m \ll p$, supposed that the signal set $\rset$ carries some low-dimensional structure, e.g., a variant of \emph{sparsity}, while the measurement ensemble $\{\a_i\}_{i = 1}^m$ follows an appropriate random design. The \emph{uniformity} over $\rset$ plays an important role in this context, since the measurement device---determined by $\{\a_i\}_{i = 1}^m$ in our case---is typically fixed in applications and should allow for the reconstruction of \emph{all} (or most) signals in $\rset$.
But also apart from this practical relevance, the above quest for uniform recovery is an interesting mathematical problem in its own right. It is significantly more involved than its non-uniform counterpart, where the $\{\a_i\}_{i = 1}^m$ may vary for each $\grtr \in \rset$.

The present work is devoted to Problem~\ref{prob:intro} in the much less explored situation of \emph{non-linear} output functions.
In fact, many conceptions from compressed sensing theory, such as the restricted isometry or nullspace property, are tailored to linear models and do not carry over directly to the non-linear case.
As we will see in the next subsection, the presence of non-linear measuring components is not merely an academic concern but affects many problems in signal processing and high-dimensional statistics.

\subsection{Prior Art}
\label{subsec:intro:literature}

There exist two branches of research that are particularly relevant to this work.
The first one is based in the field of (memoryless) \emph{quantized compressed sensing}, which deals with the fact that analog (linear) measurements often need to be quantized before further processing in practice.
A common scenario in this respect is \emph{$1$-bit compressed sensing} where only a single bit of information is retained, e.g., if \eqref{eq:intro:meas} renders observations of the form $y_i = \sign(\sp{\a_i}{\grtr})$.
Due to this considerable loss of information, it may come as a surprise that tractable recovery methods are still available and Problem~\ref{prob:intro} is relatively well understood in this situation.
A solid theoretical basis as well as efficient algorithms have been developed over the last few years, including significant progress on lower bounds \cite{jlbb13,dm18b}, robustness \cite{pv13b, dm18}, advanced quantization schemes \cite{mjcdd16, bfnpw17, ksw16, xj18,jmps19}, and non-Gaussian measurements \cite{alpv14,djr17,dm18,dm18b}.
While this list of references is certainly incomplete, the surveys of \citeauthor{dir19}~\cite{dir19} and \citeauthor{bjks15}~\cite{bjks15} provide nice overviews of the field of quantized compressed sensing.

The above achievements have also given rise to several important mathematical tools. One of the most notable breakthroughs are quantized embedding results for signal recovery---a highly geometric argument based on uniform, random hyperplane tessellations, e.g., see \cite{jlbb13,pv13b,pv14,or15,dm18}.
Closer to the original ideas of compressed sensing are reconstruction guarantees relying on variants of the restricted isometry property, e.g., see \cite{jc17,fou17,djr17,xj18}.
However, these techniques are strongly tailored to quantized measurements and it remains unclear how they could be extended to other instances of Problem~\ref{prob:intro}.

The second branch of related literature is much less restrictive with respect to the underlying observation model. It allows \eqref{eq:intro:meas} to take the form $y_i = \fobs(\sp{\a_i}{\grtr}) + \noise_i$, where $\fobs \colon \R \to \R$ can be non-linear and random.
A pioneering work on these so-called \emph{single-index models} is the one of \citeauthor{pv16}~\cite{pv16} (inspired by ideas of \citeauthor{bri82}~\cite{bri82}), who study the \emph{generalized Lasso} as reconstruction method:
\begin{equation} \label{eq:intro:klasso}\tag{$\mathsf{P}_{\sset,\Y}$}
	\min_{\pv \in \sset} \ \tfrac{1}{m} \sum_{i = 1}^m (y_i - \sp{\a_i}{\pv})^2.
\end{equation}
Here, $\sset \subset \R^p$ is a \emph{convex} constraint set, serving as an appropriate relaxation of the actual signal set $\rset$ and making \eqref{eq:intro:klasso} tractable in many situations of interest.
Although this ``linearization'' strategy might appear very coarse, it was shown to produce satisfactory outcomes for Gaussian measurements, even when $\fobs$ is highly non-linear.
A key benefit of \eqref{eq:intro:klasso} is that it does not require any (explicit) knowledge of the observation model, which enables various applications to signal processing and statistical estimation, e.g., see \cite[Chap.~3~\&~4]{gen19}.
The article of \citeauthor{pv16}~\cite{pv16} is just an example of a whole line of research with many related and follow-up works, e.g., see \cite{pvy16,os16,tah15,gen16,gmw18,gs20,so19,tr18}; remarkably, this approach also extends to phase-retrieval-like problems where $\fobs$ is an even function \cite{ybwl17,tr19,gk20}. 
For a more detailed discussion of the literature, we refer to \cite[Sec.~6]{pvy16} and \cite[Sec.~4.2]{gen19}.

At first sight, the aforementioned works seem to provide a general solution to Problem~\ref{prob:intro}, but they lack a crucial feature, namely uniformity.
Indeed, most of these results are based on concentration inequalities over high-dimensional signal sets, exploiting their ``local geometry'' around a fixed $\grtr \in \rset$. This strategy naturally leads to \emph{non-uniform} recovery guarantees, and we are not aware of a uniform extension in that regard.
Finally, we point out that the two research areas discussed above are not independent but exhibit certain overlaps in terms of their main achievements and proof techniques, for instance, see \cite{gs20,tr18}.

\subsection{Contributions and Overview}
\label{subsec:intro:contrib}

The primary objective of this work is to fill the striking gap between the lines of research discussed in the previous subsection.
On the one hand, we show that uniform recovery is not only limited to linear and quantized measurement schemes but applies to a much larger family of observation models.
On the other hand, it turns out that under mild assumptions, known non-uniform guarantees for single-index and related models naturally extend to the uniform case.
Apart from a unification, our theoretical approach will contribute to each of these research directions through new and improved results.

On the conceptual side, we stick to the methodology suggested by the second branch of literature from Subsection~\ref{subsec:intro:literature}, analyzing the performance of the generalized Lasso \eqref{eq:intro:klasso}.
Our main result, Theorem~\ref{thm:results:main} in Subsection~\ref{subsec:results:guarantee}, establishes a uniform error bound for \eqref{eq:intro:klasso} under very general conditions, including all observation models mentioned above.
This finding implies that there exists a ``universal'' recovery strategy, which often yields satisfactory outcomes.
In this context, it is worth pointing out that we have focused on \eqref{eq:intro:klasso} mainly because of its simplicity and popularity, but similar results could be shown for other reconstruction methods as well (cf.~Remark~\ref{rmk:results:exten}\ref{rmk:results:exten:metrics}).
Therefore, the present paper is a contribution to Problem~\ref{prob:intro} in the first place, whereas the estimator of choice plays a subordinate role.\footnote{In particular, we will not discuss any details about possible algorithmic implementations of \eqref{eq:intro:klasso}, which is an important subject in its own right.}

The recovery guarantee of Theorem~\ref{thm:results:main} is formulated in a quite abstract setting, which requires some technical preparation.
In order to familiarize the reader with the results of this paper, we now state a special case for single-index models with a Lipschitz continuous output function and Gaussian measurements (see Subsection~\ref{subsec:app:appl:sim} for a proof).
Note that parts of the notation from Subsection~\ref{subsec:intro:notation} is already used here; the complexity parameters $\meanwidth[t]{\cdot}$ and $\meanwidth{\cdot}$ correspond to the common notion of (local) mean width, which is formally introduced in Definition~\ref{def:results:meanwidth}.
\begin{theorem}\label{thm:intro:sim}
	There exist universal constants $c, C > 0$ for which the following holds.
	
	Let $\a_1, \dots, \a_m \in \R^p$ be independent copies of a standard Gaussian random vector $\a \distributed \Normdistr{\vnull}{\I{p}}$.
	Let $\fobs \colon \R \to \R$ be $\Lip$-Lipschitz, and for $\gaussianuniv \distributed \Normdistr{0}{1}$, we set
	\begin{equation}
		\scalfac \coloneqq \mean[\fobs(\gaussianuniv)\gaussianuniv] \quad \text{and} \quad \multiplDia \coloneqq \normsubg{\fobs(\gaussianuniv) - \scalfac\gaussianuniv}.
	\end{equation}
	Moreover, let $\rset \subset \S^{p-1}$ and let $\sset \subset \R^p$ be a convex set such that $\scalfac\rset \subset \sset$.
	For $\probsuccess \geq 1$ and a desired reconstruction accuracy $t \geq 0$, we assume that
	\begin{equation} \label{eq:intro:sim:m}
		m \geq C \cdot \Big( (1 + t^{-2} \multiplDia^2 ) \cdot \big(\effdim[t]{\sset - \scalfac\rset}+ \probsuccess^2 \big) + t^{-2} (\scalfac + \Lip)^2 \cdot \effdim{\rset} \Big),
	\end{equation}
	where $\sset - \scalfac\rset \coloneqq \{ \pv - \scalfac \grtr \suchthat \pv \in \sset, \grtr \in \rset \}$.
	Then with probability at least $1 - \exp(-c \probsuccess^2)$ on the random draw of $\{\a_i\}_{i = 1}^m$, the following holds uniformly for all $\grtr \in \rset$:
	Let $\Y = (y_1, \dots, y_m) \in \R^m$ be given by 
	\begin{equation}\label{eq:intro:sim:meas}
	y_i = \fobs(\sp{\a_i}{\grtr}) + \noise_i, \quad i = 1, \dots, m,
	\end{equation}
	such that $\big(\tfrac{1}{m}\sum_{i = 1}^m \noise_i^2 \big)^{1/2} \leq \tfrac{t}{20}$.\footnote{Here, the $\noise_i$ correspond to \emph{adversarial noise}, i.e., as long as the $\l{2}$-constraint is satisfied, they could be arbitrary (even worst-case) perturbations. This model is very common in uniform recovery and compressed sensing, cf.~\cite{fh13}. But note that compared to statistical noise, this comes at the price of consistent estimation, i.e., the reconstruction error may not become arbitrarily small if $m \to \infty$; see Remark~\ref{rmk:results:noise} and~\ref{rmk:unif}\ref{rmk:unif:noise} for further discussion.} Then every minimizer $\pvsolu$ of \eqref{eq:results:klasso} satisfies $\lnorm{\pvsolu - \scalfac\grtr} \leq t$.
\end{theorem}
Theorem~\ref{thm:intro:sim} can be seen as a uniform version of Plan's and Vershynin's main result on single-index models \cite[Thm.~1.9]{pv16}.
Most notably, the sampling-rate condition \eqref{eq:intro:sim:m} implies that the dependence on the desired accuracy $t$ is the same as in the non-uniform case.
For more details on the interplay between $m$, $t$, and the complexity parameters, we refer to the discussion of Theorem~\ref{thm:results:main} in Subsection~\ref{subsec:results:guarantee}.
To the best of our knowledge, Theorem~\ref{thm:intro:sim} is a new result in its own right, and in particular, it is not a consequence of the approaches outlined in the previous subsection.
A novel argument is required to prove such a type of recovery guarantee; see Subsection~\ref{subsec:intro:techn} below for an overview of the major technical challenge of our work.

The main result of Theorem~\ref{thm:results:main} extends Theorem~\ref{thm:intro:sim} by several important aspects, being in line with our key achievements:
\begin{arabiclist}
\item
	\emph{Discontinuous output functions.} The Lipschitz assumption in Theorem~\ref{thm:intro:sim} prohibits popular models like quantized observations. It turns out that an extension in that regard is subtle. To this end, we introduce an \emph{approximative increment condition} on the observation variable, which is satisfied for many discontinuous non-linearities; see Subsection~\ref{subsec:results:increments} for more details and Subsection~\ref{subsec:appl:1bit}--\ref{subsec:appl:sim} for applications to non-linear compressed sensing.
\item
	\emph{Beyond Gaussian measurements.} Compared to the previous point, it is relatively straightforward to allow for i.i.d.\ sub-Gaussian measurement vectors in Theorem~\ref{thm:intro:sim}.
	However, this relaxation leads to an additional constraint term, called the \emph{target mismatch}, whose size depends on the distribution of the measurements and the output function; see Definition~\ref{def:results:mmcovar} in Subsection~\ref{subsec:results:setup}.
\item
	\emph{Beyond single-index models.} In Section~\ref{sec:results}, we study the more general situation of Problem~\ref{prob:intro}, where~$y_i$ is not necessarily representable in terms of the dot product $\sp{\a_i}{\grtr}$.
	While this requires an additional step of abstraction, it enables for more complicated non-linear models as well as refined recovery tasks, e.g., support estimation (see Subsection~\ref{subsec:appl:vs}).
	For this purpose, we introduce so-called \emph{target functions} in Subsection~\ref{subsec:results:setup}, transforming a signal $\grtr \in \rset$ in such a way that it becomes compatible with the solution vector of \eqref{eq:intro:klasso} under observations of the form \eqref{eq:intro:meas}. Note that in the setup of Theorem~\ref{thm:intro:sim}, this transformation simply corresponds to rescaling $\grtr$ by the scalar factor~$\scalfac$.
\item
	\emph{Outlier robustness.} The $\l{2}$-noise constraint $\big(\tfrac{1}{m}\sum_{i = 1}^m \noise_i^2 \big)^{1/2} \leq \tfrac{t}{20}$ in Theorem~\ref{thm:intro:sim} is standard for linear models and also remains useful in the situation of Lipschitz non-linearities. However, such a condition is overly restrictive for quantized measurement schemes, where the noise is due to (few) wrong bits instead of (small) real-valued perturbations.
	Theorem~\ref{thm:results:main} therefore involves an important relaxation in this respect, permitting a portion of ``gross'' outliers in the observation vector.
\end{arabiclist}
Based on our findings in Theorem~\ref{thm:results:main}, we will also address two points of independent interest:
\begin{arabiclist}\setcounter{enumi}{4}
\item
	\emph{Uniform vs.~non-uniform recovery.} Section~\ref{sec:unif} isolates characteristics that are specifically attributable to uniformity.
	While the presence of non-linear output functions may only affect the oversampling rate (see also Section~\ref{sec:appl}), our focus here is on the quite implicit complexity parameter $\meanwidth[t]{\sset - \rset}$, cf.~\eqref{eq:intro:sim:m}.
	We show that it can be always bounded by two much more informative expressions, namely a ``fully localized'' mean width and its global counterpart (see Proposition~\ref{prop:unif:mwloc}).
	This important simplification enables us to reuse known (non-uniform) results from the literature.
	We illustrate for two examples ($\l{1}$-constraints and total variation) that there is no considerable gap between the uniform and non-uniform regime.
\item
	\emph{Plug-and-play via embeddings.} Section~\ref{sec:appl:noincr} demonstrates how to incorporate known random embedding results into our theoretical framework. 
	In this way, for instance, it is possible to obtain near-optimal error decay rates for \eqref{eq:results:klasso} with quantized measurements. 
	This achievement is due to a more general guarantee stated in Theorem~\ref{thm:appl:without_increments}.
\end{arabiclist}

The general purpose of our methodology is to enable a more systematic study of Problem~\ref{prob:intro} than before, especially for non-linear observation models.
To a certain degree, this paper promotes an alternative path to compressed sensing theory that can do without common tools like the restricted isometry property or quantized embedding results.
Despite clear overlaps, each of these approaches comes along with its own strengths and (in-)accessible regimes; see Section~\ref{sec:conclusion} for a more detailed discussion.

\subsection{Technical Challenges}
\label{subsec:intro:techn}

This subsection highlights the main technical achievements of this work.
Our principal procedure is in fact relatively standard: we decompose the excess loss associated with the estimator \eqref{eq:intro:klasso} into separate empirical processes and aim at suitable lower/upper bounds for each (cf.~\cite{men15,pv16,dm18,gen19,gk20}); see Subsection~\ref{subsec:results:increments} for an overview of this argument.

However, the quest for \emph{uniform} recovery as prescribed by Problem~\ref{prob:intro} makes this approach a challenging endeavor.
In a nutshell, the key difficulty is that the common multiplier term turns into an \emph{empirical product process} of the following type:
\begin{equation}\label{eq:intro:techn:prod}
	\tfrac{1}{m} \sum_{i=1}^m \multipl_{i}(\grtr) \cdot \sp{\a_i}{\vdir}, \qquad \text{$\grtr \in \rset$,\quad $\vdir \in (\sset - \tfunc\grtr) \intersec t\S^{p-1}$,}
\end{equation}
where the first factor takes the form $\multipl_{i}(\grtr) = \sp{\a_i}{\tfunc\grtr} - \fout(\a_i, \grtr)$ and $\tfunc \colon \rset \to \sset$ is a fixed function (think of a scalar multiplication for now).
Since we require a uniform upper bound for \eqref{eq:intro:techn:prod} over both $\grtr$ and $\vdir$, the regularity of the stochastic process $\{\multipl_{i}(\grtr)\}_{\grtr \in \rset}$ plays an important role, in the sense that it should fulfill a sub-Gaussian increment condition (see \eqref{eq:intro:incr}).\footnote{This is an important difference to the non-uniform regime, where $\grtr$ is a fixed signal. In this case, it is possible to apply known multiplier concentrations inequalities to control \eqref{eq:intro:techn:prod}, e.g., see \cite[Thm.~1.9]{men16}.}
Such a regularity assumption unfortunately does not even hold for standard models like $1$-bit quantization $\fout(\a_i, \grtr) = \sign(\sp{\a_i}{\grtr})$.
Therefore, existing chaining-based results for product processes are not directly applicable to our situation.
This may also explain why there is no unified theory for non-linear observations available so far and known uniform recovery guarantees are restricted to special cases (see Subsection~\ref{subsec:intro:literature}).

The present article provides a general remedy for this foundational problem.
Our starting point is that a given output function $\fout$ can be often approximated by a more regular one, say $\fout_t$, where $t > 0$ controls the approximation accuracy.
As such, this strategy can yield a more benign multiplier $\multipl_{t,i}(\grtr) = \sp{\a_i}{\tfunc\grtr} -\nobreak \fout_t(\a_i, \grtr)$, but it simply shifts the original problem to the resulting approximation error $\fout(\a_i, \grtr) - \fout_t(\a_i, \grtr)$.
Hence, a fundamental insight in our proof (see Step~\hyperref[step:multipl]{3} in Section~\ref{sec:proofs}) is that it suffices to consider the \emph{absolute} error $\err_{t,i} = \abs{\fout(\a_i, \grtr) - \fout_t(\a_i, \grtr)}$ instead.
Indeed, the process $\{\err_{t,i}(\grtr)\}_{\grtr \in \rset}$ fulfills an increment condition in all relevant model situations that we are aware of; see Figure~\ref{fig:appl:1bit} in Section~\ref{sec:appl} for an illustration of the above argument based on the sign-function.
With this at hand, and apart from other technicalities, we can take advantage of a recent concentration inequality due to \citeauthor{men16}~\cite[Thm.~1.13]{men16}.

The intuitive idea described in the previous paragraph is placed into a rigorous framework by means of our central Assumption~\ref{ass:results:incr}.
In this context, it is worth noting that the actual main result, Theorem~\ref{thm:results:main}, couples the approximation accuracy $t$ with the desired reconstruction error.
This might appear somewhat peculiar at first sight because the processes $\{\multipl_{t,i}(\grtr)\}_{\grtr \in \rset}$ and $\{\err_{t,i}(\grtr)\}_{\grtr \in \rset}$ usually become less regular with $t$ decreasing.
However, we demonstrate that this trade-off can be balanced out well and only affects the achievable oversampling rate.

Finally, our general technique to control empirical product processes of the form \eqref{eq:intro:techn:prod} may find application beyond signal recovery problems.
For example, it could be also used to derive a new generation of non-linear embedding results.
In fact, one of the earliest works on binary random embeddings \cite{pv14} has pointed out the usefulness of approximating non-linearities, though strongly tailored to $1$-bit quantization.
With our systematic approach, their result could not only be reproduced but further improved (cf.~\cite[Thm.~3.7]{sto19}).

\subsection{Overview and Notation}
\label{subsec:intro:notation}

The rest of this article is organized as follows. Section~\ref{sec:results} presents our findings in full generality, where the proof of Theorem~\ref{thm:results:main} is postponed to  Section~\ref{sec:proofs}. 
In Section~\ref{sec:unif}, we elaborate aspects that are specifically due to uniform recovery and highlight potential gaps to the non-uniform regime.
Section~\ref{sec:appl} and~\ref{sec:appl:noincr} are then devoted to applications and concrete examples, including a series of new guarantees for non-linear compressed sensing and high-dimensional estimation; the corresponding proofs are provided in Section~\ref{sec:app:appl} and~\ref{sec:app:appl:noincr}, respectively. Our concluding discussion can be found in Section~\ref{sec:conclusion}.

Before proceeding, let us fix some standard notations and conventions that are commonly used in this work.
The letters $c$ and $C$ are reserved for (positive) constants, whose values could change from time to time. We speak of a \emph{universal constant} if its value does not depend on any other involved parameter.
If an inequality holds up to a universal constant $C$, we usually write $A \lesssim B$ instead of $A \leq C \cdot B$. The notation $A \asymp B$ is a shortcut for $A \lesssim B \lesssim A$.

For $p \in \N$, we set $[p] \coloneqq \{1, \dots, p\}$. The \emph{cardinality} of an index set $\suppset \subset [p]$ is denoted by~$\cardinality{\suppset}$ and its \emph{set complement} in $[p]$ is $\setcompl{\suppset} \coloneqq [p] \setminus \suppset$. Vectors and matrices are denoted by lower- and uppercase boldface letters, respectively. The \mbox{$j$-th} entry of a vector $\vdir \in \R^p$ is denoted by $(\vdir)_j$, or simply by $v_j$ if there is no danger of confusion. The \emph{support} of $\vdir \in \R^p$ is defined by $\supp(\vdir) \coloneqq \{ j \in [p] \suchthat v_j \neq 0 \}$ and $\lnorm{\vdir}[0] \coloneqq \cardinality{\supp(\vdir)}$ denotes its \emph{sparsity}.
We write $\I{p} \in \R^{p \times p}$ and $\vnull \in \R^p$ for the \emph{identity matrix} and the \emph{zero vector} in $\R^p$, respectively. 
For $1 \leq q \leq \infty$, we denote the \emph{$\l{q}$-norm} on $\R^p$ by $\lnorm{\cdot}[q]$ and the associated \emph{unit ball} by $\ball[q][p]$.
The \emph{Euclidean unit sphere} is given by $\S^{p-1} \coloneqq \{ \vdir \in \R^p \suchthat \lnorm{\vdir}[2] = 1 \}$.
The spectral matrix norm is denoted by $\opnorm{\cdot}$.

Let $\ssetgen, \ssetgen' \subset \R^p$ and $\vdir \in \R^p$.
We write $\spann(\ssetgen)$ for the \emph{linear hull} of~$\ssetgen$ and the \emph{linear cone} generated by~$\ssetgen$ (not necessarily convex) is denoted by $\cone{\ssetgen} \coloneqq \{s \tilde{\vdir} \suchthat \tilde{\vdir} \in \ssetgen, s \geq 0 \}$.
By $\proj_{\ssetgen} \colon \R^p \to \R^p$, we denote the \emph{Euclidean projection} onto the closure of $\ssetgen$ if it is well-defined, and we use the shortcut $\proj_{\vdir} \coloneqq \proj_{\spann(\{\vdir\})}$.
We write $\orthcompl{\ssetgen}$ for the \emph{orthogonal complement} of $\ssetgen$.
The \emph{Minkowski difference} between $\ssetgen$ and $\ssetgen'$ is defined by $\ssetgen - \ssetgen' \coloneqq \{ \vdir_1 -\nobreak \vdir_2 \suchthat \vdir_1 \in \ssetgen, \vdir_2 \in \ssetgen' \}$, and we use the shortcut $\ssetgen - \vdir \coloneqq \ssetgen - \{\vdir\}$. For $\eps > 0$, we denote the \emph{covering number} of $\ssetgen$ at scale $\eps$ with respect to the Euclidean norm by $\covnumber[\eps]{\ssetgen}$.

The \emph{$L^q$-norm} of a real-valued random variable $a$ is given by $\norm{a}_{L^q} \coloneqq (\mean[\abs{a}^q])^{1/q}$. We call $a$ \emph{sub-Gaussian} if 
\begin{equation}
	\normsubg{a} \coloneqq \inf\big\{v > 0 \suchthat \mean[\exp(\abs{a}^2 / v^2)] \leq 2 \big\} < \infty,
\end{equation}
and $\normsubg{\cdot}$ is called the \emph{sub-Gaussian norm}. A random vector~$\a$ in $\R^p$ is called \emph{sub-Gaussian} if $\normsubg{\a} \coloneqq \sup_{\vdir \in \S^{p-1}} \normsubg{\sp{\a}{\vdir}} < \infty$.
We say that $\a$ is \emph{centered} if $\mean[\a] = \vnull$, and it is \emph{isotropic} if $\mean[\a \a^\T] = \I{p}$.
We write $\a \distributed \Normdistr{\vnull}{\I{p}}$ if $\a$ is a \emph{standard Gaussian random vector} in $\R^p$.
For a more detailed introduction to sub-Gaussian random variables and their properties, see \cite[Chap.~2~\&~3]{ver18}.
Now, let $(\rsetgen,d)$ be a pseudo-metric space and consider a real-valued stochastic process $\{a_h\}_{h \in \rsetgen}$ on $\rsetgen$.
Then $\{a_h\}_{h \in \rsetgen}$ has \emph{sub-Gaussian increments} with respect to $d$ if 
\begin{equation}\label{eq:intro:incr}
	\normsubg{a_h - a_{h'}} \leq d(h, h') \qquad \text{for all $h, h' \in \rsetgen$.}
\end{equation}

We say that a function $\fobs\colon \R^p \to \R^{p'}$ is \emph{$\Lip$-Lipschitz} if it is Lipschitz continuous with respect to the Euclidean metric and a Lipschitz constant $\Lip \geq 0$. The \emph{sign} of $v \in \R$ is denoted by $\sign(v)$, with the convention that $\sign(0) \coloneqq 1$. If $\sign(\cdot)$ is applied to a vector, the operation is understood entrywise. The \emph{ceiling} and \emph{floor function} of $v \in \R$ is denoted by $\ceil{v}$ and $\floor{v}$, respectively.

\section{Main Result}
\label{sec:results}

As preliminary steps, we introduce the formal model setup in Subsection~\ref{subsec:results:setup}, followed by our increment conditions on the observation variable in Subsection~\ref{subsec:results:increments}. The main result of Theorem~\ref{thm:results:main} is then stated and discussed in Subsection~\ref{subsec:results:guarantee}, while its proof can be found in Section~\ref{sec:proofs}.

\subsection{Model Setup}
\label{subsec:results:setup}

The following model assumption fixes the notation for the remainder of this section and specifies the instance of Problem~\ref{prob:intro} that we will study from now on:
\begin{assumption}
	\begin{asslist}
	\item\label{ass:results:meas:stat}
		Components of the observation model:
		\begin{itemize}
			\item
			\emph{Measurement vector:} a centered, isotropic, sub-Gaussian random vector $\a \in \R^p$ such that $\normsubg{\a} \leq \subgparam$ for some $\subgparam > 0$.\footnote{Note that we actually have that $\subgparam \geq \sqrt{1/\log(2)} > 1$ if $\a$ is isotropic, see \cite{jlpy20}.}
			\item
			\emph{Signal set:} a set $\rset$ (not necessarily a subset of $\R^p$).
			\item
			\emph{Output function:} a scalar function $\fout \colon \R^p \times \rset \to \R$ that may be random (not necessarily independent of $\a$).
			\item
			\emph{Observation variable:} $\ys(\grtr) \coloneqq \fout(\a, \grtr)$ for $\grtr \in \rset$.
		\end{itemize}
	\item\label{ass:results:meas:process}
		Components of the measurement process:
		\begin{itemize}
			\item
			\emph{Measurement ensemble:} $\{(\a_i, \fout_i)\}_{i = 1}^m$ are independent copies of the \emph{measurement model} $(\a, \fout)$.
			\item
			\emph{Observations:} $\ys_i(\grtr) \coloneqq \fout_i(\a_i, \grtr)$ for $i = 1, \dots, m$ and $\grtr \in \rset$. The \emph{observation vector} of~$\grtr$ is denoted by $\Ys(\grtr) \coloneqq (\ys_1(\grtr), \dots, \ys_m(\grtr)) \in \R^m$.
		\end{itemize}
	\item\label{ass:results:meas:recov}
		Components of the recovery procedure:
		\begin{itemize}
			\item
			\emph{Constraint set:} a convex subset $\sset \subset \R^p$ (not necessarily bounded).
			\item
			\emph{Target function:} a map $\tfunc \colon \rset \to \sset$, such that $\tfunc\rset \subset \sset$ is bounded.
		\end{itemize}
	\end{asslist} \label{ass:results:meas}
\end{assumption}
Part~\ref{ass:results:meas:stat} of Assumption~\ref{ass:results:meas} establishes the statistical ``template'' of our measurement model, from which i.i.d.\ observations are drawn according to part~\ref{ass:results:meas:process}.
Importantly, this defines an entire class of observation variables $\{\ys(\grtr)\}_{\grtr \in \rset}$, which corresponds to a real-valued stochastic process.
Assumption~\ref{ass:results:meas}\ref{ass:results:meas:recov} is arguably the most abstract part, but forms a crucial ingredient of our main result:
Theorem~\ref{thm:results:main} states an error bound for $\tfunc\grtr \in \sset$ instead of the actual signal $\grtr \in \rset$ (which might not even belong to $\R^p$).
The basic linearization strategy behind \eqref{eq:intro:klasso} typically calls for an appropriate transformation of the signal domain. A good example is the special case of Theorem~\ref{thm:intro:sim}, where $\tfunc$ amounts to a rescaling of $\grtr$ by the model-dependent factor~$\scalfac$.
Such a simple correction is sufficient for most applications considered in Section~\ref{sec:appl}; but there exist more complicated situations, such as the variable selection problem in Subsection~\ref{subsec:appl:vs}, where $\rset$ represents support sets on $[p]$.
At the present level of abstraction, it is useful to think of $\tfunc\grtr$ as a parameterized version of $\grtr \in \rset$ that is compatible with the estimation procedure of \eqref{eq:intro:klasso} and its constraint set $\sset \subset \R^p$.\footnote{One should bear in mind that the target function $\tfunc$ is a purely \emph{theoretical} object that does not affect the solution of \eqref{eq:intro:klasso} and may be (partially) unknown in practice. Nevertheless, it is an essential analysis tool for Problem~\ref{prob:intro}, relating the underlying observation model to the actual recovery method \eqref{eq:intro:klasso}, see also Remark~\ref{rmk:results:main}\ref{rmk:results:main:tfunc}.}

In principle, $\tfunc$ can be arbitrary in Assumption~\ref{ass:results:meas}, but there often exists a canonical choice that is driven by the size of the following model parameter:
\begin{definition}\label{def:results:mmcovar}
	Let Assumption~\ref{ass:results:meas} be satisfied. 
	Then we define the \emph{target mismatch} of $\grtr \in \rset$ by
	\begin{equation}
	\mmcovar{\grtr} \coloneqq \lnorm[\big]{\mean[\ys(\grtr)\a] - \tfunc\grtr}.
	\end{equation}
\end{definition}
The meaning of this definition becomes clearer when taking the perspective of statistical learning for a moment.
Assuming that $y_i = \ys_i(\grtr)$, the Lasso-estimator \eqref{eq:intro:klasso} can be viewed as an empirical loss minimization problem with ``data'' $\{(\a_i, y_i)\}_{i = 1}^m$.
A central question is then under what conditions its solution can approximate (in a proper sense) the associated expected loss minimization problem (obtained in the infinite sample limit $m \to \infty$):
\begin{equation}\label{eq:results:expmin}
	\min_{\pv \in \sset} \ \mean[(\ys(\grtr) - \sp{\a}{\pv})^2].
\end{equation}
It is not hard to see that the only critical point of the objective function in \eqref{eq:results:expmin} is the vector $\grtr^\ast \coloneqq \mean[\ys(\grtr)\a]$, and if $\grtr^\ast \in \sset$, this is the global optimum.
Therefore, and as its name suggests, $\mmcovar{\grtr}$ measures the mismatch between the target vector $\tfunc\grtr$ and the (global) expected loss minimizer $\grtr^\ast$.

In the context of Theorem~\ref{thm:results:main}, the target mismatch can be seen as an upper bound for the asymptotic error of estimating $\tfunc\grtr$ via \eqref{eq:results:klasso}.
Consequently, a general rule of thumb is to select $\tfunc\grtr$ such that $\mmcovar{\grtr}$ vanishes or becomes sufficiently small.
However, simply setting $\tfunc\grtr \coloneqq \mean[\ys(\grtr)\a]$ (if contained in $\sset$) is not necessarily consistent with our wish for signal recovery, e.g., in the application in Subsection~\ref{subsec:appl:1bitdither}.
Therefore, we have left $\tfunc$ unspecified in Assumption~\ref{ass:results:meas}, even though a proper choice will be obvious for all considered examples.

We close this subsection with a short remark about the noise model considered in this work:
\begin{remark}[Noise model]\label{rmk:results:noise}
	We will adopt the \emph{adversarial} (or \emph{worst-case}) noise model, which is the common setting in the field of compressed sensing, cf.~\cite{fh13}.
	More specifically, when interested in recovery of $\grtr \in \rset$, the actual input $\Y = (y_1, \dots, y_m)$ for \eqref{eq:intro:klasso} need not exactly correspond to the observation vector $\Ys(\grtr) = (\ys_1(\grtr), \dots, \ys_m(\grtr))$ introduced in Assumption~\ref{ass:results:meas}\ref{ass:results:meas:process}.\footnote{The use of the modifier `$\sim$' in Assumption~\ref{ass:results:meas} is also due to this fact and may help to distinguish between our observation model and the actual input for \eqref{eq:intro:klasso}.}
	Instead, arbitrary perturbations are allowed as long as $d(\Y, \Ys(\grtr)) \lesssim t$ for an appropriate noise metric $d(\cdot,\cdot)$ and a desired accuracy~$t$.
	In Theorem~\ref{thm:intro:sim}, for instance, we have $\ys_i(\grtr) = \fobs(\sp{\a_i}{\grtr})$ and $y_i = \ys_i(\grtr) + \noise_i$, while~$d(\cdot,\cdot)$ corresponds to the normalized $\l{2}$-norm.
	Thus, our overall goal is to show that, given $\{(\a_i, \fout_i)\}_{i = 1}^m$, recovery is possible for \emph{any} $\grtr \in \rset$ and  \emph{any} moderate perturbation of its observation~$\Ys(\grtr)$.
	
	In this context, it is worth noting that Assumption~\ref{ass:results:meas}\ref{ass:results:meas:process} does not require the variables $\ys_i(\grtr)$ to be deterministic when conditioned on~$\a_i$, since $\fout_i$ itself may be random.
	For example, one could model \emph{additive random noise} in that way, as common in statistical estimation theory.
	In principle, such a type of noise is also admissible in our main result, Theorem~\ref{thm:results:main}, but it is not entirely compatible with the aforementioned goal of \emph{uniform} recovery.
	Indeed, the ensemble $\{(\a_i, \fout_i)\}_{i = 1}^m$ is only drawn once, so that a single noise realization has to be considered for all $\grtr \in \rset$.
	We therefore mostly stick to the adversarial perspective; see Remark~\ref{rmk:unif}\ref{rmk:unif:noise} in Section~\ref{sec:unif} for a more detailed comparison of both noise models in the situation of linear measurements.
	Finally, we point out that randomized output functions are nevertheless very useful in the uniform regime, such as in the design of dithering variables (see Subsection~\ref{subsec:appl:1bitdither}).
\end{remark}

\subsection{Analysis Strategy and Increment Conditions}
\label{subsec:results:increments}

From now on, let Assumption~\ref{ass:results:meas} be satisfied. In this subsection, we introduce the key condition for our uniform recovery guarantee, namely that the class of observation variables $\{\ys(\grtr)\}_{\grtr \in \rset}$, or at least an approximation thereof, has sub-Gaussian increments.
To better understand the relevance of this condition, it is insightful to first take a closer look at our basic analysis strategy for the generalized Lasso:
\begin{equation} \label{eq:results:klasso}\tag{$\mathsf{P}_{\sset,\Y}$}
\min_{\pv \in \sset} \ \tfrac{1}{m} \sum_{i = 1}^m (y_i - \sp{\a_i}{\pv})^2.
\end{equation}
\newcommand{\refklasso}[1]{(\hyperref[eq:results:klasso]{$\mathsf{P}_{#1}$})}%
Note that we hide the dependency on the measurement vectors $\{\a_i\}_{i = 1}^m$ when referring to \eqref{eq:results:klasso}, as this will be always clear from the context.
At the present stage, the vector $\Y = (y_1, \dots, y_m) \in \R^m$ in \eqref{eq:results:klasso} is unspecified, but taking the viewpoint of our main result, Theorem~\ref{thm:results:main}, is already useful: if $\grtr \in \rset$ is to be recovered, then $\Y$ is a noisy version of the observation vector $\Ys(\grtr) = (\ys_1(\grtr), \dots, \ys_m(\grtr))$ as defined in Assumption~\ref{ass:results:meas}\ref{ass:results:meas:process}; see also Remark~\ref{rmk:results:noise}.

We now derive a simple, yet important criterion for an error bound for a (fixed) signal $\grtr \in \rset$.
To this end, let $\lossemp[\grtr](\vdir) \coloneqq \tfrac{1}{m} \sum_{i = 1}^m (y_i - \sp{\a_i}{\vdir + \tfunc\grtr})^2$ denote the \emph{empirical loss} of $\vdir \in \R^p$ over $\grtr$. 
For the sake of notational convenience, we have fixed the anchor point $\tfunc\grtr$ here, so that \eqref{eq:results:klasso} is equivalent to minimizing $\lossemp[\grtr](\cdot)$ over $\sset - \tfunc\grtr$.
The \emph{excess loss} of $\vdir \in \R^p$ over $\grtr$ is then defined by
\begin{equation}
	\exloss[\grtr](\vdir) \coloneqq \lossemp[\grtr](\vdir) - \lossemp[\grtr](\vnull).
\end{equation}
It measures how much the empirical loss is changing when traveling from $\tfunc\grtr$ in the direction of~$\vdir$.
The following fact is an immediate consequence of the convexity of $\sset$ and $\lossemp[\grtr]$; see Figure~\ref{fig:results:fact} for an illustration.
\begin{figure}[!t]
	\centering
	\begin{tikzpicture}[scale=2]
		\coordinate (K1) at (-.5,-.3);
		\coordinate (K2) at (.5,-1);
		\coordinate (K3) at (.3,-1.7);
		\coordinate (K4) at (-.2,-2.1);
		\coordinate (K5) at (-1,-1.2);
		\coordinate[below right=.35cm and .15 of K1] (muX0);
		
		\draw[fill=gray!20!white, name path = K] (K1) -- (K2) -- (K3) -- (K4) -- (K5) -- cycle;
		\node[draw,circle,minimum size=1.4cm] at (muX0) {}; 
		\begin{scope}
		\clip (K1) -- (K2) -- (K3) -- (K4) -- (K5) -- cycle;
		\node[draw=red,circle,fill=gray!60!white,minimum size=1.4cm,ultra thick] at (muX0) {};
		\end{scope}
		\draw[thick] (K1) -- (K2) -- (K3) -- (K4) -- (K5) -- cycle;
		
		\node at (barycentric cs:K1=1,K2=1,K3=1,K4=1,K5=1) {$\sset$};

		\path (muX0) -- ++(-40:.35cm) coordinate (anchorSphere) node [midway,scale=.75,xshift=-2pt] {$\vdir$};
		\node [pin={[pin edge = {red}, pin distance=.8cm]20:$\sset \intersec (t\S^{p-1} + \tfunc\grtr) = \sset_{\grtr,t} + \tfunc\grtr$},inner sep=2pt] at (anchorSphere) {};
		
		\draw[thick,-latex] (muX0) -- ++(-80:.35cm) ;
		\node[blackdot] at (muX0) {};
		\node [pin={[pin distance=.8cm]160:$\tfunc\grtr$},inner sep=2pt] at (muX0) {};
	\end{tikzpicture}
	\caption{\emph{Illustration of the localization argument behind Fact~\ref{fact:results:recovery}.} We assume that $\exloss[\grtr](\vdir) > 0$ for all $\vdir \in \sset_{\grtr,t}$, or equivalently, $\exloss[\grtr](\pv - \tfunc\grtr) > 0$ for all $\pv \in \sset_{\grtr,t} + \tfunc\grtr$ (see red arc).
	Since $\exloss[\grtr](\vnull) = 0$, the convexity of $\sset$ and $\exloss[\grtr](\cdot)$ implies that a minimizer $\pvsolu$ of \protect\eqref{eq:results:klasso} must not lie outside the dark gray intersection. In other words, we have that $\lnorm{\pvsolu - \tfunc\grtr} \leq t$.}
	\label{fig:results:fact}
\end{figure}
\begin{fact}\label{fact:results:recovery}
	For $\grtr \in \rset$ and a desired reconstruction accuracy $t > 0$, we set
	\begin{equation}
		\sset_{\grtr,t} \coloneqq \{\vdir \in \sset - \tfunc\grtr  \suchthat \lnorm{\vdir} = t \} = (\sset - \tfunc\grtr) \intersec t\S^{p-1}.
	\end{equation}
	If $\exloss[\grtr](\vdir) > 0$ for all $\vdir \in \sset_{\grtr,t}$, then every minimizer $\pvsolu$ of \eqref{eq:results:klasso} satisfies $\lnorm{\pvsolu - \tfunc\grtr} \leq t$.
\end{fact}
For a fixed accuracy $t > 0$, Fact~\ref{fact:results:recovery} implies \emph{uniform} recovery for all those $\grtr \in \rset$ satisfying $\inf_{\vdir \in \sset_{\grtr,t}} \exloss[\grtr](\vdir) > 0$.
This motivates us to establish a uniform lower bound for the excess loss for all $\grtr \in \rset$ and $\vdir \in \sset_{\grtr,t}$.
In particular, such a task is considerably more difficult than showing a bound on $\sset_{\grtr,t}$ for a fixed $\grtr \in \rset$, which would be in accordance with the non-uniform results discussed in the second part of Subsection~\ref{subsec:intro:literature}.
To get a better sense of this challenge, let us consider the following basic decomposition of the excess loss:
\begin{equation}\label{eq:results:exlossdecomp}
	\exloss[\grtr](\vdir) = \underbrace{\tfrac{1}{m} \sum_{i=1}^m\sp{\a_i}{\vdir}^2}_{\eqqcolon \quadrterm(\vdir)} {} - {} \underbrace{\tfrac{2}{m} \sum_{i=1}^m (y_i - \ys_i(\grtr)) \sp{\a_i}{\vdir}}_{\eqqcolon \advterm[\grtr](\vdir)} {} + {} \underbrace{\tfrac{2}{m} \sum_{i=1}^m(\sp{\a_i}{\tfunc\grtr} - \ys_i(\grtr)) \sp{\a_i}{\vdir}}_{\eqqcolon \multiplterm[\grtr](\vdir)}.
\end{equation}
The \emph{quadratic term} $\quadrterm(\vdir)$ and \emph{noise term} $\advterm[\grtr](\vdir)$ are rather unproblematic and can be controlled by a recent matrix deviation inequality (see Step~\hyperref[step:quadr]{1} and Step~\hyperref[step:noise]{2} in Section~\ref{sec:proofs}). 
Much more intricate is the \emph{multiplier term} $\multiplterm[\grtr](\vdir)$, since the underlying multiplier variable $\multipl(\grtr) \coloneqq \sp{\a}{\tfunc\grtr} - \ys(\grtr)$ depends on $\grtr$, so that we actually have to deal with an \emph{empirical product process}.
A major difficulty is that known concentration results for such product processes do not apply directly, since the class $\{\multipl(\grtr)\}_{\grtr \in \rset}$ need not have sub-Gaussian increments with respect to an appropriate \mbox{(pseudo-)metric}.
For example, this would happen if we would drop the Lipschitz assumption on $\fobs$ in Theorem~\ref{thm:intro:sim}.
The key idea of our approach is to approximate $\ys(\grtr)$ by a more ``regular'' observation variable $\ys_t(\grtr)$ in such a way that the resulting multiplier variable and the approximation error both have sub-Gaussian increments.
This is made precise by the following assumption:
\begin{assumption}\label{ass:results:incr}
We define a pseudo-metric on $\rset$ by $d_\tfunc(\grtr,\grtr') \coloneqq \lnorm{\tfunc\grtr - \tfunc\grtr'}$ for $\grtr,\grtr' \in \rset$.
For $t \geq 0$, assume that there exists a class of observation variables $\{\ys_t(\grtr)\}_{\grtr \in \rset}$ such that the following properties hold:
\begin{asslist}
\item\label{ass:results:incr:approx}
	\emph{Approximation error:} Setting $\err_t(\grtr) \coloneqq \abs{\ys(\grtr) - \ys_t(\grtr)}$, we assume that $\mean[\err_t(\grtr) \cdot \abs{\sp{\a}{\pv}}] \leq \tfrac{t}{64}$ for all $\grtr \in \rset$ and $\pv \in \S^{p-1}$.
\item\label{ass:results:incr:multipl}
	\emph{Multiplier increments:} Set $\multipl_t(\grtr) \coloneqq \sp{\a}{\tfunc\grtr} - \ys_t(\grtr)$ for $\grtr \in \rset$. We assume that there exist $\multiplDia \geq 0$ and $\multiplLip \geq 0$ such that
	\begin{equation}
		\normsubg{\multipl_t(\grtr) - \multipl_t(\grtr')} \leq \multiplLip \cdot d_\tfunc(\grtr,\grtr') \quad{\text{and}}\quad \normsubg{\multipl_t(\grtr)} \leq \multiplDia \qquad \text{for all $\grtr, \grtr' \in \rset$.}
	\end{equation}
\item\label{ass:results:incr:err}
	\emph{Error increments:} We assume that there exist $\errDia \geq 0$ and $\errLip \geq 0$ such that
	\begin{equation}
		\normsubg{\err_t(\grtr) - \err_t(\grtr')} \leq \errLip \cdot d_\tfunc(\grtr,\grtr') \quad{\text{and}}\quad \normsubg{\err_t(\grtr)} \leq \errDia \qquad \text{for all $\grtr, \grtr' \in \rset$.}
	\end{equation}
\end{asslist}
\end{assumption}
Assumption~\ref{ass:results:incr}\ref{ass:results:incr:multipl} and~\ref{ass:results:incr:err} imply that $\{\multipl_t(\grtr)\}_{\grtr \in \rset}$ and $\{\err_t(\grtr)\}_{\grtr \in \rset}$ have sub-Gaussian increments with respect to $\multiplLip\cdot d_\tfunc$ and $\errLip\cdot d_\tfunc$, respectively.
Similarly, $\multiplDia$ and $\errDia$ bound the (sub-Gaussian) diameter of the respective classes.
A convenient interpretation of Assumption~\ref{ass:results:incr} is as follows: choose an approximation $\ys_t(\grtr)$ for every $\grtr \in \rset$ such that the error does not become too large in the sense of part~\ref{ass:results:incr:approx}; at the same time, ensure that the increment conditions of part~\ref{ass:results:incr:multipl} and~\ref{ass:results:incr:err} are satisfied such that the parameters $\multiplLip$ and $\errLip$ do not grow too fast as $t$ becomes smaller.
A remarkable conclusion from our applications to quantized compressed sensing in Subsection~\ref{subsec:appl:1bit}--\ref{subsec:appl:mbit} is that this strategy can even succeed for observation variables with discontinuous output functions.

Finally, we emphasize that the approximation error $\err_t(\grtr)$ includes taking the absolute value, which is crucial to our proof (see Step~\hyperref[step:multipl]{3} in Section~\ref{sec:proofs}). In particular, Assumption~\ref{ass:results:incr} does not necessarily imply that the original class $\{\multipl(\grtr)\}_{\grtr \in \rset}$ has sub-Gaussian increments with respect to $\tilde{\subgparam}\cdot d_\tfunc$ for some $\tilde{\subgparam} \geq 0$.
On the other hand, it is certainly possible that $\{\multipl(\grtr)\}_{\grtr \in \rset}$ already has sub-Gaussian increments, such as in Theorem~\ref{thm:intro:sim}. 
In this case, we can simply choose $\ys_t(\grtr) \coloneqq \ys(\grtr)$, so that $\err_t(\grtr) = 0$ and Assumption~\ref{ass:results:incr}\ref{ass:results:incr:approx} and~\ref{ass:results:incr:err} are trivially fulfilled.

\subsection{Uniform Recovery Guarantee}
\label{subsec:results:guarantee}

In order to formulate the main result of this work, we require the notion of Gaussian mean width.
This geometric parameter has proven to be a useful complexity measure in high-dimen\-sion\-al signal recovery, e.g., see \cite{mpt07,rv08,sto09,crpw12,almt14} for pioneering works in that direction.
Our approach is no exception and makes use of the following localized version.
\begin{definition}\label{def:results:meanwidth}
	Let $\ssetgen \subset \R^p$ and $\gaussian \distributed \Normdistr{\vnull}{\I{p}}$.
	The \emph{(Gaussian) mean width} of $\ssetgen$ is given by
	\begin{equation}
		\meanwidth{\ssetgen} \coloneqq \mean\Big[\sup_{\vdir \in \ssetgen} \sp{\gaussian}{\vdir}\Big].
	\end{equation}
	For $t \geq 0$, we define the \emph{local mean width} of $\ssetgen$ (at scale $t$) by
	\begin{equation}
		\meanwidth[t]{\ssetgen} \coloneqq \begin{cases}
			\meanwidth{\tfrac{1}{t}\ssetgen \intersec \S^{p-1}}, & t > 0, \\
			\meanwidth{\cone{\ssetgen} \intersec \S^{p-1}}, & t = 0.
		\end{cases}
	\end{equation}
\end{definition}
As final preparatory step, we introduce two expressions that capture the size of measurement noise: For $\nout \in \{0,1,\dots,m\}$ and $\vdiralt = (w_1, \dots, w_m) \in \R^m$, let
\begin{equation}
	\norm{\vdiralt}_{[\nout]} \coloneqq \Big( \sum_{i = 1}^{\nout} \abs{w_i^*}^2 \Big)^{1/2} \quad \text{ and } \quad \sigma_{\nout}(\vdiralt)_{2} \coloneqq \Big( \sum_{i = \nout+1}^{m} \abs{w_i^*}^2 \Big)^{1/2},
\end{equation}
where $(w_1^*,\dots,w_m^*)$ is the non-increasing rearrangement of $(\abs{w_1}, \dots, \abs{w_m})$.
Obviously, we have that $\norm{\vdiralt}_{[\nout]}^2 + \sigma_{\nout}(\vdiralt)_{2}^2 = \lnorm{\vdiralt}^2$, and in particular, $\norm{\vdiralt}_{[0]} = 0$ and $\sigma_{0}(\vdiralt)_{2} = \lnorm{\vdiralt}$.
Moreover, note that $\norm{\cdot}_{[\nout]}$ simply corresponds to the \emph{$\l{2}$-norm of the $\nout$-largest entries}, while $\sigma_{\nout}(\cdot)_{2}$ is commonly known as the \emph{$\l{2}$-error of the best $\nout$-term approximation}.

We are now ready to state our main recovery guarantee, which forms the basis of all applications presented in Section~\ref{sec:appl}; see Section~\ref{sec:proofs} for a complete proof.
\begin{theorem}\label{thm:results:main}
	There exist universal constants $c, C > 0$ for which the following holds.
	
	Let Assumptions~\ref{ass:results:meas} and~\ref{ass:results:incr} be satisfied for a fixed accuracy $t \geq 0$ and let $\nout\in \{0, 1, \dots, m\}$. 
	For $\tune>0$, $\probsuccess \geq 1$, and $\probsuccess_0 \geq \sqrt{\nout \log(em/\nout)}$, we assume that\footnote{The condition \eqref{eq:results:main:m} is stated in such a way that it is convenient to handle in our specific applications (see Section~\ref{sec:appl}). However, for a basic understanding, the reader may simply set $\probsuccess = \probsuccess_0$, so that the first and second branch in \eqref{eq:results:main:m} can be merged to $m \gtrsim \subgparam^2(\log\subgparam + \subgparam^2\tune^2) \cdot \big(\effdim[t]{\sset - \tfunc\rset}+ \probsuccess^2 \big)$.}\footnote{In the case of exact recovery, i.e., $t = 0$, we follow the convention $0 \cdot \infty \coloneqq 0$, so that the condition \eqref{eq:results:main:m} requires that $\multiplDia = \errDia = \multiplLip = \errLip = 0$. Then, \eqref{eq:results:main:m} is particularly fulfilled for $m \geq C \cdot \subgparam^2 \log\subgparam  \cdot (\effdim[0]{\sset - \tfunc\rset}+ \probsuccess^2)$, where $\nout = 0$, $\tune = \subgparam^{-1} \sqrt{\log\subgparam}$, and $\probsuccess = \probsuccess_0$.}
	\begin{align}
		m \geq C \cdot \subgparam^2 \cdot \max\Big\{ &  \log\subgparam \cdot \big(\effdim[t]{\sset - \tfunc\rset}+ \probsuccess^2 \big), \\
			&\subgparam^2\tune^2 \cdot \big(\effdim[t]{\sset - \tfunc\rset}+ \probsuccess_0^2 \big),\\
			&t^{-2} (\multiplDia^2+\errDia^2) \cdot \big(\effdim[t]{\sset - \tfunc\rset}+ \probsuccess^2 \big) + t^{-2}  (\multiplLip^2+ \errLip^2) \cdot \effdim{\tfunc\rset} \Big\}. \label{eq:results:main:m}
	\end{align}
	Then with probability at least $1 - \exp(-c \probsuccess^2) - \exp(-c \probsuccess_0^2)$ on the random draw of $\{(\a_i, \fout_i)\}_{i = 1}^m$, the following holds uniformly for every $\grtr \in \rset$ with $\mmcovar{\grtr} \leq \tfrac{t}{32}$:
	Let $\Y \in \R^m$ be any input vector such that
	\begin{equation}\label{eq:results:main:rob}
	\tfrac{1}{\sqrt{m}}\norm{\Y - \Ys(\grtr)}_{[\nout]}\leq \tune t \quad \text{ and } \quad 
	\tfrac{1}{\sqrt{m}}\sigma_{\nout}(\Y - \Ys(\grtr))_{2}\leq \tfrac{t}{20}.
	\end{equation} 
	Then every minimizer $\pvsolu$ of \eqref{eq:results:klasso} satisfies $\lnorm{\pvsolu - \tfunc\grtr} \leq t$.
\end{theorem}
The constraints of \eqref{eq:results:main:rob} imply that the estimator \eqref{eq:results:klasso} is \emph{robust} against (adversarial) perturbations and outliers in the observation vector $\Ys(\grtr)$.
Here, a central role is played by the fine-tuning parameter $\nout$, controlling which coefficients of the noise vector $\Y - \Ys(\grtr)$ are captured by the first and second condition in \eqref{eq:results:main:rob}, respectively.
Enlarging $\nout$ helps to suppress ``gross'' outliers (that may appear in the $\nout$ largest coefficients of $\Y - \Ys(\grtr)$ in magnitude). 
Indeed, by adjusting the free parameter $\tune$, the first condition of \eqref{eq:results:main:rob} becomes less restrictive than the second one (which captures the remaining coefficients of $\Y - \Ys(\grtr)$).
On the other hand, choosing $\nout$ and/or $\tune$ too large may have a negative effect on the sample complexity in the second branch of \eqref{eq:results:main:m}.
The benefit of permitting outliers will become clearer in the context of quantization where the noise is due to wrong bits in $\Y$ instead of real-valued perturbations (see Subsection~\ref{subsec:appl:1bit}--\ref{subsec:appl:mbit}).
Note that the most common noise model is obtained for $\nout = 0$: \eqref{eq:results:main:rob} then simply corresponds to the baseline condition $\tfrac{1}{\sqrt{m}}\lnorm{\Y - \Ys(\grtr)} \leq \tfrac{t}{20}$, which is consistent with standard noise bounds from the literature.
We refer to Remark~\ref{rmk:unif}\ref{rmk:unif:noise} in the next section for a more detailed comparison of adversarial and statistical noise, as well as aspects of optimality.

The second important constraint of Theorem~\ref{thm:results:main} is that it does not apply to those $\grtr \in \rset$ with $\mmcovar{\grtr} > \tfrac{t}{32}$.
Hence, the target mismatch $\mmcovar{\grtr}$ can be seen as an upper bound for the asymptotic error when estimating $\tfunc\grtr$ via \eqref{eq:results:klasso}. 
In particular, the above error bound does not allow for arbitrarily high precision unless $\mmcovar{\grtr} = 0$.

Let us now turn to the key assumption \eqref{eq:results:main:m}, which relates the number of measurements $m$ to the desired accuracy $t$.
The above formulation views $m$ as a function of $t$, specifying how fast $m$ grows if $t$ decreases.
The inverse relationship is also of interest, since it describes the reconstruction error as a function of $m$.
However, this can only be made explicit in specific situations where the parameters at the right-hand side of \eqref{eq:results:main:m} are known or can be well estimated.
Of special importance in that respect are the increment parameters $\multiplLip$ and $\errLip$ from Assumption~\ref{ass:results:incr}, whose size has a significant impact on the required (over-)sampling rate.
For $1$-bit observations as studied in Subsection~\ref{subsec:appl:1bit}, for example, we achieve $\multiplLip, \errLip \lesssim t^{-1}$, so that \eqref{eq:results:main:m} would yield an error decay rate of $\asympfaster{m^{-1/4}}$---or conversely, an oversampling rate of $\asympfaster{t^{-4}}$.
Remarkably, we will derive a corollary of Theorem~\ref{thm:results:main} in Section~\ref{sec:appl:noincr} that can bypass the restrictions of Assumption~\ref{ass:results:incr} at the price of a stronger (local) stability condition on the observation model. 
This allows us to combine our approach with known embedding results from the literature and thereby to obtain near-optimal decay rates in special cases.

The dependence of \eqref{eq:results:main:m} on the complexity parameters $\meanwidth[t]{\sset - \tfunc\rset}$ and $\meanwidth{\tfunc\rset}$ is quite natural.
Similar expressions have already appeared in several of the articles discussed in Subsection~\ref{subsec:intro:literature}, e.g., see~\cite{pv13b,pvy16,dm18}.
A detailed discussion, including concrete examples and possible simplifications, can be found in Section~\ref{sec:unif} below.
A distinctive feature of Theorem~\ref{thm:results:main} is that the considered version of local mean width $\meanwidth[t]{\sset - \tfunc\rset}$ ``compares'' the parameter vectors in $\sset$ only to those in the transformed signal set $\tfunc\rset \subset \sset$.
This refinement of the common quantity $\meanwidth[t]{\sset - \sset}$ can lead to improved guarantees in certain scenarios, namely when $\tfunc\rset$ is much smaller than~$\sset$ (see Example~\ref{ex:unif:tv} for an illustration of this aspect).
\begin{remark}
\begin{rmklist}
\item\label{rmk:results:main:tfunc}
	\emph{Inversion of $\tfunc$.} In principle, the target function $\tfunc$ neither has to be injective nor does it have to be explicitly known to solve \eqref{eq:results:klasso}. However, in order to obtain a practicable statement from Theorem~\ref{thm:results:main}, the actual signal $\grtr \in \rset$ should be extractable from an approximation~$\pvsolu$ of~$\tfunc\grtr$. 
	An appropriate implementation of $\tfunc$ can differ considerably from situation to situation. 
	In the case of single-index models (see Theorem~\ref{thm:intro:sim}), for example, this would involve a rescaling of~$\pvsolu$, which requires a (rough) knowledge of the non-linearity $\fobs$. 
	In contrast, for variable selection (see Subsection~\ref{subsec:appl:vs}), it can be sufficient to perform a simple hard-thresholding step on $\pvsolu$ to obtain a good estimate of the underlying support.
\item\label{rmk:results:main:optimal}
	\emph{Optimality.} The best possible error decay rate that can result from Theorem~\ref{thm:results:main} is $\asympfaster{m^{-1/2}}$, supposed that we have $\multiplLip, \errLip \lesssim 1$.\footnote{Note that the expression $\asympfaster{m^{-1/2}}$ suppresses the dependence on $\meanwidth[t]{\sset - \tfunc\rset}$. Although the latter can be trivially bounded by $\meanwidth[0]{\sset - \tfunc\rset}$, which is independent of $t$, such an estimate might not appropriately capture the (low) complexity of $\sset$ in certain cases.}
	This rate cannot be improved \emph{in general}, or in other words, the exponent $-\tfrac{1}{2}$ is not an artifact of our proof but corresponds to a fundamental statistical barrier (see~\cite[Sec.~4]{pvy16}).
	However, it is possible to break through this barrier in specific model setups, e.g., \emph{noiseless} $1$-bit observations \cite{jlbb13,jmps19}.
	Such superior rates are usually not achieved by \eqref{eq:results:klasso}, but may require a more sophisticated estimator instead. \qedhere
\end{rmklist}\label{rmk:results:main}
\end{remark}

\begin{remark}[Possible extensions] 
	For the sake of clarity, we did not present the most general version of Theorem~\ref{thm:results:main} that we could have derived.
	There are several variants and generalizations that we expect to be practicable:
\begin{rmklist}
\item\label{rmk:results:exten:covar}
	\emph{Anisotropic measurements.} Based on an observation from \cite[Rmk.~1.7]{pv16}, \cite[Sec.~II.D]{gen16}, the isotropy of the measurement vectors in Assumption~\ref{ass:results:meas}\ref{ass:results:meas:stat} can be relaxed to any positive definite covariance matrix $\Covmatr \in \R^{p \times p}$.
	To see this, let us assume that $\a$ is centered and sub-Gaussian with $\mean[\a \a^\T] = \Covmatr$.
	Then, we may write $\a = \sqrt{\Covmatr} \bar{\a}$ where $\bar{\a}$ is isotropic, centered, and sub-Gaussian.
	If $\bar{\a}_1, \dots, \bar{\a}_m$ are i.i.d.~copies of $\bar{\a}$, a simple reformulation of \eqref{eq:results:klasso} yields
	\begin{equation}
		\argmin_{\pv \in \sset} \ \tfrac{1}{m} \sum_{i = 1}^m (y_i - \sp{\a_i}{\pv})^2 = \sqrt{\Covmatr}^{-1} \cdot \argmin_{\bar{\pv} \in \sqrt{\Covmatr}\sset} \ \tfrac{1}{m} \sum_{i = 1}^m (y_i - \sp{\bar{\a}_i}{\bar{\pv}})^2.
	\end{equation}
	In other words, \eqref{eq:results:klasso} is equivalent to a modified estimator that operates with an isotropic measurement ensemble.
	Therefore, we may apply Theorem~\ref{thm:results:main} directly to the latter one, when replacing $\sset$ and $\tfunc(\cdot)$ by $\sqrt{\Covmatr}\sset$ and $\sqrt{\Covmatr} \tfunc(\cdot)$ in Assumption~\ref{ass:results:meas}, respectively.
	Note that these adaptions are of purely theoretical nature and in practice, no explicit knowledge of $\Covmatr$ is required.
	In particular, this procedure still implies an error bound for minimizers $\pvsolu$ of \eqref{eq:results:klasso}:
	\begin{equation}
		\lnorm{\pvsolu - \tfunc\grtr} \leq \opnorm{\sqrt{\Covmatr}^{-1}} \cdot \lnorm{\sqrt{\Covmatr}\pvsolu - \sqrt{\Covmatr}\tfunc\grtr} \leq \opnorm{\sqrt{\Covmatr}^{-1}} \cdot t.
	\end{equation}
	Similarly, one can conveniently control the adapted complexity parameters (see \cite[Rmk.~1.7]{pv16}, \cite[Sec.~II.D]{gen16}):
	\begin{align}
		\meanwidth[t]{\sqrt{\Covmatr}\sset - \sqrt{\Covmatr}\tfunc\rset} &\leq \max\big\{1, \opnorm{\sqrt{\Covmatr}^{-1}}\big\} \cdot \opnorm{\sqrt{\Covmatr}} \cdot \meanwidth{\tfrac{1}{t}(\sset - \tfunc\rset) \intersec \ball[2][p]}, \\
		\meanwidth{\sqrt{\Covmatr}\tfunc\rset} &\lesssim \opnorm{\sqrt{\Covmatr}} \cdot \meanwidth{\tfunc\rset}.
	\end{align}
	Overall, we can conclude that accurate (uniform) recovery is also possible with anisotropic measurements, as long as the (unknown) covariance matrix is well-conditioned.
\item\label{rmk:results:exten:phase}
	\emph{Even output functions.} There exist certain types of non-linear models that are not directly compatible with the estimator \eqref{eq:results:klasso}. 
	For example, if $\fobs \colon \R \to \R$ is an even function in Theorem~\ref{thm:intro:sim}, it turns out that $\scalfac = \mean_{\gaussianuniv \distributed \Normdistr{0}{1}}[\fobs(\gaussianuniv)\gaussianuniv] = 0$.
	In other words, \eqref{eq:results:klasso} would simply approximate the zero vector, which is clearly not what one is aiming at.
	Notably, this scenario includes the important phase-retrieval problem $\fobs(\cdot) = \abs{\cdot}$.
	Fortunately, the linearization idea behind \eqref{eq:results:klasso} is still applicable in these situations, when combined with the \emph{phase-lifting} trick \cite{cesv13}:
	\begin{equation} \label{eq:results:liftklasso}\tag{$\mathsf{P}_{\sset,\Y}^{\text{lifted}}$}
		\min_{\vec{X} \in \sset} \ \tfrac{1}{m} \sum_{i = 1}^m \big(y_i - \sp{\a_i\a_i^\T - \mean[\a_i\a_i^\T]}{\vec{X}}_F\big)^2,
	\end{equation}
	where $\sp{\cdot}{\cdot}_F$ denotes the Hilbert-Schmidt inner product and $\sset \supset \{\grtr \grtr^\T \suchthat \grtr \in \rset \}$ is a convex subset of the positive semidefinite cone in $\R^{p \times p}$.
	The recovery performance of \eqref{eq:results:liftklasso} for even output functions has been recently analyzed by the first author in \cite[Sec.~3.5]{gk20}, inspired by earlier findings of \citeauthor{tr19}~\cite{tr19}.
	A key challenge in this regard is that the lifted measurement vectors/matrices $\a_i\a_i^\T - \mean[\a_i\a_i^\T]$ are \emph{heavier tailed} than $\a_i$, which in turn has required deeper insights from generic chaining theory.
	We expect that our results are extendable into this direction, but with a technical effort that would go beyond the scope of this work.
\item\label{rmk:results:exten:metrics}
	\emph{Different metrics.} An extension that is specific to uniform recovery concerns the choice of pseudo-metric in Assumption~\ref{ass:results:incr}.
	In principle, $d_\tfunc$ could be replaced by an arbitrary pseudo-metric~$d$ on the signal set~$\rset$.
	The claim and proof of Theorem~\ref{thm:results:main} would still hold true if the (global) complexity of $\rset$ is captured by $\talagfunc(\rset, d)$ instead of $\meanwidth{\tfunc\rset}$; see Step~\hyperref[step:multipl]{3} in Section~\ref{sec:proofs}.
	While this would make our approach slightly more flexible, we note that Talagrand's $\talagfunc$-functional is difficult to control in general. 
	
	Apart from that, the $\l{2}$-norm as \emph{error metric} in Theorem~\ref{thm:results:main} could be replaced by an arbitrary semi-norm $\norm{\cdot}$.
	Specifically, this step would lead to an error bound in terms of $\norm{\cdot}$ in Fact~\ref{fact:results:recovery}, but it also entails an adaption of the spherical intersection in the local mean width (see Definition~\ref{def:results:meanwidth}).
	Finally, different \emph{convex loss functions} are feasible for \eqref{eq:results:klasso} as well \cite{gen16}, although such an extension would require several additional technical assumptions.
	\qedhere
\end{rmklist}\label{rmk:results:exten}
\end{remark}

\section{(Non-)Uniformity and Signal Complexity}
\label{sec:unif}

In this section, we highlight important aspects that are specifically attributable to uniform recovery, compared to existing non-uniform results.
This includes a more detailed discussion of the local mean width $\meanwidth[t]{\sset - \tfunc\rset}$, which appears as the central complexity parameter in our approach.

Except for the inclusion $\tfunc\rset \subset \sset$, Theorem~\ref{thm:results:main} does not impose any restrictions on the signal set $\rset$.
If $\tfunc$ is the identity function, the choice of $\rset$ may range from a singleton $\rset = \{\grtr\}$ to the entire constraint set $\rset = \sset$.
The latter scenario corresponds to uniform recovery of all signals in~$\sset$, while the former means that one is interested in a specific $\grtr \in \sset$---which is nothing else than non-uniform recovery.
In particular, due to $\meanwidth{\{\grtr\}} = 0$, Theorem~\ref{thm:results:main} is consistent with previously known non-uniform guarantees, e.g., see~\cite[Thm.~3.6]{gen19} and \cite[Thm~1.9]{pv16}.
Our main result therefore also indicates what additional expenses may come in the case of uniform recovery.

Let us first focus on the role of \emph{non-linear} observations, whose impact is reflected by the third branch of \eqref{eq:results:main:m} in Theorem~\ref{thm:results:main}.
The key quantities in this respect are the increment parameters~$\multiplLip$ and~$\errLip$.
Their size strongly depends on the type of output function~$\fout$ considered in Assumption~\ref{ass:results:meas}.
For instance, a jump discontinuity in $\fout$ can cause a weaker oversampling rate, in the sense that $\multiplLip, \errLip \lesssim t^{-1}$;
we refer to Subsection~\ref{subsec:appl:1bit}--\ref{subsec:appl:sim} for concrete examples and Section~\ref{sec:appl:noincr} for a possible improvement.
In the non-uniform case ($\rset = \{\grtr\}$), on the other hand, this issue becomes irrelevant because one can simply set $\multiplLip = \errLip = 0$ (since Assumption~\ref{ass:results:incr}\ref{ass:results:incr:multipl} and~\ref{ass:results:incr:err} need only be satisfied for a single point).
Thus, we can draw the following informal conclusion:
\begin{highlight}
	The transition to uniform recovery with non-linear output functions may result in a worse oversampling rate (with respect to $t$).
\end{highlight}

The second foundational aspect of uniformity concerns the geometric complexity of the constraint set $\sset$ and its interplay with the actual signal set $\rset$.
To keep our exposition as clear as possible, we restrict ourselves to the situation of linear observations (though most arguments naturally carry over to non-linear models).
Applying Theorem~\ref{thm:results:main} to this special case leads to the following recovery guarantee (see Section~\ref{sec:app:unif} for a proof).
Note that we intentionally allow for two different types of noise here, based on the adversarial ($\sim \noise_i$) and statistical model ($\sim \tau_i$); see Remark~\ref{rmk:unif}\ref{rmk:unif:noise} below for further discussion.
\begin{corollary}\label{cor:unif:lin}
	There exist universal constants $c, C > 0$ for which the following holds.
	
	Let $\a_1, \dots, \a_m \in \R^p$ be independent copies of a centered, isotropic, sub-Gaussian random vector $\a\in \R^p$ with $\normsubg{\a} \leq \subgparam$.
	Let $\tau_1, \dots, \tau_m$ be independent copies of $\tau \distributed \Normdistr{0}{\sigma^2}$ for some $\sigma \geq 0$ (also independent of~$\{\a_i\}_{i=1}^m$).
	Moreover, let $\rset \subset \R^p$ be a bounded subset and let $\sset \subset \R^p$ be a convex set such that $\rset \subset \sset$.
	For $\probsuccess \geq 1$ and $t \geq 0$, we assume that
	\begin{equation}\label{eq:unif:lin:m}
		m \geq C \cdot \subgparam^2 \cdot (\log\subgparam + t^{-2} \sigma^2 ) \cdot \big(\effdim[t]{\sset - \rset} + \probsuccess^2 \big).
	\end{equation}
	Then with probability at least $1 - \exp(-c \probsuccess^2)$ on the random draw of $\{(\a_i, \tau_i)\}_{i = 1}^m$, the following holds uniformly for all $\grtr \in \rset$:
	Let $\Y = (y_1, \dots, y_m) \in \R^m$ be given by 
	\begin{equation}\label{eq:unif:lin:meas}
		y_i = \sp{\a_i}{\grtr} + \tau_i + \noise_i, \quad i = 1, \dots, m,
	\end{equation}
	such that $\big(\tfrac{1}{m}\sum_{i = 1}^m \noise_i^2 \big)^{1/2} \leq \tfrac{t}{20}$. Then every minimizer $\pvsolu$ of \eqref{eq:results:klasso} satisfies $\lnorm{\pvsolu - \grtr} \leq t$.
\end{corollary}
Except for the $\sigma$-depending summand in \eqref{eq:unif:lin:m}, the above statement does not involve any oversampling factors and it turns into an exact recovery guarantee for $t = 0$ and $\sigma = 0$.
Although Corollary~\ref{cor:unif:lin} bears resemblance with standard results from compressed sensing, it comes with a distinctive feature: the needed number of measurements in~\eqref{eq:unif:lin:m} is determined by the local mean width $\meanwidth[t]{\sset - \rset}$. 
As such, this parameter is quite implicit, so that an informative (upper) bound is required for any specific choice of $\sset$ and $\rset$.
The remainder of this section is devoted to this quest.

Arguably the most popular low-complexity model is \emph{sparsity} in conjunction with an $\l{1}$-relaxation.
In our context, this corresponds to the following scenario:
\begin{equation}\label{eq:unif:l1}
	\sset = R\ball[1][p] \quad \text{and} \quad \rset = \{\grtr \in \R^p \suchthat \lnorm{\grtr}[0] \leq s, \lnorm{\grtr}[1] = R \},
\end{equation}
where $s \in [p]$ and $R > 0$; note that the constraint $\lnorm{\grtr}[1] = R$ is a specific tuning condition for the estimator \eqref{eq:results:klasso} that could be further relaxed (see Remark~\ref{rmk:unif}\ref{rmk:unif:tuning}).
In this special case, one can derive a convenient bound for the local mean width:
\begin{equation}\label{eq:unif:mwl1}
	\meanwidth[t]{\sset - \rset} \leq \meanwidth[0]{\sset - \rset} \lesssim \sqrt{s \cdot \log(2p/s)}.
\end{equation}
Remarkably, this observation dates back to one of the earliest applications of \emph{Gordon’s escape through the mesh theorem} \cite{gor88} to signal recovery \cite{rv08}; see Section~\ref{sec:app:unif} for a proof of \eqref{eq:unif:mwl1}.
A combination of \eqref{eq:unif:mwl1} and Corollary~\ref{cor:unif:lin} ensures that uniform (exact) reconstruction is feasible with $\asympfaster{s \cdot \log(2p/s)}$ Gaussian measurements.
Therefore, our approach is indeed well in line with standard sparse recovery guarantees in compressed sensing, which are typically based on restricted isometry (cf.~\cite{fh13}).

Beyond this classical example, the situation becomes considerably more challenging and forms an important research subject in its own right.
A particular obstacle due to uniformity is that $\meanwidth[t]{\sset - \rset}$ heavily depends on the size and shape of the signal set~$\rset$. 
On a technical level, we have to deal with a (possibly uncountable) \emph{union} of spherical intersections:
\begin{equation}\label{eq:unif:mwunion}
	\meanwidth[t]{\sset - \rset} =  \meanwidth{\textstyle\bigunion_{\grtr \in \rset}[\tfrac{1}{t}(\sset - \grtr) \intersec \S^{p-1}]}.
\end{equation}
Fortunately, this cumbersome expression can be controlled through a much more convenient upper bound, as shown in the following proposition.
To the best of our knowledge, this is a new result, which could be of independent interest. 
Its proof can be found in Section~\ref{sec:app:unif} and is based on a covering argument.
\begin{proposition}\label{prop:unif:mwloc}
	Let $\sset, \rset \subset \R^p$ and $t > 0$. Then
	\begin{equation}\label{eq:unif:mwloc:bound}
		\meanwidth[t]{\sset - \rset} \lesssim \sup_{\grtr \in \rset} \ \underbrace{\meanwidth{\tfrac{1}{2t}(\sset - \grtr) \intersec \ball[2][p]}}_{\eqqcolon \meanwidthalt[t]{\sset - \grtr}} {} + {} t^{-1} \cdot \meanwidth{\rset}.
	\end{equation}
	Moreover, for $\grtr \in \rset$ and $\sset$ convex, we have that $\meanwidthalt[t]{\sset - \grtr} \lesssim \meanwidth[0]{\sset - \tilde{\grtr}} + 1$ for every $\tilde{\grtr} \in \sset$ with $\lnorm{\grtr - \tilde{\grtr}} \leq t$.
\end{proposition}
The upper bound in \eqref{eq:unif:mwloc:bound} implies a significant simplification: the first term is fully \emph{localized}, measuring the complexity of $\sset$ with respect to each individual point $\grtr \in \rset$, whereas the second term accounts for the \emph{global} size of $\rset$ (independently of the constraint set $\sset$).
In other words, the effect of $\sset$ and $\rset$ is now decoupled.
A second notable fact is that the local mean width $\meanwidthalt[t]{\sset - \grtr}$ and its conic counterpart $\meanwidth[0]{\sset - \grtr}$ are well-known complexity parameters from non-uniform recovery results, e.g., see \cite{crpw12,almt14,tro15,pv16,gen19}.
Hence, Proposition~\ref{prop:unif:mwloc} allows us to transfer any corresponding bound from the literature to the uniform regime.
The presence of the global mean width $\meanwidth{\rset}$ as an additional expense for uniformity appears natural to us.
However, this expression also entails an oversampling factor of $t^{-1}$, which prevents perfect reconstruction (when applied to Corollary~\ref{cor:unif:lin}).
Given the $\l{1}$-special case in \eqref{eq:unif:mwl1}, we suspect that this is an artifact of our proof, and it remains an open question whether such a factor could be removed in general.
Our overall conclusion is as follows:
\begin{highlight}
	Regarding signal complexity, the transition to uniform recovery requires control over the total size of the set $\rset$, measured by $\meanwidth{\rset}$.
	The constraint set $\sset$ only appears in terms of local complexity, precisely as in the non-uniform case.
\end{highlight}
As the above argumentation is fairly abstract, it is insightful to illustrate our approach by a concrete example.
In this context, we will also highlight the importance of carefully ``designed'' signal sets, in the sense that $\rset$ (and even its convex hull) is much smaller than $\sset$.
\begin{example}[Total variation in 1D]\label{ex:unif:tv}
	We consider the situation of \emph{$s$-gradient-sparse} signals in one spatial dimension, i.e., for $s \in [p-1]$ fixed, it is assumed that $\lnorm{\gradient \grtr}[0] \leq s$, where
	\begin{equation}
		\gradient  \coloneqq {\scriptsize\matr{
			-1 & 1 & 0 & \dots & 0 \\
			0 & -1 & 1 & & 0\\
			\vdots & & \ddots &\ddots & \vdots \\
			0 & \dots & 0 & -1 & 1}} \in \R^{(p-1) \times p}
	\end{equation}
	is a discrete gradient operator. 
	Geometrically, this condition simply means that the vector $\grtr$ is piecewise constant with at most $s$ jump discontinuities.
	An $\l{1}$-relaxation of gradient-sparsity then leads to the common \emph{total variation (TV)} model \cite{rof92}, which in our case corresponds to the constraint set $\sset = \{ \grtr \in \R^p \suchthat \lnorm{\gradient \grtr}[1] \leq R \}$ for some $R > 0$.
	
	A recent work of the first author \cite{gms20} has demonstrated that the reconstruction capacity of the TV model does not only depend on the number of jump discontinuities but also strongly on their position.
	More specifically, we say that an $s$-gradient-sparse signal $\grtr$ is \emph{$\Delta$-separated} for some $\Delta \in (0, 1]$ if
	\begin{equation}
		\min_{j \in [s+1]} \frac{\abs{\nu_j - \nu_{j-1}}}{p} \geq \frac{\Delta}{s+1},
	\end{equation}
	where $\supp(\gradient\grtr) = \{\nu_1,\dots,\nu_{s}\}$ with $0 \eqqcolon \nu_0 < \nu_1 < \dots < \nu_{s} < \nu_{s+1} \coloneqq p$.
	Intuitively, the constant~$\Delta$ measures how much the jump positions in $\grtr$ deviate from an equidistant pattern (where one would have $\Delta = 1$).
	Assuming that $\Delta \in (0, 1]$ is fixed (independently of $n$ and $s$), we say that~$\grtr$ is \emph{well-separated}; see \cite[Sec.~2.1]{gms20} for more details.
	This motivates us to consider the following signal set:
	\begin{equation}\label{eq:unif:tv:rset}
		\rset = \big\{\grtr \in \R^p \suchthat \lnorm{\gradient\grtr}[0] \leq s, \ \text{$\grtr$ is $\Delta$-separated}, \lnorm{\gradient\grtr}[1] = R \big\} \intersec \ball[2][p].
	\end{equation}
	Similarly to \protect\eqref{eq:unif:l1}, the constraint $\lnorm{\gradient\grtr}[1] = R$ results from a tuning condition for \protect\eqref{eq:results:klasso} (see Remark~\ref{rmk:unif}\ref{rmk:unif:tuning} below).
	
	A deep result from \cite[Thm.~2.10]{gms20} now yields the following estimate for the conic mean width:
	\begin{equation}\label{eq:unif:tv:mwloc}
		\meanwidth[0]{\sset - \grtr}  \lesssim \sqrt{\Delta^{-1} \cdot s \cdot \log^2(p)} \qquad \text{for all $\grtr \in \rset$.}
	\end{equation}
	For the global mean width, we can make use of a bound derived by \citeauthor{kks17}~\cite[Sec.~3.2]{kks17}:
	\begin{equation}\label{eq:unif:tv:mwglob}
		\meanwidth{\rset} \lesssim \sqrt{s \cdot \log(e p/ s)}.
	\end{equation}
	Note that \eqref{eq:unif:tv:mwglob} would also hold without the $\Delta$-separation condition, but it exploits that $\rset \subset \ball[2][p]$.
	Combining \eqref{eq:unif:tv:mwloc} and \eqref{eq:unif:tv:mwglob} with Proposition~\ref{prop:unif:mwloc}, we finally obtain
	\begin{equation}
		\meanwidth[t]{\sset - \rset} \lesssim (1 + t^{-1}) \sqrt{s \cdot \log^2(p)}.
	\end{equation}
	Therefore, according to Corollary~\ref{cor:unif:lin}, uniform recovery of well-separated $s$-gradient sparse signals becomes feasible with $\asympfaster{s \cdot \log^2(p)}$ Gaussian measurements.
	In particular, we have shown that there is no qualitative gap to the non-uniform guarantee derived in \cite{gms20}.

	Let us emphasize that, although the previous bounds look quite simple, their proofs build on fundamental insights into the TV model.
	For example, relying only on the basic estimate $\meanwidth[t]{\sset - \rset} \lesssim t^{-1} \cdot \meanwidth{\sset}$ would not help much, since it is unclear how $\meanwidth{\sset}$ scales compared to~$\meanwidth{\rset}$, cf.~\cite[Sec.~3.3]{kks17}.
	Even more notable is the fact that the above argument would break down if the $\Delta$-separation condition is omitted.
	Indeed, \citeauthor{cx15}~\cite{cx15} have shown that uniform recovery of \emph{all} $s$-gradient-sparse signals (including pathological examples) via TV~minimization is impossible with less than $\asympslower{\sqrt{s \cdot p}}$ Gaussian measurements.
	Thus, a striking gap emerges between the uniform and non-uniform regime in this scenario.
	Remarkably, the technical approach of \cite{cx15} is based on an adapted nullspace property, for which it is not obvious how to incorporate additional signal structure such as $\Delta$-separation.
	The presented example on the TV model therefore also underscores the merits of a geometric complexity analysis as proposed in our work.
\end{example}
While gradient-sparsity seems to be only slightly more involved than standard sparsity (with respect to an orthonormal basis), the above consideration has indicated many subtleties.
To a certain extent, Example~\ref{ex:unif:tv} is just the ``tip of the iceberg'' and similar phenomena apply in the more general context of the analysis- and synthesis-$\l{1}$-model. 
An in-depth discussion would go beyond the scope of the present paper, and we refer the interested reader to the related works~\cite{gkm20,mbkw20}.
The actual novelty of Proposition~\ref{prop:unif:mwloc} is that these previous findings may also extend to uniform recovery---a step that was not addressed so far.

We close this section with a remark on stable and robust recovery:
\begin{remark}
\begin{rmklist}
\item\label{rmk:unif:tuning}
	\emph{Tuning and stability.} The equality constraint for $\rset$ in \eqref{eq:unif:l1} and \eqref{eq:unif:tv:rset} ensures that every $\grtr \in \rset$ lies on the boundary of $\sset$.
	Such a technical step is important for bounds that are based on the \emph{conic} mean width, such as \eqref{eq:unif:mwl1} and \eqref{eq:unif:tv:mwloc}; if $\grtr$ would lie in the interior of $\sset$, then $\cone{\sset - \grtr} = \R^p$ and therefore $\meanwidth[0]{\sset - \grtr} \asymp \sqrt{p}$.
	From an algorithmic perspective, this can be seen as a tuning condition for the Lasso-estimator \eqref{eq:results:klasso}.
	The `moreover'-part of Proposition~\ref{prop:unif:mwloc} presents a convenient relaxation in this respect:
	When interested in the local complexity of some point $\grtr \in \rset \subset \sset$, one may instead consider a nearby point $\tilde{\grtr}$, whose conic mean width $\meanwidth[0]{\sset - \tilde{\grtr}}$ is smaller (non-trivial).
	The price for this simplification is a recovery error in the order of $\lnorm{\grtr - \tilde{\grtr}}$.
	This trade-off is closely related to the idea of \emph{stability} (or \emph{compressibility}) in compressed sensing theory \cite{fh13}.
\item\label{rmk:unif:noise}
	\emph{Robustness and optimality.} The statement of Corollary~\ref{cor:unif:lin} allows for a comparison of the statistical and adversarial noise model (cf.~Remark~\ref{rmk:results:noise}).
	In the absence of worst-case perturbations (i.e., $\noise_i = 0$), the condition \eqref{eq:unif:lin:m} can be translated into an error decay rate of $\asympfaster{\sigma \cdot m^{-1/2}}$. 
	Thus, \eqref{eq:results:klasso} becomes a \emph{consistent} estimator of $\grtr$ in the ordinary sense of statistics, i.e., $\lnorm{\pvsolu - \grtr} \to 0$ in probability as $m \to \infty$.
	Remarkably, the asymptotic error scaling with respect to the noise parameter $\sigma$ and sample size $m$ cannot be further improved in general---it is minimax optimal for linear observations, e.g., see \cite[Sec.~4]{pvy16}.
	
	Compared to random noise, the adversarial model is more general, since the $\noise_i$ may encode any type of perturbation, even deterministic ones.
	Given the $\l{2}$-constraint in Corollary~\ref{cor:unif:lin},
	\begin{equation}\label{eq:unif:noise:constr}
		\big(\textstyle\tfrac{1}{m}\sum_{i = 1}^m \noise_i^2 \big)^{1/2} \leq \tfrac{t}{20},
	\end{equation}
	it turns out that a recovery error in the order of $\asympfaster{t}$ is essentially optimal.
	Indeed, there exist instances of Corollary~\ref{cor:unif:lin} where the $\noise_i$ can be selected such that \eqref{eq:unif:noise:constr} holds with high probability (over $\{\a_i\}_{i = 1}^m$) and one has that $\lnorm{\pvsolu - \grtr} \asymp t$.\footnote{A simple example is as follows: For a unit vector $\grtr \in \S^{p-1}$ to be reconstructed, set $\noise_i = c \cdot t \cdot \sp{\a_i}{\grtr}$ for a constant $c > 0$ small enough. If $m$ is sufficiently large, then \eqref{eq:unif:noise:constr} holds with high probability. At the same time, we have that $y_i = \sp{\a_i}{\grtr} + \noise_i = \sp{\a_i}{(1 + ct)\grtr}$. Hence, Corollary~\ref{cor:unif:lin} can be also applied such that it certifies exact recovery of the rescaled vector $(1 + ct)\grtr$ via \eqref{eq:results:klasso}. In other words, we have that $\lnorm{\pvsolu - \grtr} = ct \cdot \lnorm{\grtr} \asymp t$.}
	The difference with the statistical regime becomes clearer when considering random noise from an adversarial perspective, i.e., $\noise_i = \tau_i \distributed \Normdistr{0}{\sigma^2}$.
	Then, it is not hard to see that with high probability, the constraint \eqref{eq:unif:noise:constr} is only attainable with $t \gtrsim \sigma$, which prevents arbitrarily high precision in terms of $t$.
	In particular, Corollary~\ref{cor:unif:lin} would not certify the consistency of \eqref{eq:results:klasso} anymore, contrary to the argument in the previous paragraph.
	
	Let us emphasize that uniformity does not play a special role in these matters, although the statistical noise model appears somewhat unnatural in this regime (see~Remark~\ref{rmk:results:noise}).
	In principle, our conclusions do also carry over to the abstract setting of Theorem~\ref{thm:results:main}, but the interpretation of the term `noise' becomes more subtle there. The reason is that the estimator \eqref{eq:results:klasso} basically treats non-linear distortions as if they would be uncorrelated statistical noise (cf.~\cite{pv16}), even if the output function is completely deterministic.
	Seen from this angle, better decay rates than $\asympfaster{\sigma \cdot m^{-1/2}}$ are possible, which however strongly depends on the specific model and estimator (see~Remark~\ref{rmk:results:main}\ref{rmk:results:main:optimal}). \qedhere
\end{rmklist}\label{rmk:unif}
\end{remark}

\section{Applications and Examples}
\label{sec:appl}

This section demonstrates how to derive ``out-of-the-box'' guarantees from Theorem~\ref{thm:results:main} for specific observation models.
Subsection~\ref{subsec:appl:1bit}--\ref{subsec:appl:mbit} are devoted to applications to quantized compressed sensing. Subsection~\ref{subsec:appl:sim} then revisits the case of single-index models, including new applications to modulo measurements and coordinate-wise distortions. Finally, Subsection~\ref{subsec:appl:vs} is concerned with a conceptually different example on the problem of variable selection.
Note that all proofs are deferred to Section~\ref{sec:app:appl}.

\subsection{\texorpdfstring{$1$}{1}-Bit Observations}
\label{subsec:appl:1bit}

As already indicated in Subsection~\ref{subsec:intro:literature}, the most basic version of \emph{$1$-bit compressed sensing} asks for the reconstruction of signals $\grtr\in \rset\subset \R^p$ from binary observations $\Y \in \{-1,1\}^m$ of the form
\begin{equation}\label{eq:appl:1bit:meas}
	y_i = \sign(\sp{\a_i}{\grtr}) + \noise_i, \quad i = 1, \dots, m.
\end{equation}
Here, the noise variables $\Noise\coloneqq(\noise_1, \dots, \noise_m)$ may take values in $\{-2, 0, 2\}$, modeling possible distortions of the linear measurement process before quantization and/or bit flips during quantization.
Importantly, the magnitude of $\grtr$ gets lost in \eqref{eq:appl:1bit:meas} due to the scaling invariance of the $\sign$-function.
Thus, the best one can hope for is to reconstruct the direction of $\grtr$.

\begin{figure}[t]
	\centering
	\begin{tikzpicture}[scale=1]
		\begin{axis}[
			x = 2cm,
			y = 2cm,
			axis lines=middle,
			xmin=-2.2,xmax=2.2,ymin=-1.2,ymax=1.2,
			xtick distance=1,
			ytick distance=1,
			tick label style={font=\scriptsize},
			xlabel=$s$,
			xlabel style={anchor=west},
			xticklabels={,,},
			yticklabels={,,},
			grid=both,
			minor tick num=1,
			grid style={thin,densely dotted,black!20}]
			\addplot[mark={}] coordinates {(0,1)} node[anchor=east]{\scriptsize$1$} ;
			\addplot[mark={}] coordinates {(0,-1)} node[anchor=west]{\scriptsize$-1$} ;
			\addplot[mark=*,mark options={scale=.5}] coordinates {(.25,0)} node[anchor=north,xshift=1pt]{\smaller$\frac{t}{128}$} ;
			\addplot[mark=*,mark options={scale=.5}] coordinates {(-.25,0)} node[anchor=north,xshift=-3pt]{\smaller$-\frac{t}{128}$} ;
			
			\addplot [blue,domain=-.25:.25,samples=2,thick] {4*x}  ;
			\addplot [blue,domain=.25:4,samples=2,thick] {.99}  ;
			\addplot [blue,domain=-4:-.25,samples=2,thick] {-.99} ;
			\node [pin={170:\small$\psi_t(s)$},inner sep=4pt] at (axis cs:-.125,-.75) {};

			\addplot [red,domain=-.25:0,samples=2,thick,dashed] {4*x+1}  ;
			\addplot [red,domain=0:.25,samples=2,thick,dashed] {-4*x+1}  ;
			\addplot [red,domain=.25:4,samples=2,thick,dashed] {0}  ;
			\addplot [red,domain=-4:-.25,samples=2,thick,dashed] {0} ;
			\node [pin={170:\small$\phi_t(s)$},inner sep=4pt] at (axis cs:-.125,.25) {};
			
			\addplot [[-,domain=0:4,samples=2,thick] {1}  ;
			\addplot [-),domain=-4:0,samples=2,thick] {-1} ;
			\node [pin={50:\small$\sign(s)$},inner sep=2pt] at (axis cs:-.20,-1) {};
		\end{axis}
	\end{tikzpicture}
	\caption{\emph{Illustration of the approximation strategy used in the proof of Corollary~\ref{cor:appl:1bit}.} The basic idea is to approximate the jump discontinuity of $\sign(s)$ (plotted in black) by a linear segment whose slope is inverse proportional to the accuracy $t$. The resulting function $\psi_t(s)$ (plotted in blue) and the absolute value of the error $\phi_t(s) = \abs{\psi_t(s) - \sign(s)}$ (plotted in dashed red) are then both $\tfrac{128}{t}$-Lipschitz. This is already enough to fulfill Assumption~\ref{ass:results:incr} with $\multiplLip \lesssim 1 + t^{-1}$ and $\errLip\lesssim t^{-1}$; note that the factor $128$ is just an appropriate constant for the proof.}
	\label{fig:appl:1bit}
\end{figure}
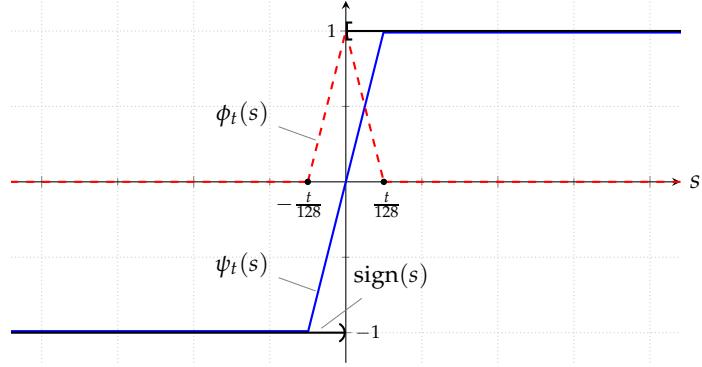

The model of \eqref{eq:appl:1bit:meas} is particularly compatible with \emph{Gaussian} measurement vectors---a related result on sub-Gaussian measurements can be found in the next subsection.
Indeed, choosing the target function $\tfunc\grtr$ proportionally to $\grtr / \lnorm{\grtr}$, one can show that the target mismatch $\mmcovar{\grtr}$ vanishes for every $\grtr\in \rset$. 
Hence, uniform recovery is possible up to arbitrarily high precision.
The following guarantee makes this claim precise and is an application of Theorem~\ref{thm:results:main} to Gaussian $1$-bit observations. 
Its proof in Subsection~\ref{subsec:app:appl:1bit} will demonstrate the usefulness of the approximation condition from Assumption~\ref{ass:results:incr}, see also Figure~\ref{fig:appl:1bit} for an illustration of the underlying argument.
\begin{corollary}\label{cor:appl:1bit}
	There exist universal constants $c, c_0, C' > 0$ for which the following holds. 
	
	Let $\a_1, \dots, \a_m \in \R^p$ be independent copies of a standard Gaussian random vector $\a \distributed \Normdistr{\vnull}{\I{p}}$.
	Let $\rset\subset \R^p$ and define $\tfunc\grtr \coloneqq \sqrt{\tfrac{2}{\pi}}\tfrac{\grtr}{\lnorm{\grtr}}$ for $\grtr \in \rset$. Moreover, let $\sset\subset \R^p$ be a convex set such that $\tfunc\rset\subset \sset$. 
	For $\probsuccess \geq 1$ and $t\in (0,1]$, we assume that
	\begin{equation}\label{eq:appl:1bit:m}
		m \geq C' \cdot t^{-2}\cdot \Big(\effdim[t]{\sset - \tfunc\rset}  + t^{-2} \cdot \effdim{\tfunc\rset}  + \probsuccess^2 \Big).
	\end{equation}
	Finally, let $\beta\in [0,1]$ be such that $\beta\sqrt{\log(e/\beta)}\leq c_0 t$. 
	Then with probability at least $1 - \exp(-c \probsuccess^2)$ on the random draw of $\{\a_i\}_{i = 1}^m$, the following holds uniformly for all $\grtr \in \rset$:
	Let $\Y \in \{-1,1\}^m$ be given by \eqref{eq:appl:1bit:meas} such that $\tfrac{1}{2m}\lnorm{\Noise}[1] \leq \beta$.
	Then every minimizer $\pvsolu$ of \eqref{eq:results:klasso} satisfies $\lnorm[\big]{\pvsolu - \sqrt{\tfrac{2}{\pi}}\tfrac{\grtr}{\lnorm{\grtr}}} \leq t$.
\end{corollary} 
Corollary~\ref{cor:appl:1bit} is in line with some of the early achievements in $1$-bit compressed sensing \cite{pv13,pv13b}.
Remarkably, the condition \eqref{eq:appl:1bit:m} translates into an error decay rate of $\asympfaster{m^{-1/4}}$ in the uniform case, which improves the original rate of $\asympfaster{m^{-1/12}}$ established by \citeauthor{pv13b} in \cite[Thm.~1.3]{pv13b}.
In fact, we are not aware of any result in the literature that implies the statement of Corollary~\ref{cor:appl:1bit}.
But we stress that the above oversampling rate is still not optimal (see \cite{jlbb13}) and can be further improved with a more specialized argument relying on random hyperplane tessellations (see Corollary~\ref{cor:appl:refined1bit} in Section~\ref{sec:appl:noincr}).
The noise constraint of Corollary~\ref{cor:appl:1bit} simply means that the fraction of wrong input bits $\tfrac{1}{2m}\lnorm{\Noise}[1]$ must not exceed $\beta$, while the latter may be in the order of $t$ (up to a log-factor).
This condition again significantly improves \cite[Thm.~1.3]{pv13b} and is a particular consequence of the outlier robustness established in Theorem~\ref{thm:results:main}; see also \cite{dm18} for a similar achievement in the situation of dithered observations.

\subsection{\texorpdfstring{$1$}{1}-Bit Observations with Dithering}
\label{subsec:appl:1bitdither}

According to \citeauthor{alpv14}~\cite{alpv14}, the conclusion of Corollary~\ref{cor:appl:1bit} cannot be extended to sub-Gaussian measurements in general, regardless of the considered reconstruction method.
A practicable remedy is the technique of \emph{dithering}, which in its most basic form, corresponds to a random shift of the quantization threshold.
Originating from quantized signal processing, e.g., see \cite{gs93,gn98,dk06}, its benefits recently also emerged in compressed sensing theory \cite{ksw16,bfnpw17,xj18,dm18,dm18b,tr18}.
For more background information, we refer to \cite{dir19} and the references therein.

Extending the original $1$-bit model \eqref{eq:appl:1bit:meas} by an additional dithering step leads to observations $\Y \in \{-1,1\}^m$ of the following form:
\begin{equation}\label{eq:appl:1bitdither:meas}
	y_i = \sign(\sp{\a_i}{\grtr} + \tau_i) + \noise_i, \quad i = 1, \dots, m,
\end{equation}
where $\Noise\coloneqq(\noise_1, \dots, \noise_m)\in \{-2,0,2\}^m$ again models noise.
The \emph{dithering variables} $\tau_i$ are independent copies of a random variable $\tau$ that is uniformly distributed on an interval $[-\lambda, \lambda]$. 
Note that a major difference between dithering and additive noise is that the parameter $\lambda > 0$ is known and adjustable in practice (while the $\tau_i$ could be in principle unknown). 
The following corollary of Theorem~\ref{thm:results:main} is based on the fact that a careful choice of $\lambda$ allows us to control the size of the (non-vanishing) target mismatch $\mmcovar{\grtr}$, where $\tfunc\grtr \coloneqq \lambda^{-1} \grtr$.
\begin{corollary}\label{cor:appl:1bitdither}
	There exist universal constants $c, c_0, \tilde{C}, C' > 0$ for which the following holds.
	
	Let $\a_1, \dots, \a_m \in \R^p$ be independent copies of a centered, isotropic, sub-Gaussian random vector $\a\in \R^p$ with $\normsubg{\a} \leq \subgparam$. Let $\tau_1, \dots, \tau_m$ be independent copies of a random variable $\tau$ that is uniformly distributed on $[-\lambda, \lambda]$ for a parameter $\lambda>0$. In addition, suppose that $\{\a_i\}_{i=1}^m$ and $\{\tau_i\}_{i=1}^m$ are independent.
	Let $\rset\subset R \ball[2][p]$ for some $R > 0$ and let $\sset\subset \R^p$ be a convex set such that $\lambda^{-1}\rset\subset \sset$. 
	For $\probsuccess \geq 1$ and $t\in (0, 1]$, we assume that
	\begin{align}
		\lambda &\geq \tilde{C} \cdot R \cdot \subgparam \cdot \sqrt{\log(e / t)}, \label{eq:appl:1bitdither:lambda}\\
		m &\geq C' \cdot \subgparam^2 \cdot \Big(  (\log\subgparam + t^{-2} )\cdot \big(\effdim[t]{\sset - \lambda^{-1}\rset}+ \probsuccess^2 \big) + \subgparam^2 t^{-4} \lambda^{-2}  \cdot \effdim{\rset} \Big). \label{eq:appl:1bitdither:m}
	\end{align}
	Finally, let $\beta\in [0,1]$ be such that $\beta\sqrt{\log(e/\beta)}\leq c_0 \subgparam^{-2} t$. 
	Then with probability at least $1 - \exp(-c \probsuccess^2)$ on the random draw of $\{(\a_i, \tau_i)\}_{i = 1}^m$, the following holds uniformly for all $\grtr \in \rset$:
	Let $\Y \in \{-1,1\}^m$ be given by \eqref{eq:appl:1bitdither:meas} such that $\tfrac{1}{2m}\lnorm{\Noise}[1] \leq \beta$. Then every minimizer $\pvsolu$ of \eqref{eq:results:klasso} satisfies $\lnorm{\pvsolu - \lambda^{-1}\grtr} \leq t$.
\end{corollary}
Corollary~\ref{cor:appl:1bitdither} exhibits many common features of known recovery guarantees based on dithering.
In particular, it can be seen as a uniform version of a recent result on the generalized Lasso by \citeauthor{tr18}~\cite[Thm.~IV.1]{tr18}.
The most notable improvement over Corollary~\ref{cor:appl:1bit} is that it is now possible to recover the actual signal $\grtr \in \rset$ up to arbitrarily high precision, and not only its direction vector $\grtr / \lnorm{\grtr}$.

Similarly to Corollary~\ref{cor:appl:1bit}, we are not aware of any result in the literature that implies the statement of Corollary~\ref{cor:appl:1bitdither}.
Nevertheless, the oversampling rate of $\asympfaster{m^{-1/4}}$ promoted by \eqref{eq:appl:1bitdither:m} can be further improved with a proof strategy that is specifically tailored to $1$-bit observations with dithering (see Corollary~\ref{cor:appl:refined1bitdither} in Section~\ref{sec:appl:noincr}).

\subsection{Multi-Bit Observations}
\label{subsec:appl:mbit}

While $1$-bit measurements are an important extreme case of quantized compressed sensing, a considerable part of the literature deals with multi-bit observation models; once again, see \cite{dir19,bjks15} for a good introduction to this subject.
In this subsection, we illustrate our approach in the prototypical situation of \emph{uniform quantization}: For a fixed $\delta > 0$, the goal is to reconstruct a signal $\grtr\in \rset\subset \R^p$ from observations $\Y \in \delta \Z^m$ of the form  
\begin{equation}\label{eq:appl:mbit:meas}
	y_i = q_\delta(\sp{\a_i}{\grtr} + \tau_i) + \noise_i, \quad i = 1, \dots, m,
\end{equation}
where $q_\delta(v) \coloneqq (2\ceil{\tfrac{v}{2\delta}}-1)\delta$ is a uniform quantizer on the grid $\delta (2\Z-1)^m$ with resolution $\delta>\nobreak0$.
The dithering variables $\tau_i$ are independent copies of a random variable $\tau$ that is uniformly distributed on $[-\delta, \delta]$.
As before, $\Noise\coloneqq(\noise_1, \dots, \noise_m)$ can describe any type of adversarial noise, but now takes values in $\delta \Z^m$.
Theorem~\ref{thm:results:main} yields the following uniform recovery guarantee for multi-bit observations with sub-Gaussian measurements and dithering.
\begin{corollary}\label{cor:appl:mbit}
	There exist universal constants $c, c_0, C' > 0$ for which the following holds. 
	
	Let $\a_1, \dots, \a_m \in \R^p$ be independent copies of a centered, isotropic, sub-Gaussian random vector $\a\in \R^p$ with $\normsubg{\a} \leq \subgparam$.
	Let $\tau_1, \dots, \tau_m$ be independent copies of a random variable $\tau$ that is uniformly distributed on $[-\delta, \delta]$ for a fixed parameter $\delta>0$. 
	In addition, suppose that $\{\a_i\}_{i=1}^m$ and $\{\tau_i\}_{i=1}^m$ are independent.
	Let $\rset\subset \R^p$ be a bounded subset and let $\sset\subset \R^p$ be a convex set such that $\rset\subset \sset$. 
	For $\probsuccess \geq 1$ and $t > 0$, we assume that
	\begin{equation}\label{eq:appl:mbit:m}
	m \geq C' \cdot \subgparam^2 \cdot \Big( (\log\subgparam + t^{-2} \delta^2) \cdot \big(\effdim[t]{\sset - \rset}+ \probsuccess^2 \big) +  \subgparam^2 t^{-4} \delta^2\cdot \effdim{\rset} \Big).
	\end{equation}
	Then with probability at least $1 - \exp(-c \probsuccess^2)$ on the random draw of $\{(\a_i, \tau_i)\}_{i = 1}^m$, the following holds uniformly for all $\grtr \in \rset$:
	Let $\Y \in \delta \Z^m$ be given by \eqref{eq:appl:mbit:meas} such that at least one of the following conditions is fulfilled:
	\begin{equation}\label{eq:appl:mbit:input}
		\text{(a) $\tfrac{1}{\sqrt{m}}\lnorm{\Noise}\leq \tfrac{t}{40}$ \quad or \quad
			(b) $\tfrac{1}{\sqrt{m}}\lnorm{\Noise}\leq \tfrac{c_0\sqrt{\delta t}}{\subgparam^2\sqrt{\max\{1, \log( \delta e/t)\}}}$ \ and \ $\tfrac{1}{m} \lnorm{\Noise}[0] \leq \tfrac{t}{\delta}$.}
	\end{equation}
	Then every minimizer $\pvsolu$ of \eqref{eq:results:klasso} satisfies $\lnorm{\pvsolu - \grtr} \leq 2t$.
\end{corollary}
Corollary~\ref{cor:appl:mbit} can be considered as a uniform version of a recent result by \citeauthor{tr18}~\cite[Thm.~III.1]{tr18}.
For a fixed quantizer resolution $\delta>0$, the condition \eqref{eq:appl:mbit:m} translates into an error decay rate of $\asympfaster{\sqrt{\delta}\cdot m^{-1/4}}$.
On the other hand, if $\delta \ll t$, then \eqref{eq:appl:mbit:m} is satisfied as soon as $m \geq C \cdot \subgparam^2 \log\subgparam  \cdot (\effdim[t]{\sset - \rset}+ \probsuccess^2 )$. 
In other words, if the quantizer resolution is much higher than the desired reconstruction accuracy, the performance of \eqref{eq:results:klasso} is the same as if the input vector would consist of linear measurements (cf. Corollary~\ref{cor:unif:lin}).
This behavior is perfectly consistent with the fact that the uniformly quantized observations \eqref{eq:appl:mbit:meas} become linear as $\delta \to 0$.
It is also worth pointing out that the constraint (b) in \eqref{eq:appl:mbit:input} implies a certain outlier robustness: if $t < \delta$ and only a fraction of $t / \delta$ bits are corrupted, then the normalized $\l{2}$-noise error may scale in the order of $\sqrt{\delta t}$ instead of~$t$.
 
To the best of our knowledge, Corollary~\ref{cor:appl:mbit} is a new result.
But we stress that there exist uniform recovery guarantees for other programs than \eqref{eq:results:klasso} in the literature, which imply better oversampling rates, e.g., see \cite[Thm.~3]{jmps19} or \cite{xj18} in the case of structured signal sets. 
Nevertheless, we expect that Corollary~\ref{cor:appl:mbit} could be easily improved in that regard, using the strategy of Section~\ref{sec:appl:noincr} in conjunction with known uniform embedding results.

\subsection{Single-Index Models and Beyond}
\label{subsec:appl:sim}

The situation of single-index models was already considered in Theorem~\ref{thm:intro:sim} in Subsection~\ref{subsec:intro:contrib}, as an appetizer for our more general approach.
As pointed out there, this guarantee can be seen as an upgrade of an earlier non-uniform result by \citeauthor{pv16}~\cite[Thm.~1.9]{pv16}, thereby demonstrating that uniform recovery is possible beyond quantized measurement schemes.
In view of Assumption~\ref{ass:results:incr}, the Lipschitz continuity of $\fobs$ in Theorem~\ref{thm:intro:sim} is certainly not necessary.
However, it verifies that a more ``regular'' observation variable can lead to a near-optimal error decay rate (see~Remark~\ref{rmk:results:main}\ref{rmk:results:main:optimal}).
Apart from that, we emphasize that (Lipschitz) continuous non-linearities are not only of academic interest but also appear in practical applications, for instance, as power amplifiers in sensor networks \cite{gj19}.

The remainder of this subsection is devoted to two new applications of our general framework, which extend Theorem~\ref{thm:intro:sim} into different directions.

\paragraph{Modulo measurements.}
The first scenario is inspired by a recent work of \citeauthor{bkr20}~\cite{bkr20} on `unlimited sampling'.
The practical motivation there is to prevent a saturation of analog-to-digital converters by applying a modulo operator in the measurement process.
Tailored to the setup of the present article, we consider \emph{modulo measurements} of the form
\begin{equation}\label{eq:appl:modulo:meas}
	y_i = \modulo_\lambda(\sp{\a_i}{\grtr}) + \noise_i, \quad i = 1, \dots, m,
\end{equation}
where the modulo function is given by $\modulo_\lambda(v)\coloneqq v - \floor[\big]{\tfrac{v+\lambda}{2\lambda}}\cdot 2\lambda$ for a fixed parameter $\lambda>0$, and $\Noise\coloneqq(\noise_1, \dots, \noise_m) \in \R^m$ models noise.
Clearly, the non-linearity $\modulo_\lambda$ neither corresponds to a quantization (cf.~Subsection~\ref{subsec:appl:1bit}--\ref{subsec:appl:mbit}) nor is it Lipschitz continuous (cf.~Theorem~\ref{thm:intro:sim}).
The following corollary of Theorem~\ref{thm:results:main} provides a uniform recovery guarantee for the observation model \eqref{eq:appl:modulo:meas}:
\begin{corollary}\label{cor:appl:modulo}
	There exist universal constants $c, C, C' > 0$ for which the following holds.
	
	Let $\a_1, \dots, \a_m \in \R^p$ be independent copies of a standard Gaussian random vector $\a \distributed \Normdistr{\vnull}{\I{p}}$. 
	For $\lambda>0$, we set $\scalfac_\lambda\coloneqq \mean[\modulo_\lambda(\gaussianuniv)\gaussianuniv]$ with $\gaussianuniv \distributed \Normdistr{0}{1}$. 
	Let $\rset\subset \S^{p-1}$ and let $\sset\subset \R^p$ be a convex set such that $\scalfac_\lambda\rset\subset \sset$.
	For $\probsuccess \geq 1$ and $t\in (0,\lambda]$, we assume that $\lambda\geq C$ and
	\begin{equation}\label{eq:appl:modulo:m}
		m \geq C'\cdot \Big((1+t^{-2})\cdot \big(\effdim[t]{\sset - \scalfac_\lambda\rset}+\probsuccess^2\big)+\lambda^2t^{-4}\cdot \effdim{\rset}\Big).
	\end{equation}
	Then $\scalfac_\lambda\in [\tfrac{1}{2},1]$ and with probability at least $1 - \exp(-c \probsuccess^2)$ on the random draw of $\{\a_i\}_{i = 1}^m$, the following holds uniformly for all $\grtr \in \rset$: Let $\Y \in \R^m$ be given by \eqref{eq:appl:modulo:meas} such that $\big(\tfrac{1}{m}\sum_{i = 1}^m \noise_i^2 \big)^{1/2} \leq \tfrac{t}{20}$. Then every minimizer $\pvsolu$ of \eqref{eq:results:klasso} satisfies $\lnorm{\pvsolu - \scalfac_\lambda\grtr} \leq t$.
\end{corollary}
Similar to our above applications to quantized compressed sensing, the (infinitely-many) discontinuities of the modulo function $\modulo_\lambda$ imply an error decay rate of $\asympfaster{m^{-1/4}}$.
To the best of our knowledge, Corollary~\ref{cor:appl:modulo} is a new result in its own right.
In particular, we are not aware of alternative proof techniques that would allow us to derive a comparable statement.
Finally, it is worth noting that the condition $\lambda \geq C$ in Corollary~\ref{cor:appl:modulo} could be further relaxed to $\lambda > 0$.
This would come at the cost of a lower bound for $\scalfac_\lambda$, since we have that $\scalfac_\lambda \to 0$ as $\lambda \to 0$ (the non-linearity~$\modulo_\lambda$ uniformly converges to $0$ as $\lambda \to 0$).

\paragraph{Coordinate-wise distortions.}
The common single-index model (see Theorem~\ref{thm:intro:sim}) assumes that a non-linear output function perturbs the linear observations $\sp{\a_i}{\grtr}$.
Instead, one can also imagine distortions that affect the computation of the dot product $\sp{\a_i}{\grtr}$ directly.
More specifically, we are interested in observations of the form
\begin{equation}\label{eq:appl:beyond:meas}
	y_i = \sumf(\a_i \circ \grtr) + \noise_i, \quad i = 1, \dots, m,
\end{equation}
where $\vec{a}\circ\vec{b}\in \R^p$ denotes the coordinate-wise product of vectors $\vec{a}, \vec{b}\in \R^p$ and $\sumf:\R^p\to \R$ can be represented as
\begin{equation}\label{eq:appl:beyond:sumf}
	\sumf(\vec{z})\coloneqq \sum_{j=1}^p f_j(z_j), \quad \vec{z}\in \R^p.
\end{equation}
Here, we assume that the functions $f_j:\R\to \R$ are odd and $\gamma$-Lipschitz, while satisfying the growth conditions
\begin{equation}\label{eq:appl:beyond:growth1}
	\alpha (v-v')\leq f_j(v)-f_j(v'), \quad \text{for all $v, v'\in \R$ with $v'\leq v$,}
\end{equation}
and
\begin{equation}\label{eq:appl:beyond:growth2}
	\beta_1 v\leq f_j(v)\leq \beta_2 v, \quad \text{for all $v\geq 0$,}
\end{equation}
with parameters $\alpha > 0$ and $\beta_2 \geq \beta_1 > 0$.
One can think of \eqref{eq:appl:beyond:meas} as computing a ``non-linear dot product'' of $\a_i$ and $\grtr$.
While models of this type are not covered by the original approach of \cite{pv16}, an application of Theorem~\ref{thm:results:main} allows us to deal with them under appropriate assumptions:
\begin{corollary}\label{cor:appl:beyond}
There exist universal constants $c, C > 0$ for which the following holds.

Let $\a_1, \dots, \a_m \in \R^p$ be independent copies of a centered, isotropic random vector $\a = (a_1, \dots, a_p)$ which has independent, symmetric, and sub-Gaussian coordinates such that $\max_{j \in [p]}\normsubg{a_j} \leq \subgparam$ for some $\subgparam > 0$. Let $\rset\subset \rad\ball[2][p]$ and 
define $\tfunc:\rset\to \R^p$ by $\tfunc\grtr\coloneqq \mean[\sumf(\a\circ \grtr)\a]$.
Let $\sset\subset \R^p$ be a convex set such that $\tfunc \rset\subset \sset$.
For $\probsuccess \geq 1$ and $t\geq 0$, we assume that
\begin{equation}\label{eq:appl:beyond:m}
m \geq C \cdot\subgparam^4\cdot \Big(\big(1+t^{-2}\rad^2(\beta_2^2+\gamma^2)\big)\cdot \big(\effdim[t]{\sset - \tfunc\rset}+ \probsuccess^2\big)+ t^{-2}(\tfrac{\gamma}{\alpha})^2\cdot \effdim{\tfunc\rset}\Big).
\end{equation}
Then with probability at least $1 - \exp(-c \probsuccess^2)$ on the random draw of $\{\a_i\}_{i = 1}^m$, the following holds uniformly for all $\grtr \in \rset$: Let $\Y \in \R^m$ be given by \eqref{eq:appl:beyond:meas} such that $\big(\tfrac{1}{m}\sum_{i = 1}^m \noise_i^2 \big)^{1/2} \leq \tfrac{t}{20}$. Then every minimizer $\pvsolu$ of \eqref{eq:results:klasso} satisfies $\lnorm{\pvsolu - \tfunc \grtr} \leq t$. Furthermore, for every $j\in [p]$, we have that
\begin{align}
	\beta_1x_j&\leq (\tfunc\grtr)_j\leq \beta_2x_j, \qquad \text{for $x_j\geq 0$,}\\
	\beta_2x_j&\leq (\tfunc\grtr)_j\leq \beta_1x_j, \qquad \text{for $x_j<0$.} \label{eq:cor:appl:beyond:target_vec}
\end{align}
\end{corollary}

The above result shows that estimation via \eqref{eq:results:klasso} is even possible in situations where instead of linear measurements $\sp{\a_i}{\grtr}=\sum_{j=1}^p(\a_i)_jx_j$ one observes distorted dot products $\sumf(\a_i\circ\grtr)=\sum_{j=1}^pf_j((\a_i)_jx_j)$. Note that the generalized Lasso may not precisely recover a scalar multiple of~$\grtr$ here, but rather a vector $\tfunc\grtr$ that lies in a hyperrectangle defined by~$\grtr$ and the parameters~$\beta_1$ and~$\beta_2$. The closer the functions $f_j$ are to the identity, the closer~$\beta_1$ and~$\beta_2$ are to~$1$, which implies that $\tfunc\grtr\approx \grtr$ according to~\eqref{eq:cor:appl:beyond:target_vec}.

\subsection{Variable Selection}
\label{subsec:appl:vs}

This subsection is devoted to an instance of Assumption~\ref{ass:results:meas} that is substantially different from the previous ones: For a fixed integer $s \leq p$ and an output function $\fobs \colon \R^p \to \R$, we ask for estimation of an index set $\suppset \subset [p]$ with $\cardinality{\suppset} \leq s$ from observations $\Y \in \R^m$ of the form
\begin{equation}\label{eq:meas:vs}
	y_i =  \fobs(\a_{i,\suppset}) + \noise_i, \quad i = 1, \dots, m.
\end{equation}
Here, $\a_{i,\suppset} \in \R^p$ is the coordinate projection of $\a_i$ onto $\suppset$ and $\noise_i \in \R$ models additive noise.
Most notably, the signals of interest are not parameters vectors in $\R^p$ anymore but correspond to (small) index sets, specifying those coefficients of a measurement vector that contribute to the observation.
From a statistical perspective, this can be seen as a \emph{variable selection model}, where $\suppset \subset [p]$ determines the set of active variables among all feature variables in $\a_i$.
In the context of uniform recovery, this leads to the following problem: Given a collection of sample data $\{\a_i\}_{i = 1}^m$, can we retrieve any possible index set $\suppset \subset [p]$ with $\cardinality{\suppset} \leq s$ from (non-linear) observations of the form \eqref{eq:meas:vs}?
The following corollary of Theorem~\ref{thm:results:main} provides a result in that direction under natural conditions on the output function~$\fobs$.
\begin{corollary}\label{cor:appl:vs}
There exist universal constants $c, C > 0$ for which the following holds.

Let $\a_1, \dots, \a_m \in \R^p$ be independent copies of a centered, isotropic random vector $\a = (a_1, \dots, a_p)$ which has independent sub-Gaussian coordinates such that $\max_{j \in [p]}\normsubg{a_j} \leq \subgparam$ for some $\subgparam > 0$. For $s \leq p$ and an output function $\fobs \colon \R^p \to \R$, let $\rset \subset \{\suppset\subset [p] \suchthat \cardinality{\suppset}\leq s\}$ and define $\tfunc\colon \rset \to \R^p$ by $\tfunc \suppset \coloneqq \mean[\fobs(\a_{\suppset})\a]$, where $\a_{\suppset} \in \R^p$ denotes the coordinate projection of $\a$ onto $\suppset$. Moreover, we assume that there exist parameters $\alpha, \beta, \Lip, \srradius > 0$ such that 
\begin{equation}\label{eq:appl:vs:target}
	\frac{\alpha}{\sqrt{s}}\leq \abs{(\tfunc \suppset)_j} \leq \frac{\beta}{\sqrt{s}}\qquad \text{for all $\suppset\in \rset$ and $j\in \suppset$}
\end{equation}
and
\begin{equation}\label{eq:appl:vs:subg}
	\normsubg{\fobs(\a_{\suppset}) - \fobs(\a_{\suppset'})} \leq \Lip\sqrt{\frac{\cardinality{\suppset\bigtriangleup \suppset'}}{s}} \quad{\text{and}}\quad \normsubg{\fobs(\a_{\suppset})} \leq \srradius \qquad \text{for all $\suppset, \suppset'\in \rset$,}
\end{equation}
where $\suppset\bigtriangleup \suppset' \subset [p]$ denotes the symmetric difference between $\suppset$ and $\suppset'$.
Let $\sset\subset \R^p$ be a convex set such that $\tfunc \rset\subset \sset$.
For $\probsuccess \geq 1$ and $t\geq 0$, we assume that
\begin{equation}\label{eq:appl:vs:m}
m \geq C \cdot \subgparam^2 \cdot \Big( \big(\log\subgparam + t^{-2} (\subgparam^2 \beta^2 + \srradius^2)\big)\cdot \big(\effdim[t]{\sset - \tfunc\rset}+ \probsuccess^2 \big) +
 t^{-2}  (\subgparam^2 + \Lip^2\alpha^{-2})\cdot \effdim{\tfunc\rset} \Big).
\end{equation}
Then with probability at least $1 - \exp(-c \probsuccess^2)$ on the random draw of $\{\a_i\}_{i = 1}^m$, the following holds uniformly for all $\suppset \in \rset$: Let $\Y \in \R^m$ be given by \eqref{eq:meas:vs} such that $\big(\tfrac{1}{m}\sum_{i = 1}^m \noise_i^2 \big)^{1/2} \leq \tfrac{t}{20}$. Then $\supp(\tfunc \suppset) = \suppset$ and every minimizer $\pvsolu$ of \eqref{eq:results:klasso} satisfies $\lnorm{\pvsolu - \tfunc \suppset} \leq t$.
\end{corollary}
The assumptions \eqref{eq:appl:vs:target} and \eqref{eq:appl:vs:subg} can be seen as natural balancing properties of the underlying observation model: \eqref{eq:appl:vs:target} requires that the coefficients of each target vector $\tfunc \suppset \in \sset$ are uniformly bounded below and above on $\suppset$ (and in particular $\alpha \leq \lnorm{\tfunc \suppset} \leq \beta$).
The increment condition \eqref{eq:appl:vs:subg} ensures that the distance between two observation variables can be controlled in terms of the symmetric difference of their associated index sets in $\rset$.

With this in mind, Corollary~\ref{cor:appl:vs} suggests the following simple procedure for variable selection: first perform a hard-thresholding step on $\pvsolu$ to extract its largest entries in magnitude; then use the corresponding indices to estimate the set of active variables $\suppset = \supp(\tfunc \suppset)$.
Note that this does not require explicit knowledge of the output function $\fobs$.
However, the (guaranteed) success of such a strategy strongly depends on the size of $\alpha$ and the accuracy $t$. In the worst case, $t$ would have to be in the order of $\alpha / \sqrt{s}$ for perfect recovery of $\suppset$, which would lead to an undesirable factor of $s$ in \eqref{eq:appl:vs:m}.
It is certainly possible to show refined versions of Corollary~\ref{cor:appl:vs}.
We suspect that sharper error bounds could be obtained by considering a different error measure than the $\l{2}$-norm (see Remark~\ref{rmk:results:exten}\ref{rmk:results:exten:metrics}). Nevertheless, a detailed elaboration would go beyond the scope of this article, and we confine ourselves with the above proof of concept.

\section{Uniform Recovery Without Increment Conditions}
\label{sec:appl:noincr}

We have seen in Subsection~\ref{subsec:appl:1bit}--\ref{subsec:appl:mbit} that the increment conditions of Assumption~\ref{ass:results:incr} can lead to inferior error decay rates for quantizing output functions, due to their points of discontinuity.
In this section, we present a workaround that can do without any (sub-Gaussian) increment conditions and thereby enables significantly better rates.
The basic idea is to cover the signal set $\rset$ by an $\eps$-net $\rset_\eps$ for some $\eps > 0$ and to apply the \emph{non-uniform} version of Theorem~\ref{thm:results:main} to each $\grtr \in \rset_\eps$ separately.
Taking the union bound then yields uniform recovery on $\rset_\eps$.
The final, and most crucial, ingredient that allows us to pass over to the entire signal set $\rset$ is the following \emph{local stability condition} on the observation model. It is based on the (outlier) noise bounds in \eqref{eq:results:main:rob} and essentially requires that close points in $\tfunc\rset$ imply close observation vectors.
\begin{assumption}\label{ass:local_robustness}
	Let Assumption~\ref{ass:results:meas} be satisfied and let $t > 0$ and $\eps > 0$. For $\nout\in \{0, 1, \dots, \floor{\tfrac{m}{2}}\}$, $\tune>0$, and  $\eta\in [0,1]$, we assume that the following holds with probability at least $1-\eta$:
	\begin{equation}\label{eq:ass:local_robustness:event}
		\sup_{\substack{\grtr,\, \grtr'\in \rset\\ \lnorm{\tfunc\grtr-\tfunc\grtr'}\leq \eps}}\tfrac{1}{\sqrt{m}}\norm{\Ys(\grtr) - \Ys(\grtr')}_{[2\nout]}\leq \tfrac{1}{2}\tune t \quad \text{ and } \quad 
		\sup_{\substack{\grtr,\, \grtr'\in \rset\\ \lnorm{\tfunc\grtr-\tfunc\grtr'}\leq \eps}}\tfrac{1}{\sqrt{m}}\sigma_{\nout}(\Ys(\grtr) - \Ys(\grtr'))_{2}\leq \tfrac{t}{40}.
	\end{equation} 
\end{assumption}
The above strategy leads to the following general uniform recovery guarantee; see Section~\ref{sec:app:appl:noincr} for a detailed proof.
\begin{theorem}\label{thm:appl:without_increments}
	There exist universal constants $c, C, C_0 > 0$ for which the following holds.
	
	Let Assumptions~\ref{ass:results:meas} and~\ref{ass:local_robustness} be satisfied, let $\multiplDia \coloneqq \sup_{\grtr\in \rset}\normsubg{\sp{\a}{\tfunc\grtr} -\nobreak \ys(\grtr)}$, and assume $\mmcovar{\grtr} \leq\nobreak \tfrac{t}{32}$ for every $\grtr \in \rset$.
	For $\probsuccess \geq 1$ and $\probsuccess_0 \geq \sqrt{2\nout \log(em/2\nout)}$, we assume that
	\begin{equation}\label{eq:appl:without_increments:m}
		m \geq C \cdot \subgparam^2 \cdot \Big((\log\subgparam+t^{-2} \multiplDia^2) \cdot \big(\sup_{\grtr \in \rset}\effdim[t]{\sset - \tfunc\grtr} +  \probsuccess^2 \big) + \subgparam^2\tune^2 \cdot \big(\sup_{\grtr \in \rset}\effdim[t]{\sset - \tfunc\grtr}+ \probsuccess_0^2 \big)\Big)
	\end{equation}
	and
	\begin{equation}\label{eq:appl:without_increments:u}
		\min\{\probsuccess^2,\probsuccess_0^2\} \geq C_0\cdot \log\covnumber[\eps]{\tfunc\rset}.
	\end{equation}
	Then with probability at least $1 - \exp(-c \probsuccess^2) - \exp(-c \probsuccess_0^2)-\eta$ on the random draw of $\{(\a_i, \fout_i)\}_{i = 1}^m$, the following holds uniformly for every $\grtr \in \rset$:
	Let $\Y \in \R^m$ be any input vector such that
	\begin{equation}\label{eq:appl:without_increments:rob}
		\tfrac{1}{\sqrt{m}}\norm{\Y - \Ys(\grtr)}_{[2\nout]}\leq \tfrac{1}{2}\tune t \quad \text{ and } \quad 
		\tfrac{1}{\sqrt{m}}\sigma_{\nout}(\Y - \Ys(\grtr))_{2}\leq \tfrac{t}{40}.
	\end{equation} 
	Then every minimizer $\pvsolu$ of \eqref{eq:results:klasso} satisfies $\lnorm{\pvsolu - \tfunc\grtr} \leq t + \eps$.
\end{theorem}
The statement of Theorem~\ref{thm:appl:without_increments} strongly resembles the one of Theorem~\ref{thm:results:main} with the important difference that the former does not rely on Assumption~\ref{ass:results:incr}. In particular, the condition \eqref{eq:appl:without_increments:m} does not depend on the increment parameters $\multiplLip$ and $\errLip$ anymore, making it less restrictive than \eqref{eq:results:main:m}.
It is worth pointing out that the global complexity of $\rset$ is now measured in terms of the covering number $\covnumber[\eps]{\tfunc\rset}$ in \eqref{eq:appl:without_increments:u} instead of the mean width $\meanwidth{\tfunc\rset}$.
Furthermore, \eqref{eq:appl:without_increments:m} establishes a refined local complexity measure, due to $\sup_{\grtr \in \rset} \effdim[t]{\sset - \tfunc\grtr} \leq \effdim[t]{\sset - \tfunc\rset}$; see also Proposition~\ref{prop:unif:mwloc}.
Nevertheless, the gain of Theorem~\ref{thm:appl:without_increments} is obviously linked to the verification of Assumption~\ref{ass:local_robustness}, which is usually a highly non-trivial task.

We now present two applications of Theorem~\ref{thm:appl:without_increments} in the prototypical case of $1$-bit observations.
In this specific situation, it turns out that Assumption~\ref{ass:local_robustness} is compatible with uniform bounds for binary embeddings, which allows us to make use of related results from the literature.
Our first application is based on the following embedding guarantee by \citeauthor{or15}~\cite{or15} for noiseless, Gaussian $1$-bit measurements (cf.~Subsection~\ref{subsec:appl:1bit}).
Hereafter, we agree on the shortcut notation $[\ssetgen]_\eps \coloneqq (\tfrac{1}{\eps}\ssetgen - \tfrac{1}{\eps}\ssetgen) \intersec \ball[2][p]$ for any subset $\ssetgen \subset \R^p$ and $\eps > 0$; also note that the parameter $\meanwidth{[\ssetgen]_\eps}$ is virtually the same as the local mean width $\meanwidth[\eps]{\ssetgen - \ssetgen}$ except that the former intersects with $\ball[2][p]$ instead of $\S^{p-1}$.
\begin{theorem}[\protect{\cite[Thm.~3.2]{or15}}]\label{thm:hyperplane:tessellation:1bit}
	There exist universal constants $c, \bar{c}, C > 0$ for which the following holds.
	
	Let $\ssetgen \subset \S^{p-1}$ and let $\A\in \R^{m\times p}$ be a random matrix with independent standard Gaussian row vectors $\a_1, \dots, \a_m\in \R^p$.
	For $\beta\in (0,1)$ and $\eps\leq \bar{c}\beta / \sqrt{\log(e/\beta)}$, we assume that
	\begin{equation}
		m\geq C\cdot \Big(\eps^2 \beta^{-3} \cdot \effdim{[\ssetgen]_\eps} + \beta^{-1} \cdot \log\covnumber[\eps]{\ssetgen}\Big).
	\end{equation}
	Then the following holds with probability at least $1-\exp(-cm\beta)$:
	\begin{equation}
		\sup_{\substack{\grtr,\, \grtr'\in \ssetgen\\ \lnorm{\grtr - \grtr'}\leq \eps}}\tfrac{1}{2m}\lnorm{\sign(\A\grtr)-\sign(\A\grtr')}[1]\leq \beta.
	\end{equation}
\end{theorem}	
Combining Theorem~\ref{thm:hyperplane:tessellation:1bit} with Theorem~\ref{thm:appl:without_increments} yields an improved version of Corollary~\ref{cor:appl:1bit} (see Section~\ref{sec:app:appl:noincr} for a proof):
\begin{corollary}\label{cor:appl:refined1bit}
	There exist universal constants $c, c', c_0, C', C_0 > 0$ for which the following holds. 
	
	Let $\a_1, \dots, \a_m \in \R^p$ be independent copies of a standard Gaussian random vector $\a \distributed \Normdistr{\vnull}{\I{p}}$.
	Let $\rset\subset \R^p$ and define $\tfunc\grtr \coloneqq \sqrt{\tfrac{2}{\pi}}\tfrac{\grtr}{\lnorm{\grtr}}$ for $\grtr \in \rset$. Moreover, let $\sset\subset \R^p$ be a convex set such that $\tfunc\rset\subset \sset$. 
	Fix $t\in (0,1)$ and $\eps\leq c't / \log(e/t)$. For $\probsuccess^2\geq \max\{1, C_0 \cdot \log\covnumber[\eps]{\tfunc\rset}\}$,
	we assume that
	\begin{align}\label{eq:appl:refined1bit:m}
		m &\geq C' \cdot \Big( t^{-2} \cdot \big(\sup_{\grtr \in \rset}\effdim[t]{\sset - \tfunc\grtr} + \probsuccess^2 \big)+ \eps^2 t^{-3}\log^{3/2}(e/t)\cdot\effdim{[\tfunc\rset]_\eps}\Big).
	\end{align}
	Finally, let $\beta\in [0,1]$ be such that $\beta\sqrt{\log(e/\beta)}\leq c_0 t$. 
	Then with probability at least $1 - \exp(-c \probsuccess^2)$ on the random draw of $\{\a_i\}_{i = 1}^m$, the following holds uniformly for all $\grtr \in \rset$:
	Let $\Y \in \{-1,1\}^m$ be given by \eqref{eq:appl:1bit:meas} such that $\tfrac{1}{2m}\lnorm{\Noise}[1]\leq \tfrac{\beta}{2}$.
	Then every minimizer $\pvsolu$ of \eqref{eq:results:klasso} satisfies $\lnorm[\big]{\pvsolu - \sqrt{\tfrac{2}{\pi}}\tfrac{\grtr}{\lnorm{\grtr}}} \leq 2t$.
\end{corollary}
To better understand the oversampling behavior of this result, let us consider the situation where $\eps \asymp t / \log(e/t)$.
Due to Sudakov minoration, we have that $\log\covnumber[\eps]{\tfunc\rset} \lesssim \eps^{-2} \cdot \effdim{\tfunc\rset}$ (e.g., see \cite[Thm.~7.4.1]{ver18}). 
Hence, in the \emph{worst case}, Corollary~\ref{cor:appl:refined1bit} would imply an error decay rate of $\asympfaster{m^{-1/4}}$ up to log factors.
However, for many low-dimensional signal sets, such as subspaces or sparse vectors, it is possible to establish a much stronger bound of the form $\log\covnumber[\eps]{\tfunc\rset} \lesssim \log(\eps^{-1}) \cdot \effdim{\tfunc\rset}$, e.g., see \cite[Sec.~2]{or15}.\footnote{Here, it is important to note that, in contrast to $\sset$, the (transformed) signal set $\tfunc\rset$ may be highly non-convex.}
In this case, Corollary~\ref{cor:appl:refined1bit} would yield an error decay rate of $\asympfaster{m^{-1/2}}$ up to log factors.
This is superior to the achievement of Corollary~\ref{cor:appl:1bit} and matches the best possible rate that can be expected for \eqref{eq:results:klasso} in general (see Remark~\ref{rmk:results:main}\ref{rmk:results:main:optimal}).

Our second application of Theorem~\ref{thm:appl:without_increments} concerns the setup of sub-Gaussian $1$-bit observations with dithering, as considered in Subsection~\ref{subsec:appl:1bitdither}.
In this case, the stability condition of Assumption~\ref{ass:local_robustness} can be related to a recent embedding result of \citeauthor{dm18}~\cite{dm18}, which is based on hyperplane tessellations for sub-Gaussian vectors.
\begin{theorem}[\protect{\cite[Thm.~2.9]{dm18}}]\label{thm:hyperplane:tessellation:1bitdither}
	For every $\subgparam > 0$, there exist constants $c, \bar{c}, C, \tilde{C}>0$ only depending on $\subgparam$ for which the following holds.
	
	Let $\rset\subset R\ball[2][p]$ for some $R > 0$ and let $\A\in \R^{m\times p}$ be a random matrix with independent, isotropic, sub-Gaussian row vectors $\a_1, \dots, \a_m\in \R^p$.
	Moreover, assume that $\max_{i\in [m]}\normsubg{\a_i}\leq \subgparam$. Let $\ttau \coloneqq (\tau_1, \dots, \tau_m) \in \R^m$ be a random vector with independent entries that are uniformly distributed on $[-\lambda, \lambda]$ for a parameter $\lambda \geq \tilde{C}\cdot R$.
	In addition, suppose that $\A$ and $\ttau$ are independent.
	For $\beta\in (0,1]$ and $\eps\leq \bar{c}\beta / \sqrt{\log(e/\beta)}$, we assume that
	\begin{equation}
	m \geq C \cdot  \Big(\eps^2 \beta^{-3} \cdot \effdim{[(2\lambda)^{-1}\rset]_\eps} + \beta^{-1} \cdot \log\covnumber[\eps]{\lambda^{-1}\rset}\Big).
	\end{equation}
	Then the following holds with probability at least $1-\exp(-cm\beta)$:
	\begin{equation}
	\sup_{\substack{\grtr,\, \grtr'\in \rset\\ \lambda^{-1}\lnorm{\grtr - \grtr'}\leq \eps}}\tfrac{1}{2m}\lnorm{\sign(\A\grtr+\ttau)-\sign(\A\grtr'+\ttau)}[1]\leq \beta.
	\end{equation}
\end{theorem}
Combining Theorem~\ref{thm:hyperplane:tessellation:1bitdither} with Theorem~\ref{thm:appl:without_increments} now yields an improved version of Corollary~\ref{cor:appl:1bitdither} (see Section~\ref{sec:app:appl:noincr} for a proof):
\begin{corollary}\label{cor:appl:refined1bitdither}
	For every $\subgparam > 0$, there exist universal constants $c_0, C_0>0$ and constants $c, c', \tilde{C}, C'>\nobreak 0$ only depending on $\subgparam$ for which the following holds.
	
	Let $\a_1, \dots, \a_m \in \R^p$ be independent copies of a centered, isotropic, sub-Gaussian random vector $\a\in \R^p$ with $\normsubg{\a} \leq \subgparam$. Let $\tau_1, \dots, \tau_m$ be independent copies of a random variable $\tau$ that is uniformly distributed on $[-\lambda, \lambda]$ for a parameter $\lambda>0$. In addition, suppose that $\{\a_i\}_{i=1}^m$ and $\{\tau_i\}_{i=1}^m$ are independent.
	Let $\rset\subset R \ball[2][p]$ for some $R > 0$ and let $\sset\subset \R^p$ be a convex set such that $\lambda^{-1}\rset\subset \sset$. Fix $t\in (0,1]$ and $\eps\leq c't / \log(e/t)$. 
	For $\probsuccess^2\geq \max\{1, C_0\cdot \log\covnumber[\eps]{\lambda^{-1}\rset}\}$, we assume that
	\begin{align}
		\lambda &\geq \tilde{C} \cdot R \cdot \sqrt{\log(e/t)}, \label{eq:cond:refined1bitdither:lambda}\\
		m &\geq C' \cdot\Big( t^{-2} \cdot \big(\sup_{\grtr \in \rset}\effdim[t]{\sset - \lambda^{-1}\grtr}+ \probsuccess^2 \big) 
		+ \eps^2 t^{-3} \log^{3/2}(e/t)\cdot \effdim{[(2\lambda)^{-1}\rset]_\eps}
		\Big).\label{eq:cond:refined1bitdither:m}
	\end{align}
	Finally, let $\beta\in [0,1]$ be such that $\beta\sqrt{\log(e/\beta)}\leq c_0 \subgparam^{-2} t$. 
	Then with probability at least $1 - \exp(-c \probsuccess^2)$ on the random draw of $\{(\a_i, \tau_i)\}_{i = 1}^m$, the following holds uniformly for all $\grtr \in \rset$:
	Let $\Y \in \{-1,1\}^m$ be given by \eqref{eq:appl:1bitdither:meas} such that $\tfrac{1}{2m}\lnorm{\Noise}[1]\leq \tfrac{\beta}{2}$. Then every minimizer $\pvsolu$ of \eqref{eq:results:klasso} satisfies $\lnorm{\pvsolu - \lambda^{-1}\grtr} \leq 2t$.	
\end{corollary}
Analogously to noiseless $1$-bit observations, Corollary~\ref{cor:appl:refined1bitdither} implies a decay rate of $\asympfaster{m^{-1/4}}$ up to log factors in the worst case, which may be improved to $\asympfaster{m^{-1/2}}$ for structured signal sets. 
This is in line with a recent finding of \citeauthor{dm18} \cite[Thm.~1.7]{dm18}, who have analyzed a different estimator.

We close this section by pointing out that the error bounds achieved by Corollary~\ref{cor:appl:refined1bit} and~\ref{cor:appl:refined1bitdither} do still not match the information-theoretic optimal rate of $\asympfaster{m^{-1}}$, e.g., see \cite{jlbb13,dm18b}.
We suspect that this gap is not an artifact of our proof, but rather due to a fundamental performance limit of the generalized Lasso \eqref{eq:results:klasso}.
A potential remedy is to consider a more specialized reconstruction method, e.g., see \cite{jmps19}.

\section{Conclusion}
\label{sec:conclusion}

This section highlights several key aspects of this work and discusses them in the light of our initial objectives as well as remaining challenges.
\begin{arabiclist}
\item
	\emph{Observation models.} Probably the greatest benefit of our approach to Problem~\ref{prob:intro} is its flexibility.
	The setup of Assumption~\ref{ass:results:meas} allows us to implement and analyze virtually every non-linear observation model that is conceivable for the program \eqref{eq:results:klasso}.
	The examples that we have seen in Section~\ref{sec:appl} are only a selection of possible applications. Importantly, each of these results is accompanied by a conceptually simple proof that does not require any deeper insight into the underlying model mechanisms; see also the proof template at the beginning of Section~\ref{sec:app:appl}.
	Therefore, our methodology may form a general path towards competitive benchmark guarantees.
	However, the resulting oversampling rates do not always match the best possible rates from the literature, due to the increment condition of Assumption~\ref{ass:results:incr}.
	In Section~\ref{sec:appl:noincr}, we have presented a workaround for this issue, based on an additional local stability assumption.
	This strategy can indeed lead to near-optimal error bounds, but it relies on the availability of a strong embedding result in each considered case.
\item
	\emph{Non-Gaussian measurements.} Our main result reveals the impact of non-Gaussian measurements, in particular, when consistent recovery via \eqref{eq:results:klasso} can be expected and when not.
	While relaxing the conditions on isotropy and sub-Gaussian tails in Assumption~\ref{ass:results:meas} is certainly practicable (see Remark~\ref{rmk:results:exten}\ref{rmk:results:exten:covar}), our analysis is inherently limited to \emph{independent} measurement vectors.
	This excludes more structured measurement ensembles, such as partial random circulant matrices or lossless expanders \cite[Chap.~12 \&~13]{fh13}.
	On the other hand, such schemes typically satisfy a variant of the restricted isometry property, which is the basis for some recent advances in quantized compressed sensing, e.g., see \cite{djr17,fou17,jc17,xj18}.
	These findings provide evidence that at least parts of our results remain valid for a much larger family of measurement vectors.
	Thus, any guarantee with structured measurements that meets the generality of Theorem~\ref{thm:results:main} would be significant.
\item
	\emph{Complexity parameters.} The research on high-dimensional signal estimation has shown that the (local) Gaussian mean width is a natural, yet accurate complexity measure for convex programs such as \eqref{eq:results:klasso}.
	A distinctive feature of our approach is the role played by the (not necessarily convex) signal set $\rset$.
	As indicated in Section~\ref{sec:unif}, $\rset$ can be viewed as a characteristic of the observation model that allows us to study any situation between non-uniform and (fully) uniform recovery.
	Note that this refinement is also a crucial ingredient of Proposition~\ref{prop:unif:mwloc} and Theorem~\ref{thm:appl:without_increments}.
	Nevertheless, it is important to bear in mind that the Gaussian mean width is not trivial to estimate in specific situations, e.g., see~\eqref{eq:unif:mwl1} and Example~\ref{ex:unif:tv}.
\item
	\emph{(Non-)convexity and data-driven priors.} Remarkably, Fact~\ref{fact:results:recovery} is the only argument in our proof that relies on the convexity of the constraint set $\sset$. 
	In principle, it is possible to drop this assumption on $\sset$ by investigating the projected gradient descent method as an algorithmic implementation of \eqref{eq:results:klasso}, e.g., see \cite{os16,ors18,so19}.
	However, the feasibility of this approach depends on the existence of an efficient projection onto $\sset$.\footnote{The existence of an efficient projection and the (non-)convexity of $\sset$ are not equivalent. There are many examples of efficient projections onto non-convex sets, while the projection onto convex sets can be NP-hard, e.g., see \cite{rov14}.}
	A modern line of research on signal processing with non-convex optimization advocates the use of data-driven priors---a natural consequence of many recent advances on \emph{generative models} in machine learning research, e.g., see \cite{bjpd17,ls20} and the references therein.
	Although the algorithmic strategy behind these methods bears resemblance to \eqref{eq:results:klasso}, we believe that the complexity of learned signal priors leads to a more difficult mathematical problem, whose understanding is still in its infancy.
	This is particularly underpinned by the fact that even the situation of convex priors is not sufficiently well understood yet (see \cite{gkm20,gms20,mbkw20}).
\end{arabiclist}

\section{Proof of the Main Result (Theorem~\ref{thm:results:main})}
\label{sec:proofs}

Recall that according to Fact~\ref{fact:results:recovery}, it suffices to show that for all $\grtr \in \rset$ and $\Y \in \R^m$ satisfying $\mmcovar{\grtr} \leq \tfrac{t}{32}$ and \eqref{eq:results:main:rob}, we have that
\begin{equation}
	\inf_{\vdir \in \sset_{\grtr,t}}\exloss[\grtr](\vdir) > 0.
\end{equation}
To this end, we make use of the decomposition in \eqref{eq:results:exlossdecomp}, i.e.,
\begin{equation}
	\exloss[\grtr](\vdir) = \quadrterm(\vdir) + \advterm[\grtr](\vdir) + \multiplterm[\grtr](\vdir),
\end{equation}
and continue by showing separate bounds for all three terms, where each bound holds uniformly for all $\grtr \in \rset$ and $\vdir \in \sset_{\grtr,t}$.
Unless stated otherwise, we assume that the hypotheses of Theorem~\ref{thm:results:main} are fulfilled throughout this section and we have that $t > 0$; the case of $t = 0$ is in fact much simpler, see Step~\hyperref[step:proof_tzero]{4b}.
Moreover, we will use the notation $\sset_{\rset,t} \coloneqq \union_{\grtr\in \rset}\sset_{\grtr,t}=(\sset - \tfunc\rset) \intersec t\S^{p-1}$.

\paragraph{Step 1: Bounding the quadratic term.} \label{step:quadr} 
Let $\A\in \R^{m\times p}$ denote the matrix with row vectors $\a_1, \dots, \a_m$. In order to control the random variable
\begin{equation}
	\quadrterm(\vdir)=\tfrac{1}{m} \sum_{i=1}^m\sp{\a_i}{\vdir}^2=\lnorm[\big]{\tfrac{1}{\sqrt{m}}\A\vdir}^2
\end{equation}
uniformly for all $\vdir \in \sset_{\rset,t}$, we make use of the following recent matrix deviation inequality for sub-Gaussian matrices.
\begin{theorem}[{\cite[Cor.~1.2]{jlpy20}}]\label{thm:proofs:matrixdev}
		There exists a universal constant $C_{\quadrterm} > 0$ for which the following holds.
	
		Let $\ssetgen\subset \R^p$ and let $\A\in \R^{m\times p}$ be a random matrix with independent, isotropic, sub-Gaussian row vectors $\a_1, \dots, \a_m\in \R^p$.
		Moreover, assume that $\max_{i\in [m]}\normsubg{\a_i}\leq \subgparam$ for some $\subgparam > 0$.
		Then for every $\probsuccess >0$, we have the following with probability at least $1-3\exp(-\probsuccess^2)$:
		\begin{equation}
			\sup_{\vdir \in \ssetgen} \abs[\Big]{ \lnorm[\big]{\tfrac{1}{\sqrt{m}}\A\vdir} - \lnorm{\vdir}}\leq C_{\quadrterm} \cdot \subgparam \sqrt{\log\subgparam} \cdot \frac{\meanwidth{\ssetgen} +\probsuccess \cdot \sup_{\vdir \in \ssetgen}\lnorm{\vdir}}{\sqrt{m}}.
		\end{equation}
\end{theorem}
We now apply this result to $\ssetgen \coloneqq \sset_{\rset,t} \subset t \S^{p-1}$ within the setting of Theorem~\ref{thm:results:main}:
According to the first branch of the assumption~\eqref{eq:results:main:m}, we have that $m \geq C \cdot \subgparam^2 \log\subgparam \cdot \big( \effdim[t]{\sset - \tfunc\rset} + \probsuccess^2 \big)$.
Hence, by adjusting the universal constant $C$ (only depending on $C_{\quadrterm}$), Theorem~\ref{thm:proofs:matrixdev} implies that the following holds with probability at least $1-3\exp(-\probsuccess^2)$:
\begin{equation}\label{eq:proofs:bnd_quadr}
	\tfrac{3t^2}{2}\geq \quadrterm(\vdir)\geq \tfrac{t^2}{2} \qquad \text{for all $\vdir \in  \sset_{\rset,t}$}.
\end{equation}

\paragraph{Step 2: Bounding the noise term.} \label{step:noise} 
In order to control the noise term $\advterm[\grtr](\vdir)$, we require the following uniform upper bound for subsums of the quadratic term.
\begin{theorem}[{\cite[Thm.~2.10]{dm18}}]\label{thm:supmaxprocess}
There exist universal constants $c, C_{\advterm}> 0$ for which the following holds.	
	
Let $\ssetgen\subset \R^p$ and let $\a_1, \dots, \a_m\in \R^p$ be independent, isotropic, sub-Gaussian random vectors such that $\max_{i\in [m]}\normsubg{\a_i}\leq \subgparam$ for some $\subgparam > 0$. Then for every $\nout\in \{0,1, \dots, m\}$ and $\probsuccess_0 \geq \sqrt{\nout\log(em/\nout)}$, we have the following with probability at least $1-2\exp(-c\probsuccess_0^2)$:
\begin{equation}
	\sup_{\vdir \in \ssetgen} \ \max_{\substack{\iset\subset [m]\\ \cardinality{\iset}\leq \nout}} \Big(\tfrac{1}{m}\sum_{i\in \iset}\abs{\sp{\a_i}{\vdir}}^2\Big)^{1/2} 
	\leq C_{\advterm} \cdot \subgparam^2\cdot \frac{\meanwidth{\ssetgen} + \probsuccess_0 \cdot\sup_{\vdir \in \ssetgen}\lnorm{\vdir}}{\sqrt{m}}.
\end{equation} 	
\end{theorem}	
For $\vdiralt\in \R^m$, we denote by $\iset_{\nout}(\vdiralt)\subset [m]$ any (possibly non-unique) index set that corresponds to the $\nout$ largest entries of $\vdiralt$ in magnitude, i.e., for all $i\in \iset_{\nout}(\vdiralt)$ and $i' \in \setcompl{\iset_{\nout}(\vdiralt)}$, we have that $\abs{w_i}\geq \abs{w_{i'}}$; note that for $\nout=0$, we simply have $\iset_{\nout}(\vdiralt)=\emptyset$.
With this notation at hand, observe that
\begin{equation}
	\norm{\vdiralt}_{[\nout]} = \Big(\sum_{i \in \iset_{\nout}(\vdiralt)} \abs{w_i}^2 \Big)^{1/2} \quad \text{and} \quad \sigma_{\nout}(\vdiralt)_2 = \Big(\sum_{i \in \setcompl{\iset_{\nout}(\vdiralt)}} \abs{w_i}^2 \Big)^{1/2}.
\end{equation}
Now, consider the specific choice $\vdiralt \coloneqq \Y-\Ys(\grtr)$. The Cauchy-Schwarz inequality then implies
\begin{align}
\sup_{\vdir \in \sset_{\grtr,t}}\abs{\advterm[\grtr](\vdir)} &\leq 
	\begin{multlined}[t]
		\tfrac{2}{m} \sup_{\vdir \in \sset_{\grtr,t}} \ \sum_{i\in \iset_{\nout}(\vdiralt)} \abs{w_i} \cdot \abs{\sp{\a_i}{\vdir}} + \tfrac{2}{m} \sup_{\vdir \in \sset_{\grtr,t}} \ \sum_{i\in \setcompl{\iset_{\nout}(\vdiralt)}} \abs{w_i} \cdot \abs{\sp{\a_i}{\vdir}}
	\end{multlined} \\
	&\leq
	\begin{multlined}[t] 
		\tfrac{2}{\sqrt{m}} \norm{\Y - \Ys(\grtr)}_{[\nout]} \cdot \sup_{\vdir \in \sset_{\rset,t}} \ \max_{\substack{\iset\subset [m]\\ \cardinality{\iset}\leq \nout}}\Big(\tfrac{1}{m}\sum_{i\in \iset}\abs{\sp{\a_i}{\vdir}}^2\Big)^{1/2}\\
		+ \tfrac{2}{\sqrt{m}} \sigma_{\nout}(\Y - \Ys(\grtr))_2 \cdot \sup_{\vdir \in \sset_{\rset,t}} \sqrt{\quadrterm(\vdir)}.
	\end{multlined}
\end{align} 
Let us now estimate the first summand of this bound. 
According to Theorem~\ref{thm:supmaxprocess}, there exist universal constants $c, C_{\advterm} > 0$ such that for every $\probsuccess_0 \geq \sqrt{\nout\log(em/\nout)}$, we have the following with probability at least $1-2\exp(-c \probsuccess_0^2)$:
\begin{equation}\label{eq:proofs:bnd_supmax}
	\sup_{\vdir \in \sset_{\rset,t}} \ \max_{\substack{\iset\subset [m]\\ \cardinality{\iset}\leq \nout}}\Big(\tfrac{1}{m}\sum_{i\in \iset}\abs{\sp{\a_i}{\vdir}}^2\Big)^{1/2}
	\leq C_{\advterm} \cdot \subgparam^2\cdot \frac{\meanwidth{\sset_{\rset,t}} + \probsuccess_0 \cdot t}{\sqrt{m}} \\
	\leq \tfrac{1}{16} \tune^{-1} t ,
\end{equation} 
where the last inequality follows from the second branch of \eqref{eq:results:main:m}.
Hence, taking a union bound with the event of \eqref{eq:proofs:bnd_quadr}, it follows that with probability at least $1-3\exp(-\probsuccess^2) - 2\exp(-c\probsuccess_0^2)$, the following holds uniformly for every $\grtr \in \rset$ and every $\Y \in \R^m$ satisfying the condition \eqref{eq:results:main:rob}:
\begin{equation}\label{eq:proofs:bnd_noise}
\sup_{\vdir \in \sset_{\grtr,t}}\abs{\advterm[\grtr](\vdir)}\leq 2 \cdot \tfrac{t}{16} \cdot t + 2 \cdot \sqrt{\tfrac{3}{2}} \cdot t \cdot \tfrac{t}{20}\leq \tfrac{t^2}{4}.
\end{equation} 

\paragraph{Step 3: Bounding the multiplier term.} \label{step:multipl} 
Our goal is to bound the random variable 
\begin{equation}
	\tfrac{1}{2}\multiplterm[\grtr](\vdir) = \tfrac{1}{m} \sum_{i=1}^m(\sp{\a_i}{\tfunc\grtr} - \ys_i(\grtr)) \sp{\a_i}{\vdir}
\end{equation}
uniformly from below for all $\vdir \in \sset_{\grtr,t}$ and $\grtr \in \rset$ that satisfy $\mmcovar{\grtr} \leq \tfrac{t}{32}$. 
Let us begin by adding and subtracting the expected value:
\begin{equation} \label{eq:proofs:decomp_multpl1}
	\tfrac{1}{2}\multiplterm[\grtr](\vdir)\geq \Big(\inf_{\vdir \in \sset_{\grtr,t}}\tfrac{1}{2}\mean[\multiplterm[\grtr](\vdir)]\Big) - \tfrac{1}{2}\abs[\big]{\multiplterm[\grtr](\vdir)- \mean[\multiplterm[\grtr](\vdir)]}.
\end{equation}
Since $\a$ is isotropic, we observe that $\tfrac{1}{2}\mean[\multiplterm[\grtr](\vdir)]=\sp{\tfunc\grtr - \mean[\ys(\grtr) \a]}{\vdir}$. 
Combining this with Definition~\ref{def:results:mmcovar}, we obtain a lower bound in terms of the target mismatch:
\begin{align}
	\inf_{\vdir \in \sset_{\grtr,t}}\tfrac{1}{2}\mean[\multiplterm[\grtr](\vdir)] 
	&\geq t \cdot \inf_{\vdir \in \S^{p-1}}\sp{\tfunc\grtr - \mean[\ys(\grtr) \a]}{\vdir}\\
	&= - t \cdot \sup_{\vdir \in \S^{p-1}}\sp{\mean[\ys(\grtr) \a] - \tfunc\grtr}{\vdir} \\
	&= - t \cdot \lnorm[\big]{\mean[\ys(\grtr) \a] - \tfunc\grtr} = - t \cdot \mmcovar{\grtr}. \label{eq:proofs:bnd_mmcovar}
\end{align}
We now turn to the centered multiplier term in \eqref{eq:proofs:decomp_multpl1}.
To this end, recall Assumption~\ref{ass:results:incr} and let $\{\ys_{t,i}(\grtr)\}_{\grtr \in \rset}$ be independent copies of the class $\{\ys_{t}(\grtr)\}_{\grtr \in \rset}$ for $i = 1, \dots, m$.\footnote{To be more precise, we assume that the tuples $(\a_i, \fout_i, \ys_{t,i})$ are independent copies of $(\a, \fout, \ys_{t})$ for $i = 1, \dots, m$. In particular, the respective conditions of Assumption~\ref{ass:results:incr}\ref{ass:results:incr:approx}--\ref{ass:results:incr:err} are also satisfied for $\{\ys_{t,i}(\grtr)\}_{\grtr \in \rset}$.}
Inserting these observation variables and applying the triangle inequality then yields
\begin{multline}
	\tfrac{1}{2}\abs{\multiplterm[\grtr](\vdir)- \mean[\multiplterm[\grtr](\vdir)]}
	\leq \tfrac{1}{m} \abs[\Big]{\sum_{i=1}^m\underbrace{(\sp{\a_i}{\tfunc\grtr} - \ys_{t,i}(\grtr))}_{\eqqcolon \multipl_{t,i}(\grtr)} \sp{\a_i}{\vdir} - \mean[(\sp{\a_i}{\tfunc\grtr} - \ys_{t,i}(\grtr)) \sp{\a_i}{\vdir}]}\\
	+ {} \mean{}[\underbrace{\abs{\ys(\grtr) - \ys_{t}(\grtr)}}_{= \err_t(\grtr)} {}\cdot{} \abs{\sp{\a}{\vdir}}] + \tfrac{1}{m} \sum_{i=1}^m \underbrace{\abs{\ys_{t,i}(\grtr) - \ys_i(\grtr)}}_{\eqqcolon \err_{t,i}(\grtr)} {}\cdot{} \abs{\sp{\a_i}{\vdir}}. \label{eq:proofs:decomp_multpl2}
\end{multline}
Note that the second empirical product process in \eqref{eq:proofs:decomp_multpl2} is not centered yet due to pulling in the absolute values---a crucial step to ensure that the resulting factors have sub-Gaussian increments.
Adding and subtracting the expected value of the last summand in \eqref{eq:proofs:decomp_multpl2} and using Assumption~\ref{ass:results:incr}\ref{ass:results:incr:approx}, 
we obtain the following upper bound:
\begin{multline}
	\tfrac{1}{2}\abs{\multiplterm[\grtr](\vdir)- \mean[\multiplterm[\grtr](\vdir)]} \leq \tfrac{1}{m} \abs[\Big]{\sum_{i=1}^m\multipl_{t,i}(\grtr) \sp{\a_i}{\vdir} - \mean[\multipl_{t,i}(\grtr) \sp{\a_i}{\vdir}]}\\*
	+ {} \tfrac{1}{m} \abs[\Big]{\sum_{i=1}^m \err_{t,i}(\grtr) \cdot \abs{\sp{\a_i}{\vdir}} - \mean[\err_{t,i}(\grtr) \cdot \abs{\sp{\a_i}{\vdir}}]} + \underbrace{2 \mean[\err_t(\grtr) \cdot \abs{\sp{\a}{\vdir}}]}_{\leq \tfrac{t^2}{32}}. \label{eq:proofs:decomp_multpl3}
\end{multline}
This estimate in conjunction with \eqref{eq:proofs:bnd_mmcovar} implies that the following holds uniformly for all $\grtr \in \rset$ with $\mmcovar{\grtr} \leq \tfrac{t}{32}$:
\begin{alignat}{3}
	\inf_{\vdir \in \sset_{\grtr,t}}\multiplterm[\grtr](\vdir)
	&\geq - \frac{t^2}{8} && - 2\sup_{\vdir \in \sset_{\grtr,t}} \tfrac{1}{m} \abs[\Big]{\sum_{i=1}^m\multipl_{t,i}(\grtr) \sp{\a_i}{\vdir} - \mean[\multipl_{t,i}(\grtr) \sp{\a_i}{\vdir}]}\\
		&&& - 2\sup_{\vdir \in \sset_{\grtr,t}}  \tfrac{1}{m} \abs[\Big]{\sum_{i=1}^m \err_{t,i}(\grtr) \cdot \abs{\sp{\a_i}{\vdir}} - \mean[\err_{t,i}(\grtr) \cdot \abs{\sp{\a_i}{\vdir}}]}\\
	&\geq - \frac{t^2}{8} && - 2\sup_{\substack{\grtr \in \rset \\ \vdir \in \sset_{\rset,t}}} \tfrac{1}{m} \abs[\Big]{\sum_{i=1}^m\multipl_{t,i}(\grtr) \sp{\a_i}{\vdir} - \mean[\multipl_{t,i}(\grtr) \sp{\a_i}{\vdir}]}\\
		&&& - 2\sup_{\substack{\grtr \in \rset \\ \vdir \in \sset_{\rset,t}}} \tfrac{1}{m} \abs[\Big]{\sum_{i=1}^m \err_{t,i}(\grtr) \cdot \abs{\sp{\a_i}{\vdir}} - \mean[\err_{t,i}(\grtr) \cdot \abs{\sp{\a_i}{\vdir}}]}. \label{eq:proofs:decomp_multpl4}
\end{alignat}
By Assumption~\ref{ass:results:incr}\ref{ass:results:incr:multipl} and~\ref{ass:results:incr:err}, all factors of the product processes 
\begin{equation}\label{eq:proofs:productproc}
	\bigg\{\tfrac{1}{m}\sum_{i=1}^m\multipl_{t,i}(\grtr) \sp{\a_i}{\vdir}\bigg\}_{\grtr\in \rset,\ \vdir \in \sset_{\rset,t}}, \quad \bigg\{\tfrac{1}{m}\sum_{i=1}^m \err_{t,i}(\grtr) \cdot \abs{\sp{\a_i}{\vdir}}\bigg\}_{\grtr\in \rset,\ \vdir \in \sset_{\rset,t}}
\end{equation}
have sub-Gaussian increments. 
A key ingredient for controlling these empirical processes is a powerful concentration inequality due to \citeauthor{men16}~\cite{men16}.
The following result is adapted from \cite[Thm.~1.13]{men16}.
\begin{theorem}\label{thm:proofs:product}
There exist universal constants $c, C_{\multiplterm} > 0$ for which the following holds. 

Let $\{g_{a}\}_{a\in \clA}$ and $\{h_{b}\}_{b\in \clB}$ be stochastic processes indexed by two sets $\clA$ and $\clB$, both defined on a common probability space $(\probset, \sigmaalg, \prob)$.
We assume that there exist $r_{\clA}, r_{\clB} \geq 0$ and pseudo-metrics $d_{\clA}$ on $\clA$ and $d_{\clB}$ on $\clB$ such that
\begin{align}
	&\normsubg{g_{a} - g_{a'}}\leq d_{\clA}(a, a') \quad {\text{and}} \quad \normsubg{g_{a}}\leq r_{\clA} \qquad \text{for all $a, a' \in \clA$,}\\
	&\normsubg{h_{b} - h_{b'}}\leq d_{\clB}(b, b') \quad {\text{and}} \quad \normsubg{h_{b}}\leq r_{\clB} \qquad \text{for all $b, b' \in \clB$.}
\end{align}
Finally, let $X_1, \dots, X_m$ be independent copies of a random variable $X\distributed \prob$. Then for every $\probsuccess\geq 1$, we have the following with probability at least $1- 2 \exp(-c\probsuccess^2)$:
\begin{multline}
	\sup_{\substack{a\in \clA \\ b\in \clB}} \ \tfrac{1}{m}\abs[\Big]{\sum_{i=1}^m g_{a}(X_i) h_{b}(X_i) - \mean[g_{a}(X_i) h_{b}(X_i)]}\\
			\leq C_{\multiplterm} \cdot \bigg(\frac{\big(\talagfunc(\clA, d_{\clA}) + \probsuccess \cdot r_{\clA}\big)\cdot\big(\talagfunc(\clB, d_{\clB}) + \probsuccess \cdot r_{\clB}\big)}{m} \\
			+ \frac{r_{\clA} \cdot \talagfunc(\clB, d_{\clB}) + r_{\clB} \cdot \talagfunc(\clA, d_{\clA}) + \probsuccess \cdot r_{\clA} r_{\clB}}{\sqrt{m}} 
			\bigg),
\end{multline}
where $\talagfunc(\cdot,\cdot)$ denotes Talagrand's $\talagfunc$-functional (e.g., see~\cite[Def.~1.2]{men16}).
\end{theorem}
Let us now apply Theorem~\ref{thm:proofs:product} to the first product process in \eqref{eq:proofs:productproc}.
To this end, we observe that the class $\{\sp{\a}{\vdir}\}_{\vdir \in \sset_{\rset,t}}$ has sub-Gaussian increments with respect to the pseudo-metric induced by $\subgparam \lnorm{\cdot}$, while Assumption~\ref{ass:results:incr}\ref{ass:results:incr:multipl} implies that $\{\multipl_t(\grtr)\}_{\grtr\in \rset}$ has sub-Gaussian increments with respect to $\multiplLip \cdot d_\tfunc$.
Hence, Theorem~\ref{thm:proofs:product} implies that the following holds with probability at least $1- 2 \exp(-c\probsuccess^2)$:\footnote{To be slightly more precise, we apply Theorem~\ref{thm:proofs:product} to the index sets $\clA \coloneqq \rset$ and $\clB \coloneqq \sset_{\rset,t}$, while $X \distributed \prob$ corresponds to the random tuple $(\a, \fout, \ys_{t})$.}
\begin{multline}
	\sup_{\substack{\grtr \in \rset \\ \vdir \in \sset_{\rset,t}}} \tfrac{1}{m} \abs[\Big]{\sum_{i=1}^m\multipl_{t,i}(\grtr) \sp{\a_i}{\vdir} - \mean[\multipl_{t,i}(\grtr) \sp{\a_i}{\vdir}]}\\*
	\leq  C_{\multiplterm} \cdot \bigg(\frac{\big(\multiplLip\talagfunc(\rset, d_\tfunc) + \probsuccess \cdot \multiplDia\big)\cdot \big(\subgparam \talagfunc(\sset_{\rset,t}, \lnorm{\cdot}) + \probsuccess \cdot \subgparam t\big)}{m} \\*
	+ \frac{\multiplDia \cdot \subgparam \talagfunc(\sset_{\rset,t}, \lnorm{\cdot}) + \subgparam t \cdot \multiplLip\talagfunc(\rset, d_\tfunc) + \probsuccess \cdot \multiplDia \subgparam t}{\sqrt{m}} \bigg).
\end{multline}
According to Talagrand's Majorizing Measure Theorem \cite[Thm.~2.4.1]{tal14}, we have that
\begin{equation}
\meanwidth{\ssetgen} \asymp \talagfunc(\ssetgen, \lnorm{\cdot}) \qquad \text{for every $\ssetgen \subset \R^p$.}
\end{equation}
Moreover, it is not hard to see that $\talagfunc(\rset, d_\tfunc) = \talagfunc(\tfunc\rset, \lnorm{\cdot})$. 
Consequently, the following bound holds with probability at least $1- 2 \exp(-c\probsuccess^2)$:
\begin{multline}
	\sup_{\substack{\grtr \in \rset \\ \vdir \in \sset_{\rset,t}}} \tfrac{1}{m} \abs[\Big]{\sum_{i=1}^m\multipl_{t,i}(\grtr) \sp{\a_i}{\vdir} - \mean[\multipl_{t,i}(\grtr) \sp{\a_i}{\vdir}]}\\ 
	\leq  C_{\multiplterm}' \cdot t \cdot \subgparam\cdot   \bigg(\frac{\big(\multiplLip \meanwidth{\tfunc\rset} + \probsuccess \cdot \multiplDia \big) \cdot \big(  \meanwidth[t]{\sset - \tfunc\rset} + \probsuccess \big)}{m} \\ 
	+ \frac{\multiplDia \cdot \meanwidth[t]{\sset - \tfunc\rset} +   \multiplLip\meanwidth{\tfunc\rset} + \probsuccess \cdot \multiplDia  }{\sqrt{m}} \bigg).\label{eq:proofs:bnd_multipl1}
\end{multline}
The second product process in \eqref{eq:proofs:productproc} can be treated similarly: The class $\{\abs{\sp{\a}{\vdir}}\}_{\vdir \in \sset_{\rset,t}}$ has sub-Gaussian increments with respect to $\subgparam \lnorm{\cdot}$, while Assumption~\ref{ass:results:incr}\ref{ass:results:incr:err} implies that $\{\err_t(\grtr)\}_{\grtr\in \rset}$ has sub-Gaussian increments with respect to $\errLip \cdot d_\tfunc$. 
An analogous application of Theorem~\ref{thm:proofs:product} shows that with probability at least $1- 2 \exp(-c\probsuccess^2)$, we have that
\begin{multline}
	\sup_{\substack{\grtr \in \rset \\ \vdir \in \sset_{\rset,t}}} \tfrac{1}{m} \abs[\Big]{\sum_{i=1}^m \err_{t,i}(\grtr) \cdot \abs{\sp{\a_i}{\vdir}} - \mean[\err_{t,i}(\grtr) \cdot \abs{\sp{\a_i}{\vdir}}]}\\
	\leq  C_{\multiplterm}' \cdot t \cdot \subgparam\cdot   \bigg(\frac{\big(\errLip \meanwidth{\tfunc\rset} + \probsuccess \cdot \errDia\big) \cdot \big(  \meanwidth[t]{\sset - \tfunc\rset} + \probsuccess \big)}{m} \\
	+ \frac{\errDia \cdot  \meanwidth[t]{\sset - \tfunc\rset} +   \errLip\meanwidth{\tfunc\rset} + \probsuccess \cdot \errDia  }{\sqrt{m}} \bigg).\label{eq:proofs:bnd_multipl2}
\end{multline}
We are ready to prove our main result.
\paragraph{Step 4a: Proof of Theorem~\ref{thm:results:main} ($t > 0$).} \label{step:proof_tpos}
	We now assume that the events corresponding to \eqref{eq:proofs:bnd_quadr}, \eqref{eq:proofs:bnd_noise}, \eqref{eq:proofs:bnd_multipl1}, and \eqref{eq:proofs:bnd_multipl2} have occurred jointly with probability at least $1- 7\exp(-c\probsuccess^2) - 2\exp(-c\probsuccess_0^2)$ for an appropriate constant $c > 0$; note that the factors $7$ and $2$ can be removed by slightly adjusting~$c$.
	Combining these bounds with \eqref{eq:proofs:decomp_multpl4}, the following holds uniformly for every $\grtr \in \rset$ with $\mmcovar{\grtr} \leq \tfrac{t}{32}$ and every $\Y \in \R^m$ satisfying the condition \eqref{eq:results:main:rob}:
	\begin{align}
	\inf_{\vdir \in \sset_{\grtr,t}}\exloss[\grtr](\vdir)
		&\geq \inf_{\vdir\in \sset_{\rset,t}}\quadrterm(\vdir) - \sup_{\vdir \in \sset_{\grtr,t}}\abs{\advterm[\grtr](\vdir)} + \inf_{\vdir \in \sset_{\grtr,t}}\multiplterm[\grtr](\vdir)\\*
		&\geq \begin{aligned}[t]
			\tfrac{t^2}{2} - \tfrac{t^2}{4} - \tfrac{t^2}{8} &- C_{\multiplterm}' \cdot t \cdot \subgparam\cdot \bigg(\frac{( \multiplDia+\errDia) \cdot \meanwidth[t]{\sset - \tfunc\rset} +   (\multiplLip+ \errLip) \cdot \meanwidth{\tfunc\rset} + \probsuccess \cdot (\multiplDia+\errDia)  }{\sqrt{m}}\bigg)  \\
			 &- C_{\multiplterm}' \cdot t \cdot \subgparam\cdot   \bigg(\frac{\big((\multiplLip+ \errLip) \cdot \meanwidth{\tfunc\rset} + \probsuccess \cdot ( \multiplDia + \errDia)\big) \cdot \big(  \meanwidth[t]{\sset - \tfunc\rset} + \probsuccess \big)}{m}\bigg).
		\end{aligned} \\* \label{eq:proofs:bnd_excess}
	\end{align}
	If we could show that this lower bound is strictly positive, the claim of Theorem~\ref{thm:results:main} would follow directly from Fact~\ref{fact:results:recovery}.
	To conclude this argument, it is enough to have that
	\begin{align}
		m &\geq C' \cdot \subgparam^2  t^{-2} \cdot \Big( (\multiplDia+\errDia) \cdot \meanwidth[t]{\sset - \tfunc\rset} + (\multiplLip+ \errLip) \cdot \meanwidth{\tfunc\rset} + \probsuccess \cdot (\multiplDia+\errDia)\Big)^2, \label{eq:proofs:multipl_cond1} \\
		m &\geq C' \cdot \subgparam  t^{-1} \cdot \big((\multiplLip+ \errLip) \cdot \meanwidth{\tfunc\rset} + \probsuccess \cdot ( \multiplDia+\errDia)\big) \cdot \big(\meanwidth[t]{\sset - \tfunc\rset} + \probsuccess\big), \label{eq:proofs:multipl_cond2}
	\end{align}
	where $C' > 0$ is an appropriate universal constant.
	Indeed, both \eqref{eq:proofs:multipl_cond1} and \eqref{eq:proofs:multipl_cond2} are consequences of \eqref{eq:results:main:m}:
	The bound of \eqref{eq:proofs:multipl_cond1} is equivalent to the third branch of \eqref{eq:results:main:m}, while \eqref{eq:proofs:multipl_cond2} follows from the multiplication of the first and third branch of \eqref{eq:results:main:m} and then taking the square root. Note that the previous argument also makes use of the basic fact that $(v + w)^2 \geq v^2 + w^2 \geq \tfrac{1}{2}(v + w)^2$ for all $v, w \geq 0$, and that $\subgparam^2 \gtrsim \subgparam$ due to the isotropy of $\a$.
	
\paragraph{Step 4b: Proof of Theorem~\ref{thm:results:main} ($t = 0$).} \label{step:proof_tzero}
We may assume that $\multiplDia = \errDia = \multiplLip = \errLip = 0$, $\nout = 0$, $\tune = \subgparam^{-1}\sqrt{\log\subgparam}$, and $\probsuccess = \probsuccess_0$, while only considering those $\grtr \in \rset$ and $\Y \in \R^m$ with $\Y = \Ys(\grtr)$ and $\mmcovar{\grtr} = 0$.
Consequently, we have a noiseless linear model, i.e., $\Y = \Ys(\grtr) = \sp{\a}{\grtr}$.
It follows that $\advterm[\grtr](\cdot) = \multiplterm[\grtr](\cdot) = 0$. Analogously to \eqref{eq:proofs:bnd_quadr}, we can conclude that with probability at least $1- 3\exp(-\probsuccess^2)$, the quadratic term satisfies $\quadrterm(\vdir) \geq \tfrac{1}{2}$ for all $\vdir \in \sset_{\rset, 0} \coloneqq \cone{\sset - \tfunc\rset} \intersec \S^{p-1}$.
On this event, let $\pvsolu \in \sset$ be any minimizer of \eqref{eq:results:klasso} with input $\Y = \Ys(\grtr)$ for some $\grtr \in \rset$.
If we would not have exact recovery, i.e., $\solu{\vdir} \coloneqq \pvsolu - \tfunc\grtr \neq \vnull$, then 
\begin{equation}
	\exloss[\grtr](\solu{\vdir}) = \quadrterm(\solu{\vdir}) = \lnorm{\solu{\vdir}}^2 \cdot \quadrterm\Big(\underbrace{\frac{\solu{\vdir}}{\lnorm{\solu{\vdir}}}}_{\mathclap{\in \sset_{\rset, 0}}}\Big) \geq \frac{\lnorm{\solu{\vdir}}^2}{2} > 0,
\end{equation}
which contradicts the fact that $\pvsolu$ is a solution to \eqref{eq:results:klasso}. \qed

\section{Proofs for Section~\ref{sec:appl}}
\label{sec:app:appl}

Each result in Section~\ref{sec:appl} is an application of Theorem~\ref{thm:results:main} and follows the same proof template:
\begin{enumerate}[label={\textbf{Step~\arabic*.}}, itemindent=2.5em, leftmargin=0em, align=left]
\item
	How the prerequisites of Theorem~\ref{thm:results:main} are met:
	\begin{enumerate}[label={\textbf{Step~1\alph*.}}, itemindent=2.5em, leftmargin=3em, align=left]
	\item
		Implementation of Assumption~\ref{ass:results:meas}.
	\item
		Controlling the target mismatch $\mmcovar{\grtr}$ for every $\grtr \in \rset$.
	\item
		Controlling the increment parameters of Assumption~\ref{ass:results:incr}.
	\end{enumerate}
\item
	Proof of the actual statement via Theorem~\ref{thm:results:main}.
\item
	Verification of Step 1b.
\item
	Verification of Step 1c.
\end{enumerate}

\subsection{Proofs for Subsection~\ref{subsec:appl:1bit}}
\label{subsec:app:appl:1bit}

\paragraph{Proof of Corollary~\ref{cor:appl:1bit}.}
We follow the proof template from the beginning of Section~\ref{sec:app:appl}:
\begin{tmpl}
\item
The model setup of Corollary~\ref{cor:appl:1bit} fits into Assumption~\ref{ass:results:meas} as follows:
\begin{itemize}
	\item
		We have that $\a \distributed \Normdistr{\vnull}{\I{p}}$ and therefore $\normsubg{\a} \lesssim 1$. The signal set $\rset$ is an arbitrary subset of~$\R^p$.
		The output function $\fout \colon \R^p \times \rset \to \R$ takes the form $\fout(\a, \grtr) \coloneqq \sign(\sp{\a}{\grtr})$.
	\item 
		The target function $\tfunc \colon \rset \to \sset$ corresponds to the (scaled) normalization $\tfunc\grtr \coloneqq \sqrt{\tfrac{2}{\pi}}\tfrac{\grtr}{\lnorm{\grtr}}$. In particular, we have that $d_\tfunc(\grtr,\grtr')=\sqrt{\tfrac{2}{\pi}}\lnorm[\big]{\tfrac{\grtr}{\lnorm{\grtr}} - \tfrac{\grtr'}{\lnorm{\grtr'}}}$.
\end{itemize}

\item
The target mismatch $\mmcovar{\grtr}$ vanishes for every $\grtr\in \rset$.

\item
There exists an approximation $\ys_t(\grtr)$ of the observation variable $\ys(\grtr)$ such that the conditions of Assumption~\ref{ass:results:incr} are fulfilled with
\begin{equation}\label{eq:app:appl:1bit:incr}
	\multiplLip \lesssim 1 + t^{-1},\quad \errLip\lesssim t^{-1},\quad \multiplDia\lesssim 1,\quad \errDia \lesssim 1.
\end{equation} 

\item
The first and third branch of the condition \eqref{eq:results:main:m} are implied by \eqref{eq:appl:1bit:m} for a sufficiently large universal constant $C'>0$.
Since $\phi(\beta) \coloneqq \beta\sqrt{\log(e/\beta)}$ defines a continuous and non-decreasing function on $[0, 1]$ with $\phi(1)=1$ and $\phi(0)\coloneqq 0$ (by continuous extension), we may assume without loss of generality that $\beta\sqrt{\log(e/\beta)} = c_0 t \in (0,1]$.
Now, we set
\begin{equation}
	\tune^2 \coloneqq \frac{1}{t\sqrt{\log(e/\beta)}}, \quad \nout \coloneqq \floor{\beta m}, \quad \probsuccess_0 \coloneqq \sqrt{2 m \beta \log(e/\beta)},
\end{equation}
implying that $\probsuccess_0\geq \sqrt{\nout \log(em/\nout)}$ and $\probsuccess_0^2\geq 2 c_0 t m$.
Combining the latter inequality with \eqref{eq:appl:1bit:m} for $C'$ sufficiently large particularly implies that $\probsuccess_0^2 \geq \probsuccess^2$.
Furthermore, observe that $\tune^2 \probsuccess_0^2=2c_0 m$ and $\tune^2 \cdot \effdim[t]{\sset - \tfunc\rset}\leq t^{-1} \cdot \effdim[t]{\sset - \tfunc\rset}$.
Hence, the second branch of the condition \eqref{eq:results:main:m} is satisfied if $c_0$ is chosen small enough and $C'$ in \eqref{eq:appl:1bit:m} large enough.

Next, we show that every $\Y \in \{-1,1\}^m$ given by \eqref{eq:appl:1bit:meas} with $\tfrac{1}{2m}\lnorm{\Noise}[1] \leq \beta$ also satisfies \eqref{eq:results:main:rob}.
Since $\Noise = \Y - \Ys(\grtr)\in \{-2,0,2\}^m$ and $\tfrac{1}{2}\lnorm{\Y - \Ys(\grtr)}[1] \leq \beta m$, it follows that $\cardinality{\{i\in [m] \suchthat y_i \neq \ys_i(\grtr)\}}\leq \floor{\beta m}$. Consequently, $\sigma_{\nout}(\Y - \Ys(\grtr))_{2}=0$ and
\begin{equation}
	\tfrac{1}{\sqrt{m}}\norm{\Y - \Ys(\grtr)}_{[\nout]}= \tfrac{1}{\sqrt{m}}\lnorm{\Y - \Ys(\grtr)}\leq 2\sqrt{\beta}= 2 \Big(\tfrac{c_0 t}{\sqrt{\log(e/\beta)}}\Big)^{1/2}\leq \tune t,
\end{equation} 
where the last inequality holds for $c_0\leq \frac{1}{4}$.
Theorem~\ref{thm:results:main} now implies the claim of Corollary~\ref{cor:appl:1bit}.

\item
Let $\grtr\in \rset$ and consider the orthogonal decomposition of the standard Gaussian random vector $\a$ along $\grtr$:
\begin{equation}
	\a = \sp{\a}{\bar\grtr}\bar\grtr + \proj_{\orthcompl{\grtr}}(\a),
\end{equation}
where $\bar\grtr \coloneqq \tfrac{\grtr}{\lnorm{\grtr}}$ and $\proj_{\orthcompl{\grtr}} \coloneqq \proj_{\orthcompl{\{\grtr\}}}$.
Since $\sign(\sp{\a}{\grtr})$ is centered and $\sp{\a}{\bar\grtr} \distributed \Normdistr{0}{1}$ is independent of $\proj_{\orthcompl{\grtr}}(\a)$, we have that 
\begin{align}
	\mean[\ys(\grtr)\a] 
	&=\mean[\sign(\sp{\a}{\grtr}) (\sp{\a}{\bar\grtr}\bar\grtr + \proj_{\orthcompl{\grtr}}(\a))]\\
	&=\mean[\abs{\sp{\a}{\bar\grtr}}]\cdot \bar\grtr + \mean[\sign(\sp{\a}{\grtr})] \cdot \mean[\proj_{\orthcompl{\grtr}}(\a)] = \sqrt{\tfrac{2}{\pi}} \cdot \bar\grtr = \tfunc\grtr,
\end{align}
which implies that $\mmcovar{\grtr}=0$.

\item
Using the shortcut $\bar\grtr \coloneqq \tfrac{\grtr}{\lnorm{\grtr}}$ again, we approximate the observation variable $\ys(\grtr) = \sign(\sp{\a}{\grtr})$ by $\ys_t(\grtr) \coloneqq \psi_t(\sp{\a}{\bar\grtr})$ for $t\in (0, 1]$ and $\grtr \in \rset$ where
\begin{equation}
	\psi_t(s) \coloneqq
	\begin{cases}
		\tfrac{128}{t}\cdot s, & \abs{s}\leq \tfrac{t}{128},\\
		\sign(s), &\text{otherwise,}
	\end{cases} \quad s \in \R.
\end{equation}
See also Figure~\ref{fig:appl:1bit} for an illustration. Since $\ys(\grtr) = \sign(\sp{\a}{\bar\grtr})$, the absolute value of the approximation error then takes the form
\begin{equation}
	\err_t(\grtr) = \abs{\sign(\sp{\a}{\bar\grtr}) - \psi_t(\sp{\a}{\bar\grtr})}\cdot \indset{[0, t^\circ]}(\abs{\sp{\a}{\bar\grtr}}),
\end{equation}
where we have used $t^\circ \coloneqq \frac{t}{128}$ for the sake of notational convenience.
We now show that for this choice of approximation, the conditions of Assumption~\ref{ass:results:incr} are indeed fulfilled with \eqref{eq:app:appl:1bit:incr}:

\emph{On Assumption~\ref{ass:results:incr}\ref{ass:results:incr:approx}.}
Let $\grtr \in \rset$, $\pv \in \S^{p-1}$, and consider the orthogonal decomposition 
\begin{equation}
	\pv = \sp{\pv}{\bar\grtr}\bar\grtr + \proj_{\orthcompl{\grtr}}(\pv).
\end{equation}
This implies
\begin{align}
	\mean[\err_t(\grtr) \cdot \abs{\sp{\a}{\pv}}]
	&\leq \mean[\indset{[0, t^\circ]}(\abs{\sp{\a}{\bar\grtr}}) \cdot \abs{\sp{\a}{\pv}}]\\*
	&\leq \abs{\sp{\pv}{\bar\grtr}}\cdot \mean[\indset{[0, t^\circ]}(\abs{\sp{\a}{\bar\grtr}}) \cdot \abs{\sp{\a}{\bar\grtr}}] + \mean[\indset{[0, t^\circ]}(\abs{\sp{\a}{\bar\grtr}}) \cdot \abs{\sp{\a}{\proj_{\orthcompl{\grtr}}(\pv)}}].
\end{align}
Clearly, $\abs{\sp{\pv}{\bar\grtr}}\leq 1$ and $\mean[\indset{[0, t^\circ]}(\abs{\sp{\a}{\bar\grtr}}) \cdot \abs{\sp{\a}{\bar\grtr}}]\leq t^\circ$.
Since the random variables $\sp{\a}{\bar\grtr}$ and $\sp{\a}{\proj_{\orthcompl{\grtr}}(\pv)}$ are independent, and $\sp{\a}{\bar\grtr}$ is standard Gaussian, we obtain
\begin{equation}
	\mean[\indset{[0, t^\circ]}(\abs{\sp{\a}{\bar\grtr}}) \cdot \abs{\sp{\a}{\proj_{\orthcompl{\grtr}}(\pv)}}] = \prob[\abs{\sp{\a}{\bar\grtr}}\leq t^\circ] \cdot \mean[\abs{\sp{\a}{\proj_{\orthcompl{\grtr}}(\pv)}}] \leq t^\circ \cdot   \mean[\abs{\sp{\a}{\proj_{\orthcompl{\grtr}}(\pv)}}].
\end{equation}
Moreover, by Jensen's inequality and the isotropy of $\a$, it follows that
\begin{equation}
	\mean[\abs{\sp{\a}{\proj_{\orthcompl{\grtr}}(\pv)}}]\leq (\mean[\abs{\sp{\a}{\proj_{\orthcompl{\grtr}}(\pv)}}^2])^{1/2}
	= \lnorm{\proj_{\orthcompl{\grtr}}(\pv)}\leq 1.
\end{equation}
Putting everything together, this shows that Assumption~\ref{ass:results:incr}\ref{ass:results:incr:approx} is satisfied.

\emph{On Assumption~\ref{ass:results:incr}\ref{ass:results:incr:multipl}.} Since $\psi_t$ is $\tfrac{128}{t}$-Lipschitz, the following holds for every $\grtr, \grtr'\in \rset$ (with $\bar\grtr' \coloneqq \tfrac{\grtr'}{\lnorm{\grtr'}}$):
\begin{align}
	\normsubg{\multipl_t(\grtr) - \multipl_t(\grtr')} 
	&\leq \sqrt{\tfrac{2}{\pi}}\normsubg{\sp{\a}{\bar\grtr-\bar\grtr'}} + \normsubg{\psi_t(\sp{\a}{\bar\grtr}) - \psi_t(\sp{\a}{\bar\grtr'})}\\
	&\leq \big(\sqrt{\tfrac{2}{\pi}} + \tfrac{128}{t}\big) \cdot \normsubg{\sp{\a}{\bar\grtr-\bar\grtr'}} \lesssim (1 + t^{-1}) \cdot d_\tfunc(\grtr,\grtr').
\end{align} 
This implies $\multiplLip \lesssim 1 + t^{-1}$. Furthermore, observe that $\abs{\ys_t(\grtr)} \leq 1$, so that
\begin{equation}
	\normsubg{\multipl_t(\grtr)}\leq \sqrt{\tfrac{2}{\pi}}\normsubg{\sp{\a}{\bar\grtr}}  + \normsubg{\ys_t(\grtr)}\lesssim 1
\end{equation}
for every $\grtr\in \rset$. This shows $\multiplDia\lesssim 1$.

\emph{On Assumption~\ref{ass:results:incr}\ref{ass:results:incr:err}.} We observe that the function $s\mapsto \phi_t(s) \coloneqq \abs{\sign(s)-\psi_t(s)}$ is $\tfrac{128}{t}$-Lipschitz. Therefore, for every $\grtr, \grtr'\in \rset$, we obtain
\begin{equation}
	\normsubg{\err_t(\grtr) - \err_t(\grtr')}\leq \tfrac{128}{t} \cdot\normsubg{\sp{\a}{\bar\grtr-\bar\grtr'}}\lesssim t^{-1} \cdot d_\tfunc(\grtr,\grtr'),
\end{equation}
which implies $\errLip\lesssim t^{-1}$. Finally, since $\abs{\err_t(\grtr)}\leq 1$ for every $\grtr\in \rset$, it follows that $\errDia\lesssim 1$. \qed
\end{tmpl}

\subsection{Proofs for Subsection~\ref{subsec:appl:1bitdither}}
\label{subsec:app:appl:1bitdither}

\paragraph{Proof of Corollary~\ref{cor:appl:1bitdither}.}

We follow the proof template from the beginning of Section~\ref{sec:app:appl}:
\begin{tmpl}
\item
The model setup of Corollary~\ref{cor:appl:1bitdither} fits into Assumption~\ref{ass:results:meas} as follows:
\begin{itemize}
\item
	The measurement vector $\a \in \R^p$ is centered, isotropic, and sub-Gaussian with $\normsubg{\a} \leq \subgparam$.
	The signal set $\rset$ satisfies $\rset \subset R \ball[2][p]$.	
	The output function $\fout \colon \R^p \times \rset \to \R$ takes the form $\fout(\a, \grtr) \coloneqq \sign(\sp{\a}{\grtr} +\tau)$, where $\tau$ is uniformly distributed on $[-\lambda, \lambda]$ and independent of~$\a$. In particular, $\fout$ is a random function.
\item 
	The target function $\tfunc \colon \rset \to \sset$ corresponds to rescaling by a factor of $\lambda^{-1}$, i.e., $\tfunc\grtr \coloneqq \lambda^{-1}\grtr$. In particular, we have that $d_\tfunc(\grtr,\grtr')= \lambda^{-1}\lnorm{\grtr - \grtr'}$.
\end{itemize}

\item
There exists an absolute constant $\tilde{C} \geq e$ such that if 
\begin{equation}\label{eq:1bitdither_cond_lambda}
	\lambda\geq \tilde{C}\cdot R \cdot \subgparam \cdot \sqrt{\log(e / t)},
\end{equation}
the target mismatch satisfies $\mmcovar{\grtr}\leq \tfrac{t}{32}$ for every $\grtr\in \rset$.

\item
There exists an approximation $\ys_t(\grtr)$ of the  observation variable $\ys(\grtr)$ such that the conditions of Assumption~\ref{ass:results:incr} are fulfilled with
\begin{equation}\label{eq:app:appl:1bitdither:incr}
	\multiplLip \leq \subgparam \cdot (1 + 64t^{-1}),\quad \errLip \leq 64 \subgparam t^{-1},\quad \multiplDia\lesssim R \subgparam \lambda^{-1} + 1, \quad \errDia \lesssim 1.
\end{equation} 

\item
The first and third branch of the condition \eqref{eq:results:main:m} are implied by \eqref{eq:appl:1bitdither:m} for a sufficiently large universal constant $C'>0$.
Since $\phi(\beta) \coloneqq \beta\sqrt{\log(e/\beta)}$ defines a continuous and non-decreasing function on $[0, 1]$ with $\phi(1)=1$ and $\phi(0)\coloneqq 0$ (by continuous extension), we may assume without loss of generality that $\beta\sqrt{\log(e/\beta)}= c_0 \subgparam^{-2} t \in (0,1]$.
Now, we set
\begin{equation}
	\tune^2 \coloneqq \frac{1}{t\subgparam^2\sqrt{\log(e/\beta)}}, \quad \nout \coloneqq \floor{\beta m}, \quad \probsuccess_0 \coloneqq \sqrt{2 m \beta \log(e/\beta)},
\end{equation}
implying that $\probsuccess_0\geq \sqrt{\nout \log(em/\nout)}$ and $\probsuccess_0^2\geq 2 c_0 m \subgparam^{-2} t$.
Combining the latter inequality with \eqref{eq:appl:1bitdither:m} for $C'$ sufficiently large particularly implies that $\probsuccess_0^2 \geq \probsuccess^2$.
Furthermore, we may assume that $c_0\leq \frac{1}{4}C^{-1}$ and that \eqref{eq:appl:1bitdither:m} holds with $C'\geq 2C$, where $C>0$ denotes the universal constant from \eqref{eq:results:main:m}. Then the second branch of the condition \eqref{eq:results:main:m} is satisfied, since
\begin{equation}
	C\subgparam^4\tune^2\probsuccess_0^2= \frac{C\subgparam^2}{t\sqrt{\log(e/\beta)}} \cdot 2 \beta m \log(e/\beta)=
	2c_0C\cdot m\leq \tfrac{m}{2} 
\end{equation}
and
\begin{equation}
	C\subgparam^4\tune^2 \cdot \effdim[t]{\sset - \lambda^{-1}\rset} = \frac{C\subgparam^2}{t\sqrt{\log(e/\beta)}} \cdot \effdim[t]{\sset - \lambda^{-1}\rset}\leq  \tfrac{1}{2} C' \cdot \subgparam^2 t^{-1} \cdot \effdim[t]{\sset - \lambda^{-1}\rset}\leq \tfrac{m}{2}.
\end{equation}

Next, we show that every $\Y \in \{-1,1\}^m$ given by \eqref{eq:appl:1bitdither:meas} with $\tfrac{1}{2m}\lnorm{\Noise}[1] \leq \beta$ also satisfies \eqref{eq:results:main:rob}.
Since $\Noise=\Y - \Ys(\grtr)\in \{-2,0,2\}^m$ and $\tfrac{1}{2}\lnorm{\Y - \Ys(\grtr)}[1]\leq \beta m$, it follows that $\cardinality{\{i\in [m] \suchthat y_i \neq \ys_i(\grtr)\}}\leq \floor{\beta m}$. Consequently, $\sigma_{\nout}(\Y - \Ys(\grtr))_{2}=0$ and
\begin{equation}
	\tfrac{1}{\sqrt{m}}\norm{\Y - \Ys(\grtr)}_{[\nout]}= \tfrac{1}{\sqrt{m}}\lnorm{\Y - \Ys(\grtr)}\leq 2\sqrt{\beta}= 2\Big(\tfrac{c_0 t}{\subgparam^2 \sqrt{\log(e/\beta)}} \Big)^{1/2} \leq \tune t,
\end{equation} 
where the last inequality holds for $c_0\leq \frac{1}{4}$.
Theorem~\ref{thm:results:main} now implies the claim of Corollary~\ref{cor:appl:1bitdither}.

\item
Since $\a$ is isotropic, we have that
\begin{equation}
	\mmcovar{\grtr} = \lnorm[\big]{\mean[\ys(\grtr)\a] - \lambda^{-1} \grtr}
	\leq \lnorm[\big]{\mean\big[\big(\sign(\sp{\a}{\grtr} + \tau)-\sp{\a}{\lambda^{-1}\grtr}\big)\a\big]}.
\end{equation}
Therefore, it suffices to show that the following holds for all $\grtr\in \rset$ and $\pv\in \S^{p-1}$:
\begin{equation}
	\mean\big[\big(\sign(\sp{\a}{\grtr} + \tau)-\sp{\a}{\lambda^{-1}\grtr}\big)\sp{\a}{\pv}\big]\leq \tfrac{t}{32}.
\end{equation}
The following identity explains why adding a uniformly distributed dither $\tau \in [-\lambda, \lambda]$ before quantization is useful:
\begin{equation}
	\mean_{\tau}[\sign(s + \tau)] = \lambda^{-1}s \cdot \indset{[-\lambda, \lambda]}(s) + \sign(s) \cdot \indset{\R \setminus [-\lambda, \lambda]}(s), \quad s \in \R.
\end{equation}
In other words, for $s$ small enough, integrating over the dithering variable $\tau$ allows us to ``smooth out'' the discontinuity of the sign function. As $\tau$ and $\a$ are independent, we can apply the above identity as follows:
\begin{align}
	& \mean\big[\big(\sign(\sp{\a}{\grtr} + \tau)-\sp{\a}{\lambda^{-1}\grtr}\big)\sp{\a}{\pv}\big]\\*
	= {} & \mean_{\a}\mean_{\tau}\big[\big(\sign(\sp{\a}{\grtr} + \tau)-\sp{\a}{\lambda^{-1}\grtr}\big)\sp{\a}{\pv}\big]\\
	= {} & \mean\big[\big(\lambda^{-1}\sp{\a}{\grtr} \cdot \indset{[-\lambda, \lambda]}(\sp{\a}{\grtr}) + \sign(\sp{\a}{\grtr}) \cdot \indset{\R \setminus [-\lambda, \lambda]}(\sp{\a}{\grtr})-\lambda^{-1}\sp{\a}{\grtr}\big)\sp{\a}{\pv}\big]\\
	= {} & \mean\big[\big(\sign(\sp{\a}{\grtr}) - \lambda^{-1}\sp{\a}{\grtr} \big) \cdot \indset{\R \setminus [-\lambda, \lambda]}(\sp{\a}{\grtr})\cdot \sp{\a}{\pv}\big].
\end{align}	
Using the Cauchy-Schwarz inequality twice, the isotropy of $\a$, and the triangle inequality, we now obtain 
\begin{align}	
	& \mean\big[\big(\sign(\sp{\a}{\grtr}) - \lambda^{-1}\sp{\a}{\grtr} \big) \cdot \indset{\R \setminus [-\lambda, \lambda]}(\sp{\a}{\grtr})\cdot \sp{\a}{\pv}\big]\\*
	\leq {} & \Big(\mean\big[\big(\sign(\sp{\a}{\grtr}) - \lambda^{-1}\sp{\a}{\grtr} \big)^2 \cdot \indset{\R \setminus [-\lambda, \lambda]}(\sp{\a}{\grtr})\big] \Big)^{1/2}
	\cdot \big(\mean [\sp{\a}{\pv}^2]\big)^{1/2}\\
	= {} & \Big(\mean\big[\big(\sign(\sp{\a}{\grtr}) - \lambda^{-1}\sp{\a}{\grtr} \big)^2 \cdot \indset{\R \setminus [-\lambda, \lambda]}(\sp{\a}{\grtr})\big] \Big)^{1/2}\\
	\leq {} & \norm{\sign(\sp{\a}{\grtr}) - \lambda^{-1}\sp{\a}{\grtr}}_{L^4}\cdot \big(\prob(\abs{\sp{\a}{\grtr}}>\lambda)\big)^{1/4}\\
	\leq {} & \big(1+ \lambda^{-1}\norm{\sp{\a}{\grtr}}_{L^4}\big) \cdot \big(\prob(\abs{\sp{\a}{\grtr}}>\lambda)\big)^{1/4}.
\end{align}
Since $\a$ is sub-Gaussian with $\normsubg{\a}\leq \subgparam$, there exist absolute constants
$C''\geq 1$, $c''>0$ such that $\norm{\sp{\a}{\grtr}}_{L^4}\leq C''\cdot \normsubg{\sp{\a}{\grtr}}\leq C''\cdot\subgparam\cdot \lnorm{\grtr}$
and $\prob(\abs{\sp{\a}{\grtr}}>\lambda)\leq 2\exp\big(-c''\lambda^2 (\subgparam\lnorm{\grtr})^{-2}\big)$.
Finally, using that every $\grtr\in \rset$ satisfies $\lnorm{\grtr}\leq R$, we can conclude that 
\begin{equation}
	\mean\big[\big(\sign(\sp{\a}{\grtr} + \tau)-\sp{\a}{\lambda^{-1}\grtr}\big)\sp{\a}{\pv}\big]\leq C'' \cdot (1+ R \subgparam \lambda^{-1}) \cdot \exp\big(-c'' \lambda^2(R\subgparam)^{-2}\big) \leq \tfrac{t}{32},
\end{equation}
where the last estimate is due to \eqref{eq:1bitdither_cond_lambda} for $\tilde{C}\geq e$ large enough.

\item
We approximate the observation variable $\ys(\grtr) = \sign(\sp{\a}{\grtr} + \tau)$ by $\ys_t(\grtr) \coloneqq \psi_t(\sp{\a}{\grtr} +\nobreak \tau)$ for $t\in (0, 1]$ and $\grtr \in \rset$ where
\begin{equation}
	\psi_t(s) = 
	\begin{cases}
		\tfrac{64}{t}\cdot \lambda^{-1}s, &\abs{s}\leq \tfrac{t}{64} \cdot \lambda,\\
		\sign(s), &\text{otherwise,}
	\end{cases} \quad s \in \R.
\end{equation}
The absolute value of the approximation error then takes the form
\begin{equation}
	\err_t(\grtr) = \abs{\sign(\sp{\a}{\grtr} + \tau) - \psi_t(\sp{\a}{\grtr} + \tau)}\cdot \indset{[0, t^\circ\lambda]}(\abs{\sp{\a}{\grtr} + \tau}),
\end{equation}
where we have used $t^\circ \coloneqq \frac{t}{64}$ for the sake of notational convenience.
We now show that for this choice of approximation, the conditions of Assumption~\ref{ass:results:incr} are indeed fulfilled with \eqref{eq:app:appl:1bitdither:incr}:

\emph{On Assumption~\ref{ass:results:incr}\ref{ass:results:incr:approx}.}
For $\grtr \in \rset$ and $\pv \in \S^{p-1}$, we have that
\begin{equation}
	\mean[\err_t(\grtr) \cdot \abs{\sp{\a}{\pv}}] \leq \mean[\indset{[0, t^\circ\lambda]}(\abs{\sp{\a}{\grtr} + \tau}) \cdot \abs{\sp{\a}{\pv}}] \leq \mean_{\a}\big[\abs{\sp{\a}{\pv}} \cdot \mean_{\tau} [\indset{[0, t^\circ\lambda]}(\abs{\sp{\a}{\grtr} + \tau})]\big].
\end{equation}
Since $\mean_{\tau}[\indset{[0, t^\circ\lambda]}(\abs{\sp{\a}{\grtr} + \tau})]\leq t^\circ$, it follows that
\begin{equation}
	\mean[\err_t(\grtr) \cdot \abs{\sp{\a}{\pv}}]\leq t^\circ \cdot \mean[\abs{\sp{\a}{\pv}}]\leq t^\circ \cdot (\mean[\abs{\sp{\a}{\pv}}^2])^{1/2}=t^\circ, 
\end{equation}	
implying the condition of Assumption~\ref{ass:results:incr}\ref{ass:results:incr:approx}.

\emph{On Assumption~\ref{ass:results:incr}\ref{ass:results:incr:multipl}.} Since $\psi_t$ is $\tfrac{64}{t \lambda}$-Lipschitz, the following holds for every $\grtr, \grtr'\in \rset$:
\begin{align}
	\normsubg{\multipl_t(\grtr) - \multipl_t(\grtr')} 
	&\leq \lambda^{-1}\normsubg{\sp{\a}{\grtr-\grtr'}} + \normsubg{\psi_t(\sp{\a}{\grtr} + \tau) - \psi_t(\sp{\a}{\grtr'} + \tau)}\\
	&\leq \big(\lambda^{-1} + \tfrac{64}{t \lambda}\big) \cdot \normsubg{\sp{\a}{\grtr-\grtr'}} \leq \lambda^{-1} \cdot (1 + \tfrac{64}{t}) \cdot \subgparam\cdot \lnorm{\grtr - \grtr'} \\
	&= \subgparam \cdot (1 + 64 t^{-1}) \cdot d_\tfunc(\grtr,\grtr').
\end{align} 
This implies $\multiplLip \leq \subgparam \cdot (1 + 64t^{-1})$. Furthermore, observe that $\abs{\ys_t(\grtr)} \leq 1$, so that
\begin{equation}
	\normsubg{\multipl_t(\grtr)}\leq \lambda^{-1}\normsubg{\sp{\a}{\grtr}}  + \normsubg{\ys_t(\grtr)}\lesssim R \subgparam \lambda^{-1} + 1
\end{equation}
for every $\grtr\in \rset$. This shows $\multiplDia\lesssim R \subgparam \lambda^{-1} + 1$.

\emph{On Assumption~\ref{ass:results:incr}\ref{ass:results:incr:err}.} We observe that the function $s\mapsto \phi_t(s) \coloneqq \abs{\sign(s)-\psi_t(s)}$ is $\tfrac{64}{t\lambda}$-Lipschitz. Therefore, for every $\grtr, \grtr'\in \rset$, we obtain
\begin{align}
	\normsubg{\err_t(\grtr) - \err_t(\grtr')}\leq \tfrac{64}{t\lambda} \cdot\normsubg{\sp{\a}{\grtr - \grtr'}}\leq 64 \subgparam t^{-1} \cdot d_\tfunc(\grtr,\grtr').
\end{align}
Hence, $\errLip\leq 64 \subgparam t^{-1}$. Finally, since $\abs{\err_t(\grtr)}\leq 1$ for every $\grtr\in \rset$, it follows that $\errDia\lesssim 1$. \qed
\end{tmpl}

\subsection{Proofs for Subsection~\ref{subsec:appl:mbit}}
\label{subsec:app:appl:mbit}

\paragraph{Proof of Corollary~\ref{cor:appl:mbit}.}

We follow the proof template from the beginning of Section~\ref{sec:app:appl}:

\begin{tmpl}
\item
The model setup of Corollary~\ref{cor:appl:mbit} fits into Assumption~\ref{ass:results:meas} as follows:
\begin{itemize}
\item
	The measurement vector $\a \in \R^p$ is centered, isotropic, and sub-Gaussian with $\normsubg{\a} \leq\nobreak \subgparam$.
	The signal set $\rset$ is a bounded subset of $\R^p$.
	The output function $\fout \colon \R^p \times \rset \to \R$ takes the form $\fout(\a, \grtr) \coloneqq q_\delta(\sp{\a}{\grtr} + \tau)$, where $\tau$ is uniformly distributed on $[-\delta, \delta]$ and independent of~$\a$. In particular, $\fout$ is a random function.
	Moreover, the observation vector of $\grtr$ is given by $\Ys(\grtr) = q_\delta(\A\grtr+\ttau)$, where $\A\in \R^{m\times p}$ denotes the sub-Gaussian random matrix with row vectors $\a_1, \dots, \a_m\in \R^p$ and $\ttau \coloneqq (\tau_1, \dots, \tau_m)$.
	
\item
	The target function $\tfunc \colon \rset \to \sset$ is the canonical embedding into $\sset$, i.e., $\tfunc\grtr \coloneqq \grtr$. In particular, we have that $d_\tfunc(\grtr,\grtr')= \lnorm{\grtr - \grtr'}$.
\end{itemize}

\item
The target mismatch $\mmcovar{\grtr}$ vanishes for every $\grtr\in \rset$.
	
\item 
If $t\leq 128\delta$, there exists an approximation $\ys_t(\grtr)$ of the observation variable $\ys(\grtr)$ such that the conditions of Assumption~\ref{ass:results:incr} are fulfilled with
\begin{equation} \label{eq:proof:mbit:assincr:tsmall}
	\multiplLip \lesssim \subgparam \delta t^{-1},\quad \errLip \lesssim \subgparam \delta t^{-1},\quad \multiplDia\lesssim \delta,\quad \errDia \lesssim \delta.
\end{equation} 
If $t> 128\delta$, we choose any maximal $\tfrac{t}{256}$-packing $\rset_t$ for $\rset$ (i.e., a maximal subset of $\rset$ such that $\lnorm{\grtr-\grtr'}> \tfrac{t}{256}$ for all $\grtr, \grtr'\in \rset_t$ with $\grtr\neq \grtr'$) and show that for the trivial choice $\ys_t(\grtr)\coloneqq\ys(\grtr)$, the conditions of Assumption~\ref{ass:results:incr} are fulfilled on $\rset_t$ with
\begin{equation}\label{eq:proof:mbit:assincr:tlarge}
	\multiplLip \lesssim \delta t^{-1},\quad \errLip =0,\quad \multiplDia\lesssim \delta,\quad \errDia = 0.
\end{equation} 

\item
We first note that for $t\geq \delta$, the inequality $c_0 \subgparam^{-2} \sqrt{\delta t} / \sqrt{\max\{1, \log( \delta e/t)\}}\leq \tfrac{t}{40}$ holds if $c_0\leq \tfrac{1}{40}$.
Furthermore, the condition $\tfrac{1}{m}\lnorm{\Noise}[0] \leq t \delta^{-1}$ is trivially fulfilled. 
Consequently, it suffices to prove the statement of Corollary~\ref{cor:appl:mbit} for all those input vectors $\Y\in \delta\Z^m$ that satisfy the condition (a) in \eqref{eq:appl:mbit:input}, and all those $\Y\in \delta\Z^m$ that satisfy condition (b) in \eqref{eq:appl:mbit:input} in the case $t\leq \delta$. We now distinguish between the two cases $t\leq 128\delta$ and $t>128\delta$:

\emph{The case  $t\leq 128\delta$.}
From \eqref{eq:proof:mbit:assincr:tsmall} it follows that the first and third branch of the condition \eqref{eq:results:main:m} are implied by \eqref{eq:appl:mbit:m} for a sufficiently large universal constant $C'>0$. 

For the choice $\tune \coloneqq \subgparam^{-1} \sqrt{\log\subgparam}$, $\nout \coloneqq 0$, and $\probsuccess \coloneqq \probsuccess_0$, we observe that the second branch of \eqref{eq:results:main:m} follows from the first one. 
Therefore, the claim of Corollary~\ref{cor:appl:mbit} follows from Theorem~\ref{thm:results:main} for all those input vectors $\Y \in \delta\Z^m$ that satisfy the condition (a) in \eqref{eq:appl:mbit:input}.

Next, let us assume that $t\leq \delta$ and consider the choice 
\begin{equation}
	\tune^2 \coloneqq \frac{\delta}{4C t\subgparam^4 \log(\delta e/ t)}, \quad \nout \coloneqq \floor{m t \delta^{-1}}, \quad \probsuccess_0 \coloneqq \sqrt{2m t \delta^{-1}\log(\delta e / t)},
\end{equation}
where $C>0$ denotes the universal constant from \eqref{eq:results:main:m}.
This implies $\probsuccess_0\geq \sqrt{\nout \log(em/\nout)}$ as well as $\probsuccess_0^2\geq \probsuccess^2$ if the universal constant $C'$ in \eqref{eq:appl:mbit:m} is large enough.
Furthermore, the second branch of \eqref{eq:results:main:m} is satisfied: we have that $C\subgparam^4\tune^2\probsuccess_0^2= \tfrac{m}{2}$, and assuming that \eqref{eq:appl:mbit:m} holds with $C'\geq \frac{1}{2}$, it follows that
\begin{equation}
	C\subgparam^4\tune^2 \cdot \effdim[t]{\sset - \tfunc\rset}\leq \tfrac{1}{4}\delta t^{-1} \cdot \effdim[t]{\sset - \rset}\leq   \tfrac{1}{4} \subgparam^2 \delta^2 t^{-2} \cdot \effdim[t]{\sset - \rset}\leq \tfrac{m}{2}.
\end{equation}
It remains to show that every $\Y \in \delta\Z^m$ given by \eqref{eq:appl:mbit:meas} such that
\begin{equation}
	\tfrac{1}{\sqrt{m}}\lnorm{\Noise}\leq \tfrac{c_0\sqrt{\delta t}}{\subgparam^2\sqrt{\log(\delta e/t)}}\quad \text{and} \quad \tfrac{1}{m}\lnorm{\Noise}[0]\leq \tfrac{t}{\delta}
\end{equation}
also satisfies the condition \eqref{eq:results:main:rob}.
Indeed, since $\Noise=\Y - \Ys(\grtr)\in \delta \Z^m$ and $\lnorm{\Noise}[0]\leq m t \delta^{-1}$, it follows that $\cardinality{\{i\in [m] \suchthat y_i \neq \ys_i(\grtr)\}}\leq \floor{m t \delta^{-1}}$. 
Consequently, $\sigma_{\nout}(\Y - \Ys(\grtr))_{2}=0$ and
\begin{equation}
	\tfrac{1}{\sqrt{m}}\norm{\Y - \Ys(\grtr)}_{[\nout]}= \tfrac{1}{\sqrt{m}}\lnorm{\Y - \Ys(\grtr)}\leq \tfrac{c_0\sqrt{\delta t}}{\subgparam^2\sqrt{\log(\delta e/t)}}\leq\tune t,
\end{equation} 
where the last inequality holds for $c_0>0$ sufficiently small. 
Theorem~\ref{thm:results:main} now implies the claim of Corollary~\ref{cor:appl:mbit} for all those input vectors $\Y \in \delta\Z^m$ that satisfy the condition (b) in \eqref{eq:appl:mbit:input}.

\emph{The case  $t> 128\delta$.}
Setting $\nout \coloneqq 0$, $\tune \coloneqq \subgparam^{-1} \sqrt{\log\subgparam}$, and $\probsuccess \coloneqq \probsuccess_0$, the second branch of \eqref{eq:results:main:m} is implied by the first one.
Due to \eqref{eq:proof:mbit:assincr:tlarge}, Theorem~\ref{thm:results:main} applied to the signal set $\rset_t$ now implies that there exist universal constants $c_1,C_1>0$ such that if
\begin{equation}\label{eq:multibit:aboveresolution:cond1}
	m \geq C_1 \cdot \subgparam^2 \cdot \Big(  (\log\subgparam + t^{-2} \delta^2) \cdot \big(\effdim[t]{\sset - \rset_t}+ \probsuccess^2 \big)  
	+ t^{-4}  \delta^2 \cdot \effdim{\rset_t} \Big),
\end{equation}
the following event $\mathcal{A}_1$ occurs with probability at least $1 - \exp(-c_1 \probsuccess^2)$: For every $\grtr' \in \rset_t$ and $\Y \in \delta \Z^m$ such that $\tfrac{1}{\sqrt{m}}\lnorm{\Y-q_{\delta}(\A\grtr'+\ttau)} \leq \tfrac{t}{20}$, every minimizer $\pvsolu$ of \eqref{eq:results:klasso} satisfies $\lnorm{\pvsolu - \grtr'} \leq t$.
	
Moreover, according to Theorem~\ref{thm:proofs:matrixdev}, there exists a universal constant $C_2>0$ such that if 
\begin{equation}\label{eq:multibit:aboveresolution:cond2}
	m\geq C_2\cdot \subgparam^2 \log\subgparam \cdot \big(\effdim[t]{\rset-\rset}  + \probsuccess^2\big),
\end{equation}
the following event $\mathcal{A}_2$ occurs with probability at least $1-3\exp(-\probsuccess^2)$:
\begin{equation}
	\sup_{\substack{\grtr\in \rset,\, \grtr'\in \rset_t\\ \lnorm{\grtr-\grtr'}\leq \tfrac{t}{256}}}  \lnorm[\big]{\tfrac{1}{\sqrt{m}}\A(\grtr-\grtr')} \leq \tfrac{t}{128}.
\end{equation}
We claim that Corollary~\ref{cor:appl:mbit} holds on the intersection of the events $\mathcal{A}_1$ and $\mathcal{A}_2$. To this end, let $\grtr\in \rset$ be arbitrary and assume that $\Y\in \delta \Z^m$ satisfies $\tfrac{1}{\sqrt{m}}\lnorm{\Y-q_{\delta}(\A\grtr+\ttau)} \leq \tfrac{t}{40}$. 
Since $\rset_t$ is a maximal $\tfrac{t}{256}$-packing for $\rset$, there exists $\grtr'\in \rset_t$ with $\lnorm{\grtr-\grtr'}\leq \tfrac{t}{256}$ (otherwise $\rset_t$ would not be maximal).
On the event $\mathcal{A}_2$ and by the triangle inequality, we obtain 
\begin{align}
	\tfrac{1}{\sqrt{m}}\lnorm[\big]{\Y-q_{\delta}(\A\grtr'+\ttau)}&\leq \tfrac{1}{\sqrt{m}}\lnorm[\big]{\Y-q_{\delta}(\A\grtr+\ttau)}
	+ \tfrac{1}{\sqrt{m}}\lnorm[\big]{q_{\delta}(\A\grtr+\ttau)-q_{\delta}(\A\grtr'+\ttau)}\\
	&\leq \begin{multlined}[t] \tfrac{t}{40} + \tfrac{1}{\sqrt{m}}\lnorm[\big]{(\A\grtr + \ttau) -(\A\grtr' + \ttau) } \\
		+ \tfrac{1}{\sqrt{m}}\lnorm[\big]{q_\delta(\A\grtr + \ttau) - (\A\grtr + \ttau)} + \tfrac{1}{\sqrt{m}}\lnorm[\big]{q_\delta(\A\grtr' + \ttau) - (\A\grtr' + \ttau)} \end{multlined} \\
	&\leq \tfrac{t}{40} + \tfrac{t}{128}+ 2\delta \leq \tfrac{t}{20}.
\end{align}
On the event $\mathcal{A}_1$, every minimizer $\pvsolu$ of \eqref{eq:results:klasso} satisfies $\lnorm{\pvsolu - \grtr'} \leq t$, and therefore 
\begin{equation}
	\lnorm{\pvsolu - \grtr} \leq \lnorm{\pvsolu - \grtr'} + \lnorm{\grtr' - \grtr} \leq t + \tfrac{t}{256}\leq 2t.
\end{equation}
Finally, for $C'>0$ sufficiently large, we conclude that the condition \eqref{eq:appl:mbit:m} implies both \eqref{eq:multibit:aboveresolution:cond1} and 
\eqref{eq:multibit:aboveresolution:cond2}. Hence, the events $\mathcal{A}_1$ and $\mathcal{A}_2$ occur jointly with probability at least $1 - \exp(-c \probsuccess^2)$, provided that $c>0$ is chosen small enough.

\item
Let $\grtr\in \rset$. Since $\a$ is isotropic, we have that
\begin{equation}
	\mmcovar{\grtr} = \lnorm[\big]{\mean[\ys(\grtr)\a] - \grtr}
	\leq \lnorm[\big]{\mean\big[\big(q_\delta(\sp{\a}{\grtr} + \tau)-\sp{\a}{\grtr}\big)\a\big]},
\end{equation}
so that it suffices to show that 
\begin{equation}
	\mean\big[\big(q_\delta(\sp{\a}{\grtr} + \tau)-\sp{\a}{\grtr}\big)\sp{\a}{\pv}\big]=0 \qquad \text{for all $\pv\in \S^{p-1}$}.
\end{equation}
It is straightforward to check that
\begin{equation}\label{eq:multi:dither:identity}
	\mean[q_\delta(s + \tau)] = s \qquad \text{for all $s\in \R$}.
\end{equation}
In other words, integrating over the dithering variable $\tau$ allows us to ``eliminate'' the quantizer. 
For the sake of completeness, we show the identity \eqref{eq:multi:dither:identity} for $s = k \cdot (2\delta) + s'$ where $k\in \Z$ and $s'\in [0,\delta]$; the case $s = k \cdot (2\delta) + s'$ where $k\in \Z$ and $s'\in (\delta, 2\delta)$ can be treated analogously. First, observe that
\begin{align}
	q_\delta(s + \tau)=(2\ceil{\tfrac{s + \tau}{2\delta}}-1)\delta = (2\ceil{k + \tfrac{s' + \tau}{2\delta}}-1)\delta.
\end{align}
Since $\tfrac{s'+\tau}{2\delta} \in [-\tfrac{1}{2}, 1]$, it follows that $q_\delta(s + \tau)\in \{(2k-1)\delta, (2k+1)\delta\}$ and more precisely, 
\begin{alignat}{2}
	&\tau\in [-\delta-s',-s'] &&\quad\implies\quad q_\delta(s + \tau) =(2k-1)\delta,\\
	&\tau\in (-s', 2\delta-s'] &&\quad\implies\quad q_\delta(s + \tau) =(2k+1)\delta.
\end{alignat}
Therefore, we obtain
\begin{align}
	\mean[q_\delta(s + \tau)] 
	&= \prob(\tau\in [-\delta-s',-s']) \cdot(2k-1)\delta + \prob(\tau \in (-s', 2\delta-s'])  \cdot (2k+1)\delta\\
	&= \prob(\tau\in [-\delta,-s']) \cdot(2k-1)\delta + \prob(\tau \in (-s', \delta])  \cdot (2k+1)\delta\\
	&= \big(\tfrac{-s'+\delta}{2\delta}\big) \cdot(2k-1)\delta + \big(\tfrac{\delta + s'}{2\delta}\big)  \cdot (2k+1)\delta\\
	&=\tfrac{s'}{2} + \tfrac{1}{2}(2k-1)\delta + \tfrac{s'}{2} + \tfrac{1}{2}(2k+1)\delta = 2k\delta + s' = s.
\end{align}
Since $\tau$ and $\a$ are independent, we can apply \eqref{eq:multi:dither:identity} as follows:
\begin{align}
	\mean\big[\big(q_\delta(\sp{\a}{\grtr} + \tau)-\sp{\a}{\grtr}\big)\sp{\a}{\pv}\big]
	&=\mean_{\a}\mean_\tau\big[\big(q_\delta(\sp{\a}{\grtr} + \tau)-\sp{\a}{\grtr}\big)\sp{\a}{\pv}\big]\\
	&=\mean\big[\big(\sp{\a}{\grtr}-\sp{\a}{\grtr}\big)\sp{\a}{\pv}\big] = 0.
\end{align}	

\item
Before distinguishing between the cases $t\leq 128\delta$ and $t>128\delta$ according to Step~1b, let us analyze the action of the quantizer $q_\delta$ in more detail. To this end, we partition the real axis $\R$ into half-open intervals $\mathcal{I}_k$ of length $2\delta$ given by $\mathcal{I}_k \coloneqq (e_{k-1}, e_k]$ for $k\in \Z$ with $e_k \coloneqq k\cdot 2\delta$. Then~$q_\delta$
maps every point in the interval $\mathcal{I}_k$ to its center point:
\begin{equation}
	q_\delta(s)= \frac{e_{k-1}+e_k}{2}, \quad s\in \mathcal{I}_k.
\end{equation}
In particular, $q_\delta(s)$ is discontinuous exactly at the points $\{ e_k \suchthat k\in \Z \}$. The basic idea is now to approximate the quantizer
$q_\delta$ by a Lipschitz continuous function by ``cutting out'' intervals of a certain radius $t^\circ > 0$ around each point $e_k$ and 
inserting a straight line that connects both quantization values. Assuming that $t^\circ \leq \delta$, the resulting function takes the form 
\begin{equation}
	\psi_{t^{\circ}}(s) \coloneqq 
	\begin{cases}
		\tfrac{\delta}{t^\circ}\cdot s + e_k(1-\tfrac{\delta}{t^\circ}) , &\text{if $\exists k\in \Z: s\in \mathcal{E}_{k,t^\circ}$,}\\ 
		q_\delta(s), &\text{otherwise,}
	\end{cases} \quad s \in \R,
\end{equation}
where $\mathcal{E}_{k,t^\circ} \coloneqq [e_k-t^\circ, e_k+t^\circ]$; note that if $t^\circ=\delta$, then $\psi_{t^{\circ}}$ is just the identity function.
We also define
\begin{equation}
	\phi_{t^{\circ}}(s)\coloneqq\abs{q_\delta(s)-\psi_{t^{\circ}}(s)},\quad s\in \R.
\end{equation}

Let us now show that $\psi_{t^{\circ}}$ and $\phi_{t^{\circ}}$ are both $\tfrac{\delta}{t^\circ}$-Lipschitz.

Since $q_\delta$ is locally constant on $\R\setminus \{e_k \suchthat k\in \Z\}$ and $s\mapsto \tfrac{\delta}{t^\circ}\cdot s + e_k(1-\tfrac{\delta}{t^\circ})$ is clearly $\tfrac{\delta}{t^\circ}$-Lipschitz, 
it is sufficient to show that $\psi_{t^{\circ}}$ is continuous, i.e.,
$\tfrac{\delta}{t^\circ}\cdot s + e_k(1-\tfrac{\delta}{t^\circ})=q_\delta(s)$ for $s \in \R$ with $\abs{s-e_k}=t^\circ$ for some $k \in \Z$. If $s=e_k-t^\circ$, then $\tfrac{\delta}{t^\circ}\cdot s + e_k(1-\tfrac{\delta}{t^\circ})=e_k-\delta$, and we also have that
\begin{align}
	q_\delta(s)=\big(2\ceil[\big]{\tfrac{k\cdot(2\delta) - t^\circ}{2\delta}}-1\big)\delta=(2\ceil{k  - \tfrac{t^\circ}{2\delta}}-1)\delta.
\end{align}
Since $t^\circ\in (0,\delta]$, it follows that $\ceil{k  - \tfrac{t^\circ}{2\delta}}=k$, and therefore, $q_\delta(s)=(2k-1)\delta = e_k-\delta$. The case
$s=e_k+t^\circ$ works analogously. Thus, $\psi_{t^{\circ}}$ is indeed $\tfrac{\delta}{t^\circ}$-Lipschitz.

For $\phi_{t^{\circ}}$, it is clearly sufficient to show $\tfrac{\delta}{t^\circ}$-Lipschitz continuity on $\mathcal{E}_{k,t^\circ}$ for every $k\in \Z$. 
In the case $s\in [e_k-t^\circ, e_k]$, we observe that $q_\delta(s)=e_k-\delta$ and $\tfrac{\delta}{t^\circ}\cdot s + e_k(1-\tfrac{\delta}{t^\circ})\geq e_k-\delta$, implying that
\begin{equation}
	\phi_{t^{\circ}}(s) = \tfrac{\delta}{t^\circ}\cdot s + e_k(1-\tfrac{\delta}{t^\circ}) - (e_k-\delta)= \delta - \tfrac{\delta}{t^\circ}(e_k-s).
\end{equation}
Thus, $\phi_{t^{\circ}}$ is $\tfrac{\delta}{t^\circ}$-Lipschitz continuous on $[e_k-t^\circ, e_k]$.
On the other hand, if $s\in (e_k, e_k+t^\circ]$, we have that $q_\delta(s)=e_k+\delta$ and $e_k+\delta\geq \tfrac{\delta}{t^\circ}\cdot s + e_k(1-\tfrac{\delta}{t^\circ})$, implying that
\begin{equation}
	\phi_{t^{\circ}}(s) = e_k+\delta - \big(\tfrac{\delta}{t^\circ}\cdot s + e_k(1-\tfrac{\delta}{t^\circ})\big) = \delta - \tfrac{\delta}{t^\circ}(s-e_k).
\end{equation}
Thus, $\phi_{t^{\circ}}$ is $\tfrac{\delta}{t^\circ}$-Lipschitz continuous on $(e_k, e_k+t^\circ]$.
Moreover, $\lim_{s\to e_k}\phi_{t^{\circ}}(s)=\delta=\phi_{t^{\circ}}(e_k)$, which shows that $\phi_{t^{\circ}}$ is indeed $\tfrac{\delta}{t^\circ}$-Lipschitz continuous on $\mathcal{E}_{k,t^\circ}$. Note that the above calculations also allow us to write
\begin{equation}
	\phi_{t^{\circ}}(s) = 
	\begin{cases}
		\delta - \tfrac{\delta}{t^\circ}\abs{s-e_k}, &\text{if $\exists k\in \Z: s\in [e_k-t^\circ, e_k+t^\circ]$,}\\ 
		0, &\text{otherwise,}
	\end{cases} \quad s \in \R.
\end{equation}
In particular, we have that $0\leq \phi_{t^{\circ}}(s)\leq \delta \cdot \indset{\mathcal{E}_{t^{\circ}}}(s)$, where $\mathcal{E}_{t^{\circ}}\coloneqq\union_{k\in \Z}\mathcal{E}_{k,t^\circ}$.

We now distinguish between the two cases $t\leq 128\delta$ and $t > 128\delta$:

\emph{The case $t\leq 128\delta$.}
Using the above notation, we choose $t^\circ \coloneqq \frac{t}{128}$ and approximate $\ys(\grtr) = q_\delta(\sp{\a}{\grtr} + \tau)$ by $\ys_t(\grtr) \coloneqq \psi_{t^\circ}(\sp{\a}{\grtr} + \tau)$.
The absolute value of the approximation error is then given by $\err_t(\grtr) = \phi_{t^\circ}(\sp{\a}{\grtr} +\nobreak \tau)$.
We now show that for this choice of approximation, the conditions of Assumption~\ref{ass:results:incr} are indeed fulfilled with \eqref{eq:proof:mbit:assincr:tsmall}:

\emph{On Assumption~\ref{ass:results:incr}\ref{ass:results:incr:approx}.}
We first observe that the indicator function $\indset{\mathcal{E}_{t^\circ}}$ is $2\delta$-periodic on $\R$.
Since the random variable $s + \tau$ is uniformly distributed on $[s-\delta, s+\delta]$ for every $s \in \R$, this implies
\begin{equation}\label{eq:multi:indicator}
	\mean_{\tau}[\indset{\mathcal{E}_{t^\circ}}(s + \tau)] = \mean_{\tau}[\indset{\mathcal{E}_{t^\circ}}(\tau)] = \mean_{\tau}[\indset{\mathcal{E}_{0, t^\circ}}(\tau)]=\mean_{\tau}[\indset{[-t^\circ,t^\circ ]}(\tau)] = t^\circ\delta^{-1}.
\end{equation}
Now let $\grtr \in \rset$ and $\pv \in \S^{p-1}$. Using the independence of $\a$ and $\tau$ in conjunction with the inequality \eqref{eq:multi:indicator}, we obtain
\begin{align}
	\mean[\err_t(\grtr) \cdot \abs{\sp{\a}{\pv}}]
	&=\mean[\phi_{t^\circ}(\sp{\a}{\grtr}+\tau) \cdot \abs{\sp{\a}{\pv}}] \leq \delta \cdot \mean[\indset{\mathcal{E}_{t^\circ}}(\sp{\a}{\grtr} + \tau) \cdot \abs{\sp{\a}{\pv}}]\\
	&\leq \delta \cdot \mean_{\a} \big[\abs{\sp{\a}{\pv}} \cdot \mean_{\tau} [\indset{\mathcal{E}_{t^\circ}}(\sp{\a}{\grtr} + \tau)]\big] \leq t^\circ \cdot \mean[\abs{\sp{\a}{\pv}}] \leq \tfrac{t}{64} \cdot \mean[\abs{\sp{\a}{\pv}}].
\end{align}
From Jensen's inequality and the isotropy of $\a$, it follows that
\begin{equation}\label{eq:L1:isotropy}
	\mean[\abs{\sp{\a}{\pv}}]\leq (\mean[\abs{\sp{\a}{\pv}}^2])^{1/2}=\lnorm{\pv}=1,
\end{equation}
which shows that Assumption~\ref{ass:results:incr}\ref{ass:results:incr:approx} is satisfied.

\emph{On Assumption~\ref{ass:results:incr}\ref{ass:results:incr:multipl}.} Since $\psi_{t^\circ}$ is $\tfrac{\delta}{t^\circ}$-Lipschitz, the following holds for every $\grtr, \grtr'\in \rset$:
\begin{align}
	\normsubg{\multipl_t(\grtr) - \multipl_t(\grtr')} 
	&\leq \normsubg{\sp{\a}{\grtr-\grtr'}} + \normsubg{\psi_{t^\circ}(\sp{\a}{\grtr} + \tau) - \psi_{t^\circ}(\sp{\a}{\grtr'} + \tau)}\\
	&\leq (1 + \tfrac{\delta}{t^\circ}) \cdot \normsubg{\sp{\a}{\grtr-\grtr'}} \leq (1 + \tfrac{\delta}{t^\circ}) \cdot \subgparam\cdot \lnorm{\grtr - \grtr'}\\
	&= \subgparam \cdot (1 + \tfrac{128 \delta}{t}) \cdot d_\tfunc(\grtr,\grtr').
\end{align} 
This implies $\multiplLip \lesssim \subgparam \delta t^{-1}$. Furthermore, we clearly have that $\sup_{s\in \R}\abs{s-\psi_{t^\circ}(s + \tau)}\lesssim \delta$, and therefore
\begin{equation}
	\normsubg{\multipl_t(\grtr)}=\normsubg{\sp{\a}{\grtr}-\psi_{t^\circ}(\sp{\a}{\grtr} + \tau)}\lesssim \delta
\end{equation}
for every $\grtr\in \rset$. This shows $\multiplDia\lesssim \delta$.

\emph{On Assumption~\ref{ass:results:incr}\ref{ass:results:incr:err}.} Since $\phi_{t^\circ}$ is $\tfrac{\delta}{t^\circ}$-Lipschitz, the following holds for every $\grtr, \grtr'\in \rset$:
\begin{align}
	\normsubg{\err_t(\grtr) - \err_t(\grtr')}
	&=\normsubg{\phi_{t^\circ}(\sp{\a}{\grtr}+\tau) - \phi_{t^\circ}(\sp{\a}{\grtr'}+\tau) }\\
	&\leq \tfrac{\delta}{t^\circ} \normsubg{\sp{\a}{\grtr - \grtr'}} \leq \subgparam\cdot \tfrac{128\delta}{t} \cdot d_\tfunc(\grtr,\grtr'),
\end{align}
and therefore, $\errLip \lesssim \subgparam \delta t^{-1}$. Finally, since $0\leq \phi_{t^\circ}(s)\leq \delta$ for every $s\in \R$, it follows that $\errDia\lesssim \delta$.

\emph{The case $t> 128\delta$.} Let $\rset_t\subset \rset$ be any maximal $\tfrac{t}{256}$-packing for $\rset$. We show that for the choice $\ys_t(\grtr) \coloneqq \ys(\grtr)$, the conditions of Assumption~\ref{ass:results:incr} are fulfilled on $\rset_t$ with \eqref{eq:proof:mbit:assincr:tlarge}.
Since $\err_t(\grtr) = 0$ for all $\grtr \in \rset_t$, Assumption~\ref{ass:results:incr}\ref{ass:results:incr:approx} and~\ref{ass:results:incr:err} are trivially fulfilled with $\errLip=\errDia=0$. 
Furthermore, observing that $\multipl_t(\grtr)=\sp{\a}{\grtr}-q_\delta(\sp{\a}{\grtr} + \tau)$ and $\sup_{s\in \R} \abs{s - q_\delta(s + \tau)}\leq 2\delta$, we conclude that $\sup_{\grtr\in \rset_t}\normsubg{\multipl_t(\grtr)}\lesssim \delta$. 
Finally, using that $\lnorm{\grtr-\grtr'}> \tfrac{t}{256}$ for all $\grtr, \grtr' \in \rset_t$ with $\grtr\neq \grtr'$, we can bound the sub-Gaussian norm of the multiplier increments as follows:
\begin{equation}
	\normsubg{\multipl_t(\grtr)-\multipl_t(\grtr')} \leq \normsubg{\multipl_t(\grtr)} + \normsubg{\multipl_t(\grtr')} \lesssim \delta< \tfrac{256\delta}{t}\lnorm{\grtr-\grtr'}= \tfrac{256\delta}{t}\cdot d_\tfunc(\grtr,\grtr').
\end{equation}
This shows that Assumption~\ref{ass:results:incr}\ref{ass:results:incr:multipl} is satisfied on $\rset_t$ with $\multiplLip\lesssim \delta t^{-1}$ and $\multiplDia\lesssim \delta$. \qed
\end{tmpl}

\subsection{Proofs for Subsection~\ref{subsec:appl:sim}}
\label{subsec:app:appl:sim}

\paragraph{Proof of Theorem~\ref{thm:intro:sim}.} 

We follow the proof template from the beginning of Section~\ref{sec:app:appl}:
\begin{tmpl}
\item
The model setup of Theorem~\ref{thm:intro:sim} fits into Assumption~\ref{ass:results:meas} as follows:
\begin{itemize}
\item
	We have that $\a \distributed \Normdistr{\vnull}{\I{p}}$ and therefore $\normsubg{\a} \lesssim 1$. The signal set $\rset$ is an arbitrary subset of~$\S^{p-1}$.
	The output function $\fout \colon \R^p \times \rset \to \R$ takes the form $\fout(\a, \grtr) \coloneqq \fobs(\sp{\a}{\grtr})$.
\item
	The target function $\tfunc \colon \rset \to \sset$ corresponds to rescaling by a factor of $\scalfac = \mean[\fobs(\gaussianuniv)\gaussianuniv]$ with $\gaussianuniv \distributed \Normdistr{0}{1}$, i.e., $\tfunc\grtr \coloneqq \scalfac\grtr$. In particular, we have that $d_\tfunc(\grtr,\grtr')= \scalfac\lnorm{\grtr - \grtr'}$.
\end{itemize}

\item
The target mismatch $\mmcovar{\grtr}$ vanishes for every $\grtr\in \rset$.
	
\item
Let $\gaussianuniv \distributed \Normdistr{0}{1}$. For the trivial choice $\ys_t(\grtr)\coloneqq\ys(\grtr)$, the conditions of Assumption~\ref{ass:results:incr} are fulfilled with
\begin{equation} 
	\multiplLip \lesssim 1 + \Lip \scalfac^{-1},\quad \errLip = 0,\quad \multiplDia= \normsubg{\fobs(\gaussianuniv) - \scalfac\gaussianuniv},\quad \errDia = 0.
\end{equation}

\item
Setting $\nout \coloneqq 0$, $\tune \coloneqq 1$, and $\probsuccess_0 \coloneqq \probsuccess$, the claim of Theorem~\ref{thm:intro:sim} follows directly from Theorem~\ref{thm:results:main}.
	
\item
Let $\grtr\in \rset \subset \S^{p-1}$ and consider the orthogonal decomposition of the standard Gaussian random vector $\a$ along $\grtr$:
\begin{equation}
	\a = \sp{\a}{\grtr}\grtr + \proj_{\orthcompl{\grtr}}(\a),
\end{equation}
where $\proj_{\orthcompl{\grtr}} \coloneqq \proj_{\orthcompl{\{\grtr\}}}$.
Since $\proj_{\orthcompl{\grtr}}(\a)$ is centered and $\sp{\a}{\grtr} \distributed \Normdistr{0}{1}$ is independent of $\proj_{\orthcompl{\grtr}}(\a)$, we have that 
\begin{align}
	\mean[\ys(\grtr)\a] 
	&=\mean[\fobs(\sp{\a}{\grtr}) (\sp{\a}{\grtr}\grtr + \proj_{\orthcompl{\grtr}}(\a))]\\
	&=\scalfac\grtr + \mean[\fobs(\sp{\a}{\grtr})] \cdot \mean[\proj_{\orthcompl{\grtr}}(\a)] =\scalfac\grtr = \tfunc\grtr,
\end{align}
which implies that $\mmcovar{\grtr}=0$.

\item
We simply set $\ys_t(\grtr)\coloneqq\ys(\grtr)$. 
Then $\err_t(\grtr) = 0$, implying that Assumption~\ref{ass:results:incr}\ref{ass:results:incr:approx} and~\ref{ass:results:incr:err} are trivially fulfilled with $\errLip=\errDia=0$. 
Furthermore, the following holds for every $\grtr, \grtr'\in \rset$:
\begin{align}
	\normsubg{\multipl_t(\grtr) - \multipl_t(\grtr')} 
	&\leq \scalfac \normsubg{\sp{\a}{\grtr-\grtr'}} + \normsubg{\fobs(\sp{\a}{\grtr}) - \fobs(\sp{\a}{\grtr'})}\\*
	&\lesssim \scalfac \lnorm{\grtr - \grtr'} + \Lip \normsubg{\sp{\a}{\grtr-\grtr'}} \\*
	&\lesssim (\scalfac + \Lip) \cdot \lnorm{\grtr - \grtr'} = (1 + \Lip\scalfac^{-1}) \cdot d_\tfunc(\grtr,\grtr').
\end{align} 
This implies $\multiplLip \lesssim 1 + \Lip\scalfac^{-1}$. 
Since $\sp{\a}{\grtr} \distributed \Normdistr{0}{1}$ for every $\grtr\in \rset \subset \S^{p-1}$, we can also conclude that
\begin{equation}
	\normsubg{\multipl_t(\grtr)}=\normsubg{\scalfac \sp{\a}{\grtr} - \fobs(\sp{\a}{\grtr})}=\normsubg{\fobs(\gaussianuniv) - \scalfac\gaussianuniv}, \quad \gaussianuniv \distributed \Normdistr{0}{1},
\end{equation}
which shows that $\multiplDia \coloneqq \normsubg{\fobs(\gaussianuniv) - \scalfac\gaussianuniv}$ is a valid choice. Hence, Assumption~\ref{ass:results:incr}\ref{ass:results:incr:multipl} is satisfied as well. \qed
\end{tmpl}

\paragraph{Proof of Corollary~\ref{cor:appl:modulo}.} 

We follow the proof template from the beginning of Section~\ref{sec:app:appl}:
\begin{tmpl}
\item
The model setup of Corollary~\ref{cor:appl:modulo} fits into Assumption~\ref{ass:results:meas} as follows:
\begin{itemize}
\item
	We have that $\a \distributed \Normdistr{\vnull}{\I{p}}$ and therefore $\normsubg{\a} \lesssim 1$. The signal set $\rset$ is an arbitrary subset of~$\S^{p-1}$.
	The output function $\fout \colon \R^p \times \rset \to \R$ takes the form $\fout(\a, \grtr) \coloneqq \modulo_\lambda(\sp{\a}{\grtr})$.
\item
	The target function $\tfunc \colon \rset \to \sset$ corresponds to rescaling by a factor of $\scalfac_\lambda = \mean[\modulo_\lambda(\gaussianuniv)\gaussianuniv]$ with $\gaussianuniv \distributed \Normdistr{0}{1}$, i.e., $\tfunc\grtr \coloneqq \scalfac_\lambda\grtr$. In particular, we have that $d_\tfunc(\grtr,\grtr')= \scalfac_\lambda\lnorm{\grtr - \grtr'}$.
\end{itemize}

\item
The target mismatch $\mmcovar{\grtr}$ vanishes for every $\grtr\in \rset$.
	
\item

There exists an approximation $\ys_t(\grtr)$ of the observation variable $\ys(\grtr)$ such that the conditions of Assumption~\ref{ass:results:incr} are fulfilled with
\begin{equation}\label{eq:app:appl:modulo:incr}
	\multiplLip \lesssim \tfrac{\lambda}{t},\quad \errLip\lesssim \tfrac{\lambda}{t},\quad \multiplDia\lesssim 1,\quad \errDia\lesssim 1.
\end{equation}

\item
We begin by showing that if $\lambda\geq C$ for a sufficiently large absolute constant $C>0$, then $\scalfac_\lambda\geq \tfrac{1}{2}$. Indeed, for $\gaussianuniv \distributed \Normdistr{0}{1}$, we have that
\begin{align}
	\scalfac_\lambda = \mean[\modulo_\lambda(\gaussianuniv)\gaussianuniv]  
	&= \mean[\indset{[-\lambda, \lambda)}(\gaussianuniv)\gaussianuniv^2]	
	+ \mean[\indset{\R\backslash[-\lambda, \lambda)}(\gaussianuniv)\modulo_\lambda(\gaussianuniv)\gaussianuniv]\\
	&\geq \mean[\indset{[-\lambda, \lambda)}(\gaussianuniv)\gaussianuniv^2] - \mean[\indset{\R\backslash[-\lambda, \lambda)}(\gaussianuniv)\gaussianuniv^2]\\
	&=\mean[\gaussianuniv^2] - 2\mean[\indset{\R\backslash[-\lambda, \lambda)}(\gaussianuniv)\gaussianuniv^2]\\
	&=1-2\mean[\indset{\R\backslash[-\lambda, \lambda)}(\gaussianuniv)\gaussianuniv^2].
\end{align}
The claim now follows by observing that $\mean[\gaussianuniv^2]<\infty$, which implies $\lim_{\lambda\to \infty}\mean[\indset{\R\backslash[-\lambda, \lambda)}(\gaussianuniv)\gaussianuniv^2]=0$.
Moreover, 
\begin{align}
	\scalfac_\lambda  
	&= \mean[\indset{[-\lambda, \lambda)}(\gaussianuniv)\gaussianuniv^2]	
	+ \mean[\indset{\R\backslash[-\lambda, \lambda)}(\gaussianuniv)\modulo_\lambda(\gaussianuniv)\gaussianuniv]\\
	&\leq \mean[\indset{[-\lambda, \lambda)}(\gaussianuniv)\gaussianuniv^2] + \mean[\indset{\R\backslash[-\lambda, \lambda)}(\gaussianuniv)\lambda \abs{\gaussianuniv}] \leq \mean[\gaussianuniv^2] =1.
\end{align}

Setting $\nout \coloneqq 0$, $\tune \coloneqq 1$, and $\probsuccess_0 \coloneqq \probsuccess$, the claim of Corollary~\ref{cor:appl:modulo} follows directly from Theorem~\ref{thm:results:main}.
	
\item
Let $\grtr\in \rset \subset \S^{p-1}$ and consider the orthogonal decomposition of the standard Gaussian random vector $\a$ along $\grtr$:
\begin{equation}
	\a = \sp{\a}{\grtr}\grtr + \proj_{\orthcompl{\grtr}}(\a),
\end{equation}
where $\proj_{\orthcompl{\grtr}} \coloneqq \proj_{\orthcompl{\{\grtr\}}}$.
Since $\proj_{\orthcompl{\grtr}}(\a)$ is centered and $\sp{\a}{\grtr} \distributed \Normdistr{0}{1}$ is independent of $\proj_{\orthcompl{\grtr}}(\a)$, we have that 
\begin{align}
	\mean[\ys(\grtr)\a] 
	&=\mean[\modulo_\lambda(\sp{\a}{\grtr}) (\sp{\a}{\grtr}\grtr + \proj_{\orthcompl{\grtr}}(\a))]\\
	&=\scalfac_\lambda\grtr + \mean[\modulo_\lambda(\sp{\a}{\grtr})] \cdot \mean[\proj_{\orthcompl{\grtr}}(\a)] =\scalfac_\lambda\grtr = \tfunc\grtr,
\end{align}
which implies that $\mmcovar{\grtr}=0$.

\item

Let $t\in (0, \lambda)$. 
We define the half-open intervals $\mathcal{I}_{k,t}\coloneqq[k\lambda-(\lambda-t), k\lambda+(\lambda-t))$ for $k \in \Z$ even and $\mathcal{I}_{k,t}\coloneqq[k\lambda-t, k\lambda+t)$ for $k \in \Z$ odd. Moreover, we set
\begin{equation}
	\mathcal{I}_{\text{even},t}\coloneqq \bigunion_{\substack{k\in \Z \\ k \text{ even}}}\mathcal{I}_{k,t}, \quad 
	\mathcal{I}_{\text{odd},t}\coloneqq \bigunion_{\substack{k\in \Z \\ k \text{ odd}}}\mathcal{I}_{k,t}
\end{equation}
and observe that the intervals $\mathcal{I}_{k,t}$, $k\in \Z$, form a partition of $\R$. 
Finally, we introduce the functions
\begin{equation}
	g_t(s)\coloneqq -\big(\tfrac{\lambda -t}{t}\big)\cdot s, \qquad s\in \R, 
\end{equation} 
and
\begin{equation}
	\psi_t(s) \coloneqq \begin{cases}
		\modulo_\lambda(s),  &s\in \mathcal{I}_{k,t},\; k \text{ even},\\  
		g_t(s-k\lambda), \quad &s\in \mathcal{I}_{k,t},\; k \text{ odd},
	\end{cases}
\end{equation}
and 
\begin{equation}
	\phi_t(s)\coloneqq \abs{\modulo_\lambda(s)-\psi_t(s)}, \qquad s\in \R.
\end{equation}
Let us show that both $\psi_t$ and $\phi_t$ are $\tfrac{\lambda}{t}$-Lipschitz continuous.
We begin by verifying that $\psi_t$ is continuous. Clearly, $\psi_t$ is continuous on $\mathcal{I}_{k,t}$ for every odd integer $k$.
Let $k$ be an even integer. 
Then $s\in \mathcal{I}_{k,t}=[k\lambda-(\lambda-t), k\lambda+(\lambda-t))$, which implies $\floor[\big]{\tfrac{s+\lambda}{2\lambda}}=\tfrac{k}{2}$. 
Therefore, $\modulo_\lambda(s)=s-k\lambda$ if $s\in \mathcal{I}_{k,t}$ for $k$ an even integer. This shows that $\psi_t$ is continuous on every interval $\mathcal{I}_{k,t}$.
Let us now write $\mathcal{I}_{k,t}=[u_k, v_k)$ for $k\in \Z$. 
To show that $\psi_t$ is continuous, it remains to check that $\lim_{s\uparrow v_k}\psi_t(s)=\psi_t(u_{k+1})$ for every $k\in \Z$. 
For $k$ even, we have that
\begin{equation}
	\lim_{s\uparrow v_k}\psi_t(s) = \lim_{s\uparrow v_k}\modulo_\lambda(s)=\lim_{s\uparrow v_k} s-k\lambda = v_k-k\lambda = \lambda - t.
\end{equation}
Furthermore, it holds that $u_{k+1}=(k+1)\lambda-t\in \mathcal{I}_{k+1,t}$, which implies 
\begin{equation}
	\psi_t(u_{k+1})=g_t(u_{k+1}-(k+1)\lambda)=g_t(-t)=\lambda -t.  
\end{equation}
On the other hand, for $k$ odd, we have that
\begin{equation}
	\lim_{s\uparrow v_k}\psi_t(s)=\lim_{s\uparrow v_k}g_t(s-k\lambda)=\lim_{s\uparrow v_k} -\big(\tfrac{\lambda-t}{t}\big)(s-k\lambda)=-\big(\tfrac{\lambda-t}{t}\big)(v_k-k\lambda)=-(\lambda-t).
\end{equation}
Furthermore, it holds that $u_{k+1}=k\lambda+t\in \mathcal{I}_{k+1,t}$, which implies 
\begin{equation}
	\psi_t(u_{k+1})=\modulo_\lambda(u_{k+1})=u_{k+1}-(k+1)\lambda=-(\lambda-t).
\end{equation}
This shows that $\psi_t$ is continuous. Since $\psi_t$ is continuous and piecewise linear, it is $\max\{1, \tfrac{\lambda-t}{t}\}$-Lipschitz. Therefore, $\psi_t$ is also $\tfrac{\lambda}{t}$-Lipschitz. 

Next, we show that $\phi_t$ is $\tfrac{\lambda}{t}$-Lipschitz as well. First note that 
\begin{equation}
	\phi_t(s) = \abs{\modulo_\lambda(s)-\psi_t(s)}\cdot \indset{\mathcal{I}_{\text{odd},t}}(s),\qquad s\in \R.
\end{equation}
Let $s\in \mathcal{I}_{k,t}=[k\lambda-t, k\lambda + t)$ for $k$ an odd integer. Let us show that  
\begin{equation}
	\abs{\modulo_\lambda(s)-\psi_t(s)}=\lambda - \tfrac{\lambda}{t}\abs{k\lambda -s}. 
\end{equation}
To this end, we distinguish the cases $s\in [k\lambda -t, k\lambda)$ and $s\in [k\lambda, k\lambda+t)$. First, let $s\in [k\lambda -t, k\lambda)$. It follows $\floor[\big]{\tfrac{s+\lambda}{2\lambda}}=\tfrac{k-1}{2}$, which implies $\modulo_\lambda(s)=s-k\lambda+\lambda$. 
Furthermore, $g_t(s-k\lambda)=s-k\lambda+\tfrac{\lambda}{t}(k\lambda-s)$. 
It follows that $\modulo_\lambda(s)\geq g_t(s-k\lambda)$ and 
\begin{equation}
	\abs{\modulo_\lambda(s)-\psi_t(s)} = \modulo_\lambda(s) - g_t(s-k\lambda) = \lambda - \tfrac{\lambda}{t}\abs{k\lambda-s}.
\end{equation}	
The case $s\in [k\lambda, k\lambda+t)$ can be treated analogously. All together, this shows that for any $s\in \R$, we have that
\begin{equation}
	\phi_t(s) = \sum_{\substack{k\in \Z\\ k\text{ odd}}}\indset{\mathcal{I}_{k,t}}(s)\cdot (\lambda - \tfrac{\lambda}{t}\abs{k\lambda -s}).
\end{equation}
In particular, $\phi_t$ is $\tfrac{\lambda}{t}$-Lipschitz.

We approximate the observation variable $\ys(\grtr)=\modulo_\lambda(\sp{\a}{\grtr})$ by $\ys_t(\grtr)=\psi_{t^\circ}(\sp{\a}{\grtr})$ for $t^{\circ}\coloneqq \tfrac{t}{56\cdot 64}$. 
Observe that $t^\circ < \tfrac{\lambda}{2}$.
Let us show that the conditions of Assumption~\ref{ass:results:incr} are indeed fulfilled with \eqref{eq:app:appl:modulo:incr}:

\emph{On Assumption~\ref{ass:results:incr}\ref{ass:results:incr:approx}.}
Let $\grtr\in \rset$ and $\pv\in \S^{p-1}$, and consider the orthogonal decomposition 
\begin{equation}
	\pv = \sp{\pv}{\grtr}\grtr + \proj_{\orthcompl{\grtr}}(\pv).
\end{equation}
This implies
\begin{equation}
	\mean[\err_t(\grtr) \cdot \abs{\sp{\a}{\pv}}]\leq \abs{\sp{\pv}{\grtr}}\cdot \mean[\err_t(\grtr) \cdot\abs{\sp{\a}{\grtr}}] + \mean[\err_t(\grtr) \cdot\abs{\sp{\a}{\proj_{\orthcompl{\grtr}}(\pv)}}].
\end{equation}
Clearly, $\abs{\sp{\pv}{\grtr}}\leq 1$ and $\err_t(\grtr)=\phi_{t^\circ}(\sp{\a}{\grtr})\leq \lambda\cdot \indset{\mathcal{I}_{\text{odd}, t^{\circ}}}(\sp{\a}{\grtr})$, which implies
\begin{equation}
	\mean[\err_t(\grtr) \cdot \abs{\sp{\a}{\pv}}] \leq \underbrace{\lambda \cdot \mean[\indset{\mathcal{I}_{\text{odd}, t^{\circ}}}(\sp{\a}{\grtr})\cdot \abs{\sp{\a}{\grtr}}]}_{=:S_1} {} + {} \underbrace{\lambda\cdot \mean[\indset{\mathcal{I}_{\text{odd}, t^{\circ}}}(\sp{\a}{\grtr})\cdot \abs{\sp{\a}{\proj_{\orthcompl{\grtr}}(\pv)}}]}_{=:S_2}.
\end{equation}
Let us estimate $S_1$ and $S_2$ separately, starting with $S_2$. Since $\sp{\a}{\grtr}$ and $\sp{\a}{\proj_{\orthcompl{\grtr}}(\pv)}$ are independent, we observe that
\begin{equation}
	\mean[\indset{\mathcal{I}_{\text{odd}, t^{\circ}}}(\sp{\a}{\grtr})\cdot \abs{\sp{\a}{\proj_{\orthcompl{\grtr}}(\pv)}}] = \mean[\indset{\mathcal{I}_{\text{odd}, t^{\circ}}}(\sp{\a}{\grtr})]\cdot 
	\mean[\abs{\sp{\a}{\proj_{\orthcompl{\grtr}}(\pv)}}].
\end{equation}
Using Jensen's inequality, we obtain  
\begin{equation}
	\mean[\abs{\sp{\a}{\proj_{\orthcompl{\grtr}}(\pv)}}]\leq \Big(\mean[\abs{\sp{\a}{\proj_{\orthcompl{\grtr}}(\pv)}}^2]\Big)^{1/2}=\lnorm{\proj_{\orthcompl{\grtr}}(\pv)}\leq \lnorm{\pv}=1.
\end{equation}
Therefore, it follows that $S_2 \leq \lambda \cdot \prob(\sp{\a}{\grtr}\in \mathcal{I}_{\text{odd},t^\circ})$.
Since $\sp{\a}{\grtr}\sim \Normdistr{0}{1}$, it holds that
\begin{equation}
	\tfrac{1}{2}\prob(\sp{\a}{\grtr}\in \mathcal{I}_{\text{odd},t^\circ}) = \tfrac{1}{\sqrt{2\pi}}
	\sum_{\substack{k\in \N\\ k\text{ odd}}} \int_{k\lambda-t^\circ}^{k\lambda+t^\circ}e^{-x^2/2}\, dx \leq \tfrac{2t^\circ}{\sqrt{2\pi}}
	\sum_{\substack{k\in \N\\ k\text{ odd}}} e^{-(k\lambda-t^\circ)^2/2}.	
\end{equation}
For every $k\in \N$, 
\begin{equation}
	2\lambda e^{-((k+2)\lambda-t^\circ)^2/2}\leq \int_{k\lambda-t^\circ}^{(k+2)\lambda-t^\circ}e^{-x^2/2}\, dx.
\end{equation}
Therefore, 
\begin{align}
	\sum_{\substack{k\in \N\\ k\text{ odd}}} e^{-(k\lambda-t^\circ)^2/2}
	&= e^{-(\lambda-t^\circ)^2/2} + \sum_{\substack{k\in \N\\ k\text{ odd}}} e^{-((k+2)\lambda-t^\circ)^2/2}\\
	&\leq \tfrac{1}{\lambda-t^\circ}\int_{0}^{\lambda-t^\circ}e^{-x^2/2}\, dx + \tfrac{1}{2\lambda} \sum_{\substack{k\in \N\\ k\text{ odd}}}\int_{k\lambda-t^\circ}^{(k+2)\lambda-t^\circ}e^{-x^2/2}\, dx\\
	&\leq \tfrac{1}{\lambda-t^\circ}\int_{0}^{\infty}e^{-x^2/2}\,dx.
\end{align}
Using the identity $\tfrac{1}{\sqrt{2\pi}}\int_{0}^{\infty}e^{-x^2/2}\,dx=\tfrac{1}{2}$, we obtain  
\begin{align}
	S_2 \leq \tfrac{2 t^{\circ}\lambda}{\lambda-t^{\circ}}\leq 4t^{\circ},
\end{align}
where the last inequality follows from $t^\circ<\tfrac{\lambda}{2}$. 

Let us now estimate $S_1$.
Since $\sp{\a}{\grtr}\sim \Normdistr{0}{1}$, we have that
\begin{equation}
	\mean[\indset{\mathcal{I}_{\text{odd}, t^{\circ}}}(\sp{\a}{\grtr})\cdot \abs{\sp{\a}{\grtr}}] = 
	\tfrac{1}{\sqrt{2\pi}}
	\sum_{\substack{k\in \N\\ k\text{ odd}}} \int_{k\lambda-t^\circ}^{k\lambda+t^\circ}xe^{-x^2/2}\, dx.
\end{equation}
It holds that
\begin{align}
	&\sum_{\substack{k\in \N\\ k\text{ odd}}} \int_{k\lambda-t^\circ}^{k\lambda+t^\circ}xe^{-x^2/2}\, dx\\
	&\leq 2t^\circ\sum_{\substack{k\in \N\\ k\text{ odd}}} (k\lambda+t^\circ)e^{-(k\lambda-t^\circ)^2/2}\\
	&= 2t^\circ\Big((\lambda+t^\circ)\cdot e^{-(\lambda-t^\circ)^2/2} + 
	\sum_{\substack{k\in \N\\ k\text{ odd}}}((k+2)\lambda+t^\circ)e^{-((k+2)\lambda-t^\circ)^2/2}\, dx\Big)\\&\leq 2t^\circ\Big( \tfrac{1}{\lambda-t^\circ}\int_0^{\lambda-t^\circ}(x+\lambda+t^\circ)\cdot e^{-x^2/2}\, dx + 
	\tfrac{1}{2\lambda}\sum_{\substack{k\in \N\\ k\text{ odd}}}\int_{k\lambda-t^\circ}^{(k+2)\lambda-t^\circ}(x+2\lambda+2t^\circ)\cdot e^{-x^2/2}\, dx\Big)\\
&\leq  \tfrac{4t^\circ}{\lambda}\int_0^\infty (x+3\lambda)\cdot e^{-x^2/2}\, dx\\
&=\tfrac{4t^\circ}{\lambda}\cdot (1+3\lambda\sqrt{\tfrac{\pi}{2}}).
\end{align}
Therefore, if $\lambda\leq 2$, then 
\begin{equation}
	\sum_{\substack{k\in \N\\ k\text{ odd}}} \int_{k\lambda-t^\circ}^{k\lambda+t^\circ}xe^{-x^2/2}\, dx\leq 
	\tfrac{4t^\circ}{\lambda}\cdot (1+6\sqrt{\tfrac{\pi}{2}})\leq \tfrac{52t^\circ}{\lambda}.
\end{equation}
If $\lambda>2$, then $\lambda-t^\circ>1$. Since the function $x\mapsto x\exp(-x^2/2)$ is monotonically decreasing for $x\in [1,\infty)$, we obtain 
\begin{align}
	&\sum_{\substack{k\in \N\\ k\text{ odd}}} \int_{k\lambda-t^\circ}^{k\lambda+t^\circ}xe^{-x^2/2}\, dx\\*
	&\leq 2t^\circ\cdot\sum_{\substack{k\in \N\\ k\text{ odd}}} (k\lambda-t^\circ)\cdot e^{-(k\lambda-t^\circ)^2/2}\\*
	&=2t^\circ\cdot\Big( (\lambda-t^\circ)\cdot \exp(-(\lambda-t^\circ)^2/2)+\sum_{\substack{k\in \N\\ k\text{ odd}}} ((k+2)\lambda-t^\circ)\cdot e^{-((k+2)\lambda-t^\circ)^2/2}\Big).
\end{align}
Since $\lambda-t^\circ>1$, it follows $(\lambda-t^\circ)\cdot \exp(-(\lambda-t^\circ)^2/2)\leq \tfrac{1}{\lambda-t^\circ}$. Furthermore, 
\begin{equation}
	\sum_{\substack{k\in \N\\ k\text{ odd}}} ((k+2)\lambda-t^\circ)\cdot e^{-((k+2)\lambda-t^\circ)^2/2}
	\leq \tfrac{1}{2\lambda}\sum_{\substack{k\in \N\\ k\text{ odd}}} \int_{k\lambda - t^\circ}^{(k+2)\lambda-t^\circ} xe^{-x^2/2}\,dx \leq \tfrac{1}{2\lambda}.
\end{equation}
Therefore, if $\lambda>2$, then 
\begin{equation}
	\sum_{\substack{k\in \N\\ k\text{ odd}}} \int_{k\lambda-t^\circ}^{k\lambda+t^\circ}xe^{-x^2/2}\, dx\leq \tfrac{8t^\circ}{\lambda}.
\end{equation}
This shows $S_1\leq 52 t^\circ$.
Putting everything together, we obtain
\begin{equation}
	\mean[\err_t(\grtr) \cdot \abs{\sp{\a}{\pv}}] \leq S_1 + S_2\leq 56t^\circ=\tfrac{t}{64},
\end{equation}
which implies that Assumption~\ref{ass:results:incr}\ref{ass:results:incr:approx} is satisfied.

\emph{On Assumption~\ref{ass:results:incr}\ref{ass:results:incr:multipl}.}
For $\grtr, \grtr'\in \rset$, we have that
\begin{align}
	\normsubg{\multipl_t(\grtr) - \multipl_t(\grtr')}&\leq \scalfac_\lambda\cdot\normsubg{\sp{\a}{\grtr-\grtr'}}+\normsubg{\psi_{t^\circ}(\sp{\a}{\grtr})-\psi_{t^\circ}(\sp{\a}{\grtr'})}\\
	&\lesssim \scalfac_\lambda\cdot \lnorm{\grtr-\grtr'} + \tfrac{\lambda}{t^\circ} \cdot \lnorm{\grtr-\grtr'},
\end{align}
where we have used that $\psi_{t^\circ}$ is $\tfrac{\lambda}{t^\circ}$-Lipschitz. Since $\scalfac_\lambda\geq \tfrac{1}{2}$, $t\leq \lambda$, and $t^\circ=\tfrac{t}{56\cdot 64}$, it follows that
\begin{equation}
\normsubg{\multipl_t(\grtr) - \multipl_t(\grtr')}\lesssim \tfrac{\lambda}{t}\cdot d_\tfunc(\grtr,\grtr').
\end{equation}
Let $\grtr\in \rset\subset\S^{p-1}$. Since $\scalfac_\lambda\leq 1$ and $\abs{\psi_{t^\circ}(s)}\leq \abs{s}$ for all $s\in \R$, we observe that
\begin{equation}
\normsubg{\multipl_t(\grtr)}\leq \scalfac_\lambda\cdot \normsubg{\sp{\a}{\grtr}} + \normsubg{\psi_{t^\circ}(\sp{\a}{\grtr})}\lesssim 1.
\end{equation}
Therefore, Assumption~\ref{ass:results:incr}\ref{ass:results:incr:multipl} is satisfied with $\multiplLip\lesssim \tfrac{\lambda}{t}$ and $\multiplDia\lesssim 1$.

\emph{On Assumption~\ref{ass:results:incr}\ref{ass:results:incr:err}.}
For $\grtr, \grtr'\in \rset$, 
\begin{align}
\normsubg{\err_t(\grtr)-\err_t(\grtr')}=\normsubg{\phi_{t^\circ}(\sp{\a}{\grtr}) - \phi_{t^\circ}(\sp{\a}{\grtr'})}\lesssim \tfrac{\lambda}{t^\circ}\cdot \lnorm{\grtr-\grtr'},
\end{align}
where we have used that $\phi_{t^\circ}$ is $\tfrac{\lambda}{t^\circ}$-Lipschitz. Since $\scalfac_\lambda\geq \tfrac{1}{2}$ and $t^\circ=\tfrac{t}{56\cdot 64}$, it follows that
\begin{equation}
\normsubg{\err_t(\grtr)-\err_t(\grtr')}\lesssim \tfrac{\lambda}{t}\cdot d_\tfunc(\grtr,\grtr').
\end{equation}
Let $\grtr\in \rset\subset\S^{p-1}$. Since $\abs{\phi_{t^\circ}(s)}\leq \abs{s}$ for all $s\in \R$, we observe that
\begin{equation}
\normsubg{\err_t(\grtr)}=\normsubg{\phi_{t^\circ}(\sp{\a}{\grtr})}\lesssim 1.
\end{equation}
Therefore, Assumption~\ref{ass:results:incr}\ref{ass:results:incr:err} is satisfied with $\errLip\lesssim \tfrac{\lambda}{t}$ and $\errDia\lesssim 1$. \qed
\end{tmpl}

\paragraph{Proof of Corollary~\ref{cor:appl:beyond}.} 

We follow the proof template from the beginning of Section~\ref{sec:app:appl}:
\begin{tmpl}
\item
The model setup of Corollary~\ref{cor:appl:beyond} fits into Assumption~\ref{ass:results:meas} as follows:
\begin{itemize}
\item
	The measurement vector $\a \in \R^p$ is centered, isotropic, and it has independent, symmetric, and sub-Gaussian coordinates with $\max_{j \in [p]}\normsubg{a_j} \leq \subgparam$; in particular, we have that $\normsubg{\a}\lesssim \subgparam$ (e.g., see~\cite[Lem.~3.4.2]{ver18}). The signal set $\rset$ is a subset of~$\rad\ball[2][p]$.
	The output function $\fout \colon \R^p \times \rset \to \R$ takes the form $\fout(\a, \grtr) \coloneqq \sumf(\a\circ\grtr)$, where $\sumf:\R^p\to \R$ is given by $\sumf(\vec{z})\coloneqq \sum_{j=1}^p f_j(z_j)$, with $f_j:\R\to \R$ odd, $\gamma$-Lipschitz, and satisfying the conditions \eqref{eq:appl:beyond:growth1} and \eqref{eq:appl:beyond:growth2}.
\item
	The target function $\tfunc \colon \rset \to \R^p$ is defined by $\tfunc\grtr\coloneqq \mean[\sumf(\a\circ\grtr)\a]$.
\end{itemize}

\item
The target mismatch $\mmcovar{\grtr}$ vanishes for every $\grtr\in \rset$.
	
\item
For the trivial choice $\ys_t(\grtr)\coloneqq\ys(\grtr)$, the conditions of Assumption~\ref{ass:results:incr} are fulfilled with
\begin{equation} 
	\multiplLip \lesssim \subgparam \cdot \tfrac{\gamma}{\alpha},\quad \errLip = 0,\quad \multiplDia\lesssim \subgparam(\beta_2+\gamma)\rad,\quad \errDia = 0.
\end{equation}

\item
We begin by showing that
\begin{equation}
(\tfunc \grtr)_j = \mean[f_j(a_j x_j)a_j], \qquad \text{for all $j\in [p]$,}
\end{equation}
where $a_j$ is the $j$-th coordinate of the measurement vector $\a\in \R^p$.
Indeed, we have that
\begin{align}
	(\tfunc \grtr)_j = (\mean[\sumf(\a\circ\grtr)\a])_j = \mean[\sumf(\a\circ\grtr)a_j]
	= \sum_{j'=1}^p \mean[f_{j'}(a_{j'}x_{j'})a_j] = \mean[f_j(a_jx_j)a_j],
\end{align}
where for the last equality we have used the independence of the random variables $a_1, \ldots, a_p$
and $\mean[a_j]=0$. 

Next, let us show that $(\tfunc \grtr)_j\in [\beta_1x_j, \beta_2x_j]$ for $x_j\geq 0$ and $(\tfunc \grtr)_j\in [\beta_2x_j, \beta_1x_j]$ for $x_j<0$. Let us first assume that $x_j\geq 0$. 
Then clearly $\mean[f_j(a_jx_j)a_j\indset{[0,\infty)}(a_j)]\leq \beta_2 x_j \mean[a_j^2\indset{[0,\infty)}(a_j)]$. 
Furthermore, 
\begin{align}
	\mean[f_j(a_jx_j)a_j\indset{(-\infty,0)}(a_j)] = \mean[f_j(-a_jx_j)(-a_j)\indset{(-\infty,0)}(a_j)]
	\leq \beta_2x_j \mean[a_j^2\indset{(-\infty,0)}(a_j)],
\end{align}
where we have used that $f_j$ is an odd function for the first equality. 
Therefore, 
\begin{equation}
	\mean[f_j(a_jx_j)a_j] = \mean[f_j(a_jx_j)a_j\indset{[0,\infty)}(a_j)] + \mean[f_j(a_jx_j)a_j\indset{(-\infty,0)}(a_j)]
	\leq \beta_2x_j \mean[a_j^2]
	= \beta_2x_j,
\end{equation}
where $\mean[a_j^2]=1$ follows from the assumption that $\a$ is isotropic.
Similarly, we observe that
\begin{align}
	\mean[f_j(a_jx_j)a_j\indset{[0,\infty)}(a_j)] \geq \beta_1x_j \mean[a_j^2\indset{[0,\infty)}(a_j)] 
\end{align}
and 
\begin{align}
	\mean[f_j(a_jx_j)a_j\indset{(-\infty,0)}(a_j)] = \mean[f_j(-a_jx_j)(-a_j)\indset{(-\infty,0)}(a_j)]
	\geq \beta_1x_j\mean[a_j^2\indset{(-\infty,0)}(a_j)],
\end{align}
which implies
\begin{equation}
	\mean[f_j(a_jx_j)a_j]\geq \beta_1x_j\mean[a_j^2]=\beta_1x_j.
\end{equation}
This shows that $\beta_1x_j\leq (\tfunc \grtr)_j\leq \beta_2x_j$ for $x_j\geq 0$. Analogously, one can show that $\beta_2x_j\leq (\tfunc \grtr)_j\leq\beta_1x_j$ for $x_j<0$. 
Therefore, $\abs{(\tfunc\grtr)_j}\leq\beta_2\abs{x_j}$, which implies $\lnorm{\tfunc\grtr}\leq \beta_2\lnorm{\grtr}$ for every $\grtr\in \rset$.
As a consequence, it holds that $\tfunc \rset\subset\beta_2\rad\ball[2][p]$.

Setting $\nout \coloneqq 0$, $\tune \coloneqq 1$, and $\probsuccess_0 \coloneqq \probsuccess$, the claim of Corollary~\ref{cor:appl:beyond} follows directly from Theorem~\ref{thm:results:main}.
	
\item

By definition of the target function it follows that $\mmcovar{\grtr}=0$ for every $\grtr\in \rset$.

\item
We simply set $\ys_t(\grtr)\coloneqq\ys(\grtr)$. 
Then $\err_t(\grtr) = 0$, implying that Assumption~\ref{ass:results:incr}\ref{ass:results:incr:approx} and~\ref{ass:results:incr:err} are trivially fulfilled with $\errLip=\errDia=0$. 
Since $\multipl_t(\grtr)=\sp{\a}{\tfunc\grtr}-\sumf(\a\circ\grtr)$, the following holds for every $\grtr, \grtr'\in \rset$:
\begin{align}
	\normsubg{\multipl_t(\grtr) - \multipl_t(\grtr')} 
	&\leq \normsubg{\sp{\a}{\tfunc\grtr-\tfunc\grtr'}} + \normsubg{\sumf(\a\circ\grtr)-\sumf(\a\circ\grtr')}.
\end{align} 
Clearly, we have that $\normsubg{\sp{\a}{\tfunc\grtr-\tfunc\grtr'}}\lesssim \subgparam\cdot d_\tfunc(\grtr,\grtr')$ and
\begin{equation}
	\normsubg{\sumf(\a\circ\grtr)-\sumf(\a\circ\grtr')}
	= \normsubg{\sum_{j=1}^pf_j(a_jx_j)-f_j(a_jx_j')}.
\end{equation}
Let $j\in [p]$. Using that $f_j$ is $\gamma$-Lipschitz, we obtain
\begin{equation}
	\normsubg{f_j(a_jx_j)-f_j(a_jx_j')}
	\lesssim \gamma\cdot \normsubg{a_jx_j-a_jx_j'}
	\leq \subgparam\cdot \gamma\cdot \abs{x_j-x_j'}.
\end{equation}
Furthermore, since $a_j$ is a symmetric random variable and $f_j$ is odd, it holds $\mean[f_j(a_jx)]=0$ for every $x\in \R$. In particular, the random variable $f_j(a_jx_j)-f_j(a_jx_j')$ is centered.
Since the random variables $a_1,\ldots, a_p$ are independent, Hoeffding's inequality yields
\begin{equation}
	\normsubg{\sum_{j=1}^p f_j(a_jx_j)-f_j(a_jx_j') }^2 \lesssim \sum_{j=1}^p \normsubg{f_j(a_jx_j)-f_j(a_jx_j') }^2
	\lesssim \subgparam^2 \cdot \gamma^2 \cdot \lnorm[\big]{\grtr-\grtr'}^2.
\end{equation}
Therefore, 
\begin{align}
	\normsubg{\multipl_t(\grtr) - \multipl_t(\grtr')} 
	&\lesssim \subgparam\cdot d_\tfunc(\grtr,\grtr') + \subgparam \cdot \gamma \cdot \lnorm{\grtr-\grtr'}.
\end{align} 
We now show that $d_\tfunc(\grtr,\grtr')\geq \alpha \lnorm{\grtr-\grtr'}$. For $j\in [p]$, we have that
\begin{align}
	\abs{(\tfunc\grtr)_j - (\tfunc\grtr')_j} &= \abs{\mean[(f_j(a_jx_j)-f_j(a_jx_j'))a_j]}\\
	&= \abs{\mean[(f_j(a_jx_j)-f_j(a_jx_j'))a_j\indset{[0,\infty)}(a_j)]
	+ \mean[(f_j(a_jx_j)-f_j(a_jx_j'))a_j\indset{(-\infty,0)}(a_j)]}.
\end{align}
We may assume that $x_j\geq x_j'$. 
If $a_j\geq 0$, then $a_jx_j\geq a_jx_j'$, which implies $f_j(a_jx_j)-f_j(a_jx_j')\geq \alpha(a_jx_j-a_jx_j')$. 
Therefore, $(f_j(a_jx_j)-f_j(a_jx_j'))a_j\geq \alpha(a_jx_j-a_jx_j')a_j$, which implies 
\begin{align}
	\mean[(f_j(a_jx_j)-f_j(a_jx_j'))a_j\indset{[0,\infty)}(a_j)]&\geq \mean[\alpha(a_jx_j-a_jx_j')a_j\indset{[0,\infty)}(a_j)]\\
	&=\alpha(x_j-x_j')\mean[a_j^2\indset{[0,\infty)}(a_j)] \geq 0.
\end{align}
If $a_j<0$, then $a_jx_j\leq a_jx_j'$, which implies $f_j(a_jx_j')-f_j(a_jx_j)\geq \alpha(a_jx_j'-a_jx_j)$.
Therefore, $f_j(a_jx_j)-f_j(a_jx_j')\leq \alpha(a_jx_j-a_jx_j')$, which implies $(f_j(a_jx_j)-f_j(a_jx_j'))a_j\geq \alpha(a_jx_j-a_jx_j')a_j$. This shows
\begin{align}
	\mean[(f_j(a_jx_j)-f_j(a_jx_j'))a_j\indset{(-\infty,0)}(a_j)]&\geq \mean[\alpha(a_jx_j-a_jx_j')a_j\indset{(-\infty,0)}(a_j)]\\
	&=\alpha(x_j-x_j')\mean[a_j^2\indset{(-\infty,0)}(a_j)] \geq 0.
\end{align}
It follows that $\abs{(\tfunc\grtr)_j - (\tfunc\grtr')_j}\geq \alpha(x_j-x_j')=\alpha\abs{x_j-x_j'}$, and therefore $d_\tfunc(\grtr,\grtr')\geq \alpha \lnorm{\grtr-\grtr'}$. 
Due to this inequality, we obtain
\begin{align}
	\normsubg{\multipl_t(\grtr) - \multipl_t(\grtr')} 
	&\lesssim \subgparam\cdot d_\tfunc(\grtr,\grtr') + \subgparam \cdot \gamma \cdot \lnorm{\grtr-\grtr'}\\
	&\leq \subgparam\cdot d_\tfunc(\grtr,\grtr') + \subgparam\cdot\tfrac{\gamma}{\alpha}\cdot d_\tfunc(\grtr,\grtr')\\
	&\leq 2\cdot\subgparam\cdot\tfrac{\gamma}{\alpha}\cdot d_\tfunc(\grtr,\grtr'),
\end{align}
where the last inequality follows from $\alpha\leq \gamma$. 

Now let $\grtr\in \rset$. It holds that
\begin{align}
	\normsubg{\multipl_t(\grtr)}&=\normsubg{\sp{\a}{\tfunc\grtr}-\sumf(\a\circ\grtr)} \leq \normsubg{\sp{\a}{\tfunc\grtr}} + \normsubg{\sumf(\a\circ\grtr)}\\*
	&\lesssim \subgparam\cdot \lnorm{\tfunc\grtr} + \normsubg{\sumf(\a\circ\grtr)} \leq \subgparam\cdot \beta_2\cdot \rad + \normsubg{\sumf(\a\circ\grtr)}.
\end{align}
Since $f_j$ is $\gamma$-Lipschitz and odd for every $j\in [p]$, we obtain
\begin{align}
	\normsubg{f_j(a_jx_j)}=\normsubg{f_j(a_jx_j)-f_j(0)}\lesssim \gamma\cdot \normsubg{a_jx_j}\leq \subgparam\cdot \gamma\cdot \abs{x_j}.
\end{align}
Using Hoeffding's inequality, we conclude that  
\begin{equation}
	\normsubg{\sumf(\a\circ\grtr)}=\normsubg{\sum_{j=1}^p f_j(a_jx_j)}
	\lesssim \Big( \sum_{j=1}^p\normsubg{f_j(a_jx_j)}^2\Big)^{1/2}
	\leq \subgparam\cdot \gamma\cdot \lnorm{\grtr}.
\end{equation}
Hence, Assumption~\ref{ass:results:incr}\ref{ass:results:incr:multipl} is satisfied 
with $\multiplLip\lesssim \subgparam \cdot \tfrac{\gamma}{\alpha}$ and $\multiplDia\lesssim \subgparam(\beta_2+\gamma)\rad$. \qed
\end{tmpl}

\subsection{Proofs for Subsection~\ref{subsec:appl:vs}}
\label{subsec:app:appl:vs}

\paragraph{Proof of Corollary~\ref{cor:appl:vs}.}

We follow the proof template from the beginning of Section~\ref{sec:app:appl}:
\begin{tmpl}
\item
The model setup of Corollary~\ref{cor:appl:vs} fits into Assumption~\ref{ass:results:meas} as follows:
\begin{itemize}
\item
	The measurement vector $\a \in \R^p$ is centered, isotropic, and has independent sub-Gaussian coordinates with $\max_{j \in [p]}\normsubg{a_j} \leq \subgparam$; in particular, we have that $\normsubg{\a}\lesssim \subgparam$ (e.g., see~\cite[Lem.~3.4.2]{ver18}). The signal set $\rset$ is an arbitrary subset of  $\{\suppset\subset [p] \suchthat \cardinality{\suppset}\leq s\}$.
	The output function $\fout \colon \R^p \times \rset \to \R$ takes the form $\fout(\a, \suppset) \coloneqq \fobs(\a_{\suppset})$, where $(\a_{\suppset})_j = a_j$ for $j\in \suppset$ and $(\a_{\suppset})_j =0$ for $j\in \setcompl{\suppset}$.
\item
	The target function $\tfunc \colon \rset \to \sset$ is defined by $\tfunc \suppset \coloneqq \mean[\fobs(\a_{\suppset})\a]$.
\end{itemize}

\item
The target mismatch $\mmcovar{\suppset}$ vanishes for every $\suppset\in \rset$.

\item
For the trivial choice $\ys_t(\suppset)\coloneqq\ys(\suppset)$, the conditions of Assumption~\ref{ass:results:incr} are fulfilled with
\begin{equation} 
	\multiplLip \leq \subgparam + \Lip \alpha^{-1},\quad \errLip=0,\quad \multiplDia\leq \subgparam \beta + \srradius,\quad \errDia =0.
\end{equation} 

\item
Setting $\nout \coloneqq 0$, $\tune \coloneqq \subgparam^{-1} \sqrt{\log\subgparam}$, and $\probsuccess \coloneqq \probsuccess_0$, the claim of Corollary~\ref{cor:appl:vs} follows directly from Theorem~\ref{thm:results:main}.

\item
This is clear by the definition of the target function.

\item
We simply set $\ys_t(\suppset)\coloneqq\ys(\suppset)$. Then $\err_t(\suppset) = 0$, implying that Assumption~\ref{ass:results:incr}\ref{ass:results:incr:approx} and~\ref{ass:results:incr:err} are trivially fulfilled with $\errLip=\errDia=0$. To see that Assumption~\ref{ass:results:incr}\ref{ass:results:incr:multipl} holds as well, we first recall that the coordinates of $\a$ are centered and independent, so that
\begin{equation}
	(\tfunc \suppset)_j=\mean[\fobs(\a_{\suppset})a_j]= \mean[\fobs(\a_{\suppset})]\cdot \mean[a_j]=0 \qquad 
	\text{for all $\suppset \in \rset$ and $j\in \setcompl{\suppset}$.}
\end{equation}
Together with the lower bound in \eqref{eq:appl:vs:target}, it follows that $\supp(\tfunc \suppset) = \suppset$ for all $\suppset \in \rset$. Using the lower bound in \eqref{eq:appl:vs:target} again, we obtain the following estimate for all $\suppset, \suppset'\in \rset$:
\begin{equation}\label{eq:sr:distance}
	d_\tfunc(\suppset,\suppset')= \lnorm{\tfunc \suppset-\tfunc \suppset'}\geq
	\Big(\sum_{j\in \suppset\setminus \suppset'} (\tfunc \suppset)_j^2 + \sum_{j\in \suppset'\setminus \suppset} (\tfunc \suppset')_j^2 \Big)^{1/2}\geq 
	\tfrac{\alpha}{\sqrt{s}}\sqrt{\cardinality{\suppset\bigtriangleup \suppset'}}.
\end{equation}
Combining this with the assumption \eqref{eq:appl:vs:subg}, it follows that
\begin{align}
	\normsubg{\multipl_t(\suppset) - \multipl_t(\suppset')} 
	&\leq \normsubg{\sp{\a}{\tfunc \suppset- \tfunc \suppset'}} + \normsubg{\fobs(\a_{\suppset}) - \fobs(\a_{\suppset'})}\\
	&\leq \subgparam \cdot d_\tfunc(\suppset,\suppset') + \Lip \alpha^{-1} \cdot d_\tfunc(\suppset,\suppset') = (\subgparam + \Lip \alpha^{-1}) \cdot d_\tfunc(\suppset,\suppset').
\end{align} 
This implies $\multiplLip \leq \subgparam + \Lip \alpha^{-1}$. 
Since $\supp(\tfunc \suppset) = \suppset$ and $\cardinality{\suppset}\leq s$ for $\suppset\in \rset$, we also have that $\cardinality{\supp(\tfunc \suppset)}\leq s$. The upper bound in \eqref{eq:appl:vs:target} therefore yields $\lnorm{\tfunc \suppset}\leq \beta$ for every $\suppset\in \rset$. Combining this estimate with \eqref{eq:appl:vs:subg}, we obtain the following upper bound for the sub-Gaussian norm of $\multipl_t(\suppset)$:
\begin{equation}
	\normsubg{\multipl_t(\suppset)}=\normsubg{\sp{\a}{\tfunc \suppset} - \fobs(\a_{\suppset}) }
	\leq \subgparam\lnorm{\tfunc \suppset} + \normsubg{\fobs(\a_{\suppset})}
	\leq \subgparam \beta + \srradius.
\end{equation}
Hence, $\multiplDia \coloneqq \subgparam \beta + \srradius$ is a valid choice for Assumption~\ref{ass:results:incr}\ref{ass:results:incr:multipl}. \qed
\end{tmpl}

\section{Proofs for Section~\ref{sec:appl:noincr}}
\label{sec:app:appl:noincr}

\paragraph{Proof of Theorem~\ref{thm:appl:without_increments}.}

Let $\rset_\eps$ be a minimal subset of $\rset$ such that $\tfunc\rset_\eps$ is a minimal $\eps$-net for $\tfunc\rset$; in particular, we have that $\cardinality{\rset_\eps} = \cardinality{\tfunc\rset_\eps}=\covnumber[\eps]{\tfunc\rset}$.
The assumptions of \eqref{eq:appl:without_increments:m} and \eqref{eq:appl:without_increments:u} allow us to first apply Theorem~\ref{thm:results:main} to every $\grtr' \in \rset_\eps$ as a singleton signal set (Assumption~\ref{ass:results:incr} is trivially fulfilled here), and then to take a union bound over $\rset_\eps$. 
Consequently, there exist universal constants $c, C > 0$ such that the following event $\mathcal{A}_1$ holds with probability at least $1 - \exp(-c \probsuccess^2) - \exp(-c \probsuccess_0^2)$: For every $\grtr' \in \rset_\eps$ with $\mmcovar{\grtr'} \leq \tfrac{t}{32}$ and 
every $\Y \in \R^m$ such that
\begin{equation}
	\tfrac{1}{\sqrt{m}}\norm{\Y - \Ys(\grtr')}_{[2\nout]}\leq \tune t \quad \text{ and } \quad 
	\tfrac{1}{\sqrt{m}}\sigma_{2\nout}(\Y - \Ys(\grtr'))_{2}\leq \tfrac{t}{20},
\end{equation} 
any minimizer $\pvsolu$ of \eqref{eq:results:klasso} satisfies $\lnorm{\pvsolu - \tfunc\grtr'} \leq t$.
Note that we have applied the robustness conditions of \eqref{eq:results:main:rob} for $2\nout$ instead of $\nout$ here, which is possible due to $2\nout \leq m$.

Let $\mathcal{A}_2$ denote the event of \eqref{eq:ass:local_robustness:event} in Assumption~\ref{ass:local_robustness}. We now show that on the event $\mathcal{A}_1\intersec\nobreak\mathcal{A}_2$, which occurs with probability at least $1 - \exp(-c \probsuccess^2) - \exp(-c \probsuccess_0^2)-\eta$, the conclusion of Theorem~\ref{thm:appl:without_increments} holds:
Let $\grtr\in \rset$ be fixed and consider any input vector $\Y \in \R^m$ satisfying
\begin{equation}
	\tfrac{1}{\sqrt{m}}\norm{\Y - \Ys(\grtr)}_{[2\nout]}\leq \tfrac{1}{2}\tune t \quad \text{ and } \quad 
	\tfrac{1}{\sqrt{m}}\sigma_{\nout}(\Y - \Ys(\grtr))_{2}\leq \tfrac{t}{40}.
\end{equation}
By the definition of $\rset_\eps$, there exists $\grtr'\in \rset_\eps$ such that $\lnorm{\tfunc\grtr-\tfunc\grtr'}\leq \eps$. According to the event $\mathcal{A}_2$, we conclude that
\begin{equation}
	\tfrac{1}{\sqrt{m}}\norm{\Y - \Ys(\grtr')}_{[2\nout]}\leq 
	\tfrac{1}{\sqrt{m}}\norm{\Y - \Ys(\grtr)}_{[2\nout]} + \tfrac{1}{\sqrt{m}}\norm{\Ys(\grtr) - \Ys(\grtr')}_{[2\nout]}
	\leq \tune t
\end{equation}
and 
\begin{equation}
	\tfrac{1}{\sqrt{m}}\sigma_{2\nout}(\Y - \Ys(\grtr'))_{2}\leq 
	\tfrac{1}{\sqrt{m}}\sigma_{\nout}(\Y - \Ys(\grtr))_{2} + \tfrac{1}{\sqrt{m}}\sigma_{\nout}(\Ys(\grtr) - \Ys(\grtr'))_{2}\leq \tfrac{t}{20}.
\end{equation}
Finally, according to the event $\mathcal{A}_1$, every minimizer $\pvsolu$ of \eqref{eq:results:klasso} satisfies $\lnorm{\pvsolu - \tfunc\grtr'} \leq t$ and therefore $\lnorm{\pvsolu - \tfunc\grtr}\leq  \lnorm{\pvsolu - \tfunc\grtr'} + \lnorm{\tfunc\grtr' - \tfunc\grtr}\leq t + \eps$. \qed

\paragraph{Proof of Corollary~\ref{cor:appl:refined1bit}.}

Analogously to Corollary~\ref{cor:appl:1bit}, the model setup of Corollary~\ref{cor:appl:refined1bit} fits into Assumption~\ref{ass:results:meas} as follows:
\begin{itemize}
\item
	We have that $\a \distributed \Normdistr{\vnull}{\I{p}}$ and therefore $\normsubg{\a} \lesssim 1$. The signal set $\rset$ is an arbitrary subset of $\R^p$.
	The output function $\fout \colon \R^p \times \rset \to \R$ takes the form $\fout(\a, \grtr) \coloneqq \sign(\sp{\a}{\grtr})$.
	In particular, the resulting observation vector of $\grtr$ is given by $\Ys(\grtr) = \sign(\A\grtr)$ where 
	$\A\in\nobreak \R^{m\times p}$ is a standard Gaussian random matrix with row vectors $\a_1, \dots, \a_m\in \R^p$.
\item 
	The target function $\tfunc \colon \rset \to \sset$ corresponds to the (scaled) normalization $\tfunc\grtr \coloneqq \sqrt{\tfrac{2}{\pi}}\tfrac{\grtr}{\lnorm{\grtr}}$. 
\end{itemize}
As already shown in Subsection~\ref{subsec:app:appl:1bit} (Step~3), the target mismatch $\mmcovar{\grtr}$ vanishes for every $\grtr\in \rset$.
Moreover, we clearly have that $\multiplDia=\sup_{\grtr\in \rset}\normsubg{\sp{\a}{\tfunc\grtr} - \ys(\grtr)}\lesssim 1$.
Since $\phi(\beta) \coloneqq \beta\sqrt{\log(e/\beta)}$ defines a continuous and non-decreasing function on $[0, 1]$ with $\phi(1)=1$ and $\phi(0)\coloneqq 0$ (by continuous extension), we may assume without loss of generality that
\begin{equation} \label{eq:refined1bit:proof:beta}
	\beta\sqrt{\log(e/\beta)} = c_0 t \in (0,1].
\end{equation}
Now, we set 
\begin{equation}
	\tune^2 \coloneqq \frac{1}{t\sqrt{\log(e/\beta)}}, \quad \nout \coloneqq \floor[\big]{\tfrac{\beta}{2} m}, \quad \probsuccess_0 \coloneqq \sqrt{2 m \beta \log(e/\beta)},
\end{equation}
implying that $\probsuccess_0 \geq \sqrt{2\nout \log(em/2\nout)}$, $\nout\in \{0, 1, \dots, \floor{\tfrac{m}{2}}\}$, and $\probsuccess_0^2\geq 2c_0t m$. 
The latter inequality again implies that $\probsuccess_0^2 \geq \probsuccess^2$, and since $\probsuccess^2\geq C_0 \cdot \log\covnumber[\eps]{\tfunc\rset}$, the condition of \eqref{eq:appl:without_increments:u} is fulfilled.
Similarly, it is not hard to see that the condition of \eqref{eq:appl:without_increments:m} follows from \eqref{eq:appl:refined1bit:m} for $C'$ sufficiently large (cf.~Subsection~\ref{subsec:app:appl:1bit} (Step~2)).

According to Theorem~\ref{thm:hyperplane:tessellation:1bit} with $\ssetgen \coloneqq \sqrt{\tfrac{\pi}{2}}\tfunc\rset \subset \S^{p-1}$, there exist universal constants $c, \bar{c}, C > 0$ (possibly slightly different from those in Theorem~\ref{thm:hyperplane:tessellation:1bit}) such that if
\begin{equation}\label{eq:cond:local_robustness:1bit}
	m\geq C\cdot \Big(\eps^2 \beta^{-3} \cdot \effdim{[\tfunc\rset]_\eps} + \beta^{-1} \cdot \log\covnumber[\eps]{\tfunc\rset}\Big)
\end{equation}
for $\eps\leq \bar{c}\beta / \sqrt{\log(e/\beta)}$, the following holds with probability at least $1-\exp(-c m \beta)$:
\begin{equation}
	\sup_{\substack{\grtr,\, \grtr'\in \rset\\ \lnorm{\tfunc\grtr-\tfunc\grtr'}\leq \eps}}\tfrac{1}{2m}\lnorm{\sign(\A\grtr)-\sign(\A\grtr')}[1] \leq \tfrac{\beta}{2}.
\end{equation}
On this event and by the above choice of $\nout$, we have that
\begin{equation}
	\sup_{\substack{\grtr,\, \grtr'\in \rset\\ \lnorm{\tfunc\grtr-\tfunc\grtr'}\leq \eps}}\tfrac{1}{\sqrt{m}}\sigma_{\nout}(\Ys(\grtr) - \Ys(\grtr'))_{2}=0
\end{equation}
and 
\begin{equation}
	\sup_{\substack{\grtr,\, \grtr'\in \rset\\ \lnorm{\tfunc\grtr-\tfunc\grtr'}\leq \eps}}\tfrac{1}{\sqrt{m}}\norm{\Ys(\grtr) - \Ys(\grtr')}_{[2\nout]}\leq \sqrt{2\beta}= \Big(\tfrac{2c_0t}{\sqrt{\log(e/\beta)}}\Big)^{1/2}\leq \tfrac{1}{2}\tune t,
\end{equation}
where the last inequality holds for $c_0>0$ small enough. 
Consequently, Assumption~\ref{ass:local_robustness} would hold for $\eta \coloneqq \exp(-c m \beta)$ if we can show that \eqref{eq:cond:local_robustness:1bit} is satisfied under the hypotheses of Corollary~\ref{cor:appl:refined1bit}. 
To this end, we first note that the relationship \eqref{eq:refined1bit:proof:beta} implies that there exists a universal constant $c_0'>0$ such that $\beta\geq c_0't/\sqrt{\log(e/t)}$.
This particularly leads to the following estimates:
\begin{equation}
	\beta \geq c_0' t^2 \qquad \text{and} \qquad  \frac{\beta}{\sqrt{\log(e/\beta)}}\geq \frac{(c_0')^2t}{c_0\log(e/t)}.
\end{equation}
Combining the first one with \eqref{eq:appl:refined1bit:m}, it follows that $m \beta \geq  c_0' m t^2 \gtrsim \probsuccess^2$, while the second one yields
\begin{equation}
	\eps \leq \frac{c't}{\log(e/t)} \leq \frac{\bar{c}\beta}{\sqrt{\log(e/\beta)}}
\end{equation}
if $c'$ is chosen sufficiently small.
Hence, \eqref{eq:cond:local_robustness:1bit} is a consequence of \eqref{eq:appl:refined1bit:m} and the assumption $\probsuccess^2\geq C_0\cdot \log\covnumber[\eps]{\tfunc\rset}$. 

Since all assumptions of Theorem~\ref{thm:appl:without_increments} are satisfied, the following holds with probability at least $1 - \exp(-c \probsuccess^2)$ uniformly for every $\grtr \in \rset$:
For any input vector $\Y \in \R^m$ such that
\begin{equation}\label{eq:results:main:rob:1bit}
	\tfrac{1}{\sqrt{m}}\norm{\Y - \Ys(\grtr)}_{[2\nout]}\leq \tfrac{1}{2}\tune t \quad \text{ and } \quad 
	\tfrac{1}{\sqrt{m}}\sigma_{\nout}(\Y - \Ys(\grtr))_{2}\leq \tfrac{t}{40}, 
\end{equation} 
every minimizer $\pvsolu$ of \eqref{eq:results:klasso} satisfies $\lnorm{\pvsolu - \tfunc\grtr} \leq t + \eps\leq 2t$. 
The claim of Corollary~\ref{cor:appl:refined1bit} now follows from the fact that any input vector $\Y \in \{-1,1\}^m$ given by \eqref{eq:appl:1bit:meas} with $\tfrac{1}{2m}\lnorm{\Noise}[1] \leq \tfrac{\beta}{2}$ satisfies \eqref{eq:results:main:rob:1bit}. \qed

\paragraph{Proof of Corollary~\ref{cor:appl:refined1bitdither}.}

Analogously to Corollary~\ref{cor:appl:1bitdither}, the model setup of Corollary~\ref{cor:appl:refined1bitdither} fits into Assumption~\ref{ass:results:meas} as follows:
\begin{itemize}
\item
	The measurement vector $\a \in \R^p$ is centered, isotropic, and sub-Gaussian with $\normsubg{\a} \leq \subgparam$.
	The signal set $\rset$ satisfies $\rset \subset R \ball[2][p]$.	
	The output function $\fout \colon \R^p \times \rset \to \R$ takes the form $\fout(\a, \grtr)=\sign(\sp{\a}{\grtr} +\tau)$, where $\tau$ is uniformly distributed on $[-\lambda, \lambda]$ and independent of~$\a$. In particular, $\fout$ is a random function.
	Moreover, the observation vector of $\grtr$ is given by $\Ys(\grtr) = \sign(\A\grtr+\ttau)$, where $\A\in \R^{m\times p}$ denotes the sub-Gaussian random matrix with row vectors $\a_1, \dots, \a_m\in \R^p$ and $\ttau \coloneqq (\tau_1, \dots, \tau_m)$.
\item 
	The target function $\tfunc \colon \rset \to \sset$ corresponds to rescaling by a factor of $\lambda^{-1}$, i.e., $\tfunc\grtr \coloneqq \lambda^{-1}\grtr$.
\end{itemize}
As already shown in Subsection~\ref{subsec:app:appl:1bitdither} (Step~3), there exists a universal constant $\tilde{C}'\geq e$ such that if
\begin{equation}\label{eq:refined1bitdither:proof:lambda}
	\lambda\geq \tilde{C}' \cdot R\cdot  \subgparam \cdot \sqrt{\log(e/t)},
\end{equation}
the target mismatch satisfies $\mmcovar{\grtr}\leq \tfrac{t}{32}$ for every $\grtr\in \rset$.
Moreover, we clearly have that
\begin{equation}
	\normsubg{\sp{\a}{\lambda^{-1}\grtr} - \ys(\grtr)}\lesssim \lambda^{-1} \normsubg{\sp{\a}{\grtr}} + 1 \lesssim R\subgparam\lambda^{-1} +1
\end{equation}
for every $\grtr\in \rset$, and together with \eqref{eq:refined1bitdither:proof:lambda}, it follows that $\multiplDia=\sup_{\grtr\in \rset}\normsubg{\sp{\a}{\lambda^{-1}\grtr} - \ys(\grtr)}\lesssim\nobreak 1$. 
Since $\phi(\beta) \coloneqq \beta\sqrt{\log(e/\beta)}$ defines a continuous and non-decreasing function on $[0, 1]$ with $\phi(1)=\nobreak 1$ and $\phi(0)\coloneqq 0$ (by continuous extension), we may assume without loss of generality that
\begin{equation} \label{eq:refined1bitdither:proof:beta}
	\beta\sqrt{\log(e/\beta)}=c_0 \subgparam^{-2} t \in (0,1].
\end{equation}
Now, we set 
\begin{equation}
	\tune^2 \coloneqq \frac{1}{t\subgparam^2\sqrt{\log(e/\beta)}}, \quad \nout \coloneqq \floor[\big]{\tfrac{\beta}{2} m}, \quad \probsuccess_0 \coloneqq \sqrt{2 m \beta \log(e/\beta)},
\end{equation}
implying that $\probsuccess_0 \geq \sqrt{2\nout \log(em/2\nout)}$, $\nout\in \{0, 1, \dots, \floor{\tfrac{m}{2}}\}$, and $\probsuccess_0^2\geq 2c_0m\subgparam^{-2} t$. 
Combining the latter inequality with \eqref{eq:cond:refined1bitdither:m} for $C' \gtrsim \subgparam^2$ sufficiently large implies that $\probsuccess_0^2 \geq \probsuccess^2$, and since $\probsuccess^2\geq C_0 \cdot \log\covnumber[\eps]{\lambda^{-1}\rset}$, the condition of \eqref{eq:appl:without_increments:u} is fulfilled.
Similarly, it is not hard to see that the condition of \eqref{eq:appl:without_increments:m} follows from \eqref{eq:cond:refined1bitdither:m} for $C' \gtrsim \subgparam^2 \log\subgparam$ sufficiently large (cf.~Subsection~\ref{subsec:app:appl:1bitdither} (Step~2)).

According to Theorem~\ref{thm:hyperplane:tessellation:1bitdither}, there exist constants 
$c, \bar{c}, C, \tilde{C} > 0$ only depending on $\subgparam$ (possibly slightly different from those in Theorem~\ref{thm:hyperplane:tessellation:1bitdither}) such that if $\lambda \geq \tilde{C}\cdot R$ and
\begin{equation}\label{eq:cond:local_robustness:1bitdither}
	m \geq C \cdot  \Big(\eps^2\beta^{-3}\cdot\effdim{[(2\lambda)^{-1}\rset]_\eps} + \beta^{-1} \cdot \log\covnumber[\eps]{\lambda^{-1}\rset}\Big)
\end{equation}
for $\eps\leq \bar{c}\beta / \sqrt{\log(e/\beta)}$, the following holds with probability at least $1-\exp(-c m\beta)$:
\begin{equation}
	\sup_{\substack{\grtr,\, \grtr'\in \rset\\ \lambda^{-1} \lnorm{\grtr-\grtr'}\leq \eps}}\tfrac{1}{2m}\lnorm{\sign(\A\grtr+\ttau)-\sign(\A\grtr'+\ttau)}[1]\leq \tfrac{\beta}{2}.
\end{equation}
On this event and by the above choice of $\nout$, we have that
\begin{equation}
	\sup_{\substack{\grtr,\, \grtr'\in \rset\\ \lambda^{-1}\lnorm{\grtr-\grtr'}\leq \eps}}\tfrac{1}{\sqrt{m}}\sigma_{\nout}(\Ys(\grtr) - \Ys(\grtr'))_{2}=0
\end{equation}
and 
\begin{equation}
	\sup_{\substack{\grtr,\, \grtr'\in \rset\\ \lambda^{-1}\lnorm{\grtr-\grtr'}\leq \eps}} \tfrac{1}{\sqrt{m}}\norm{\Ys(\grtr) - \Ys(\grtr')}_{[2\nout]} \leq \sqrt{2\beta}= \Big(\tfrac{2c_0t}{\subgparam^2\sqrt{\log(e/\beta)}}\Big)^{1/2}\leq \tfrac{1}{2}\tune t,
\end{equation}
where the last inequality holds for $c_0>0$ small enough. 
Consequently, Assumption~\ref{ass:local_robustness} would hold for $\eta \coloneqq \exp(-c m \beta)$ if we can show that \eqref{eq:cond:local_robustness:1bitdither} is satisfied under the hypotheses of Corollary~\ref{cor:appl:refined1bitdither}.
To this end, we first note that the relationship \eqref{eq:refined1bitdither:proof:beta} implies that there exists a universal constant $c_0'>0$ such that $\beta\geq c_0't\subgparam^{-2}/\sqrt{\log(e\subgparam^2/t)}$.
This particularly leads to the following estimates:
\begin{equation}
	\beta \geq c_0' t^2 \subgparam^{-2} \qquad \text{and} \qquad  \frac{\beta}{\sqrt{\log(e/\beta)}}\geq \frac{(c_0')^2t}{c_0\subgparam^2\log(e\subgparam^2/t)}.
\end{equation}
Combining the first one with \eqref{eq:cond:refined1bitdither:m}, it follows that $m \beta \geq  c_0' m t^2 \subgparam^{-2} \gtrsim \probsuccess^2$ for $C' \gtrsim \subgparam^2$ sufficiently large, while the second one yields
\begin{equation}
	\eps \leq \frac{c't}{\log(e/t)} \leq \frac{\bar{c}\beta}{\sqrt{\log(e/\beta)}}
\end{equation}
if $c'$ is chosen sufficiently small (depending on $\subgparam$).
Hence, \eqref{eq:cond:local_robustness:1bitdither} is a consequence of \eqref{eq:cond:refined1bitdither:m} and the assumption $\probsuccess^2\geq C_0\cdot \log\covnumber[\eps]{\lambda^{-1}\rset}$. 

Since all assumptions of Theorem~\ref{thm:appl:without_increments} are satisfied, the following holds with probability at least $1 - \exp(-c \probsuccess^2)$ uniformly for every $\grtr \in \rset$:
For any input vector $\Y \in \R^m$ such that
\begin{equation}\label{eq:results:main:rob:1bitdither}
	\tfrac{1}{\sqrt{m}}\norm{\Y - \Ys(\grtr)}_{[2\nout]}\leq \tfrac{1}{2}\tune t \quad \text{ and } \quad 
	\tfrac{1}{\sqrt{m}}\sigma_{\nout}(\Y - \Ys(\grtr))_{2}\leq \tfrac{t}{40}, 
\end{equation} 
every minimizer $\pvsolu$ of \eqref{eq:results:klasso} satisfies $\lnorm{\pvsolu - \lambda^{-1}\grtr} \leq t + \eps\leq 2t$. 
The claim of Corollary~\ref{cor:appl:refined1bitdither} now follows from the fact that any input vector $\Y \in \{-1,1\}^m$ given by \eqref{eq:appl:1bitdither:meas} with $\tfrac{1}{2m}\lnorm{\Noise}[1] \leq \tfrac{\beta}{2}$ satisfies \eqref{eq:results:main:rob:1bitdither}. \qed

\section{Proofs for Section~\ref{sec:unif}}
\label{sec:app:unif}

\paragraph{Proof of Corollary~\ref{cor:unif:lin}.}

We follow the proof template from the beginning of Section~\ref{sec:app:appl}:
\begin{tmpl}
	\item
	The model setup of Corollary~\ref{cor:unif:lin} fits into Assumption~\ref{ass:results:meas} as follows:
	\begin{itemize}
		\item
		The measurement vector $\a \in \R^p$ is centered, isotropic, and sub-Gaussian with $\normsubg{\a} \leq \subgparam$. The signal set $\rset$ is a bounded subset of $\R^p$.
		The output function $\fout \colon \R^p \times \rset \to \R$ takes the form $\fout(\a, \grtr) \coloneqq \sp{\a}{\grtr} + \tau$, where $\tau \distributed \Normdistr{0}{\sigma^2}$.
		\item
		The target function $\tfunc \colon \rset \to \sset$ is the canonical embedding into $\sset$, i.e., $\tfunc\grtr \coloneqq \grtr$. In particular, we have that $d_\tfunc(\grtr,\grtr')= \lnorm{\grtr - \grtr'}$.
	\end{itemize}
	
	\item
	The target mismatch $\mmcovar{\grtr}$ vanishes for every $\grtr\in \rset$.
	
	\item
	For the trivial choice $\ys_t(\grtr)\coloneqq\ys(\grtr)$, the conditions of Assumption~\ref{ass:results:incr} are fulfilled with
	\begin{equation} 
		\multiplLip = 0,\quad \errLip = 0,\quad \multiplDia \lesssim \sigma, \quad \errDia = 0.
	\end{equation}
	
	\item
	Setting $\nout \coloneqq 0$, $\tune \coloneqq \subgparam^{-1} \sqrt{\log\subgparam}$, and $\probsuccess \coloneqq \probsuccess_0$, the claim of Corollary~\ref{cor:unif:lin} follows directly from Theorem~\ref{thm:results:main}.
	
	\item
	Let $\grtr\in \rset$. By the isotropy of $\a$, we have that $\mean[\ys(\grtr)\a] = \mean[\sp{\a}{\grtr} \a]=\grtr = \tfunc\grtr$, and therefore $\mmcovar{\grtr}=0$.
	
	\item
	We simply set $\ys_t(\grtr)\coloneqq\ys(\grtr)$. Then $\err_t(\grtr) = 0$, implying that Assumption~\ref{ass:results:incr}\ref{ass:results:incr:approx} and~\ref{ass:results:incr:err} are trivially fulfilled with $\errLip=\errDia=0$. Furthermore, since $\multipl_t(\grtr)= \sp{\a}{\grtr}-\ys_t(\grtr)= -\tau$, Assumption~\ref{ass:results:incr}\ref{ass:results:incr:multipl} holds with $\multiplLip=0$ and $\multiplDia \coloneqq \normsubg{\tau} \lesssim \sigma$, since $\tau \distributed \Normdistr{0}{\sigma^2}$. \qed
\end{tmpl}

\paragraph{Proof of Inequality~(\ref{eq:unif:mwl1}).} 
Since $\lnorm{\grtr}[1] = R$ for every $\grtr \in \rset$, we may use \cite[Lem.~4.5]{rv08} to observe that
\begin{equation}
	\cone{R\ball[1][p] - \grtr} \intersec \S^{p-1} \subset \ssetgen \coloneqq 2 \convhull{\vdir \in \S^{p-1} \suchthat \lnorm{\vdir}[0] \leq s },
\end{equation}
where $\convhull{\cdot}$ denotes the convex hull.
Moreover, \cite[Lem.~2.3]{pv13b} yields $\meanwidth{\ssetgen} \lesssim \sqrt{s \cdot \log(2p/s)}$.
Therefore, we obtain
\begin{align}
	\meanwidth[t]{\sset - \rset} &\leq \meanwidth[0]{\sset - \rset} \\ 
	&= \meanwidth{\cone{\sset - \rset} \intersec \S^{p-1}} \\
	&= \meanwidth{\textstyle\bigunion_{\grtr \in \rset}\cone{R\ball[1][p] - \grtr} \intersec \S^{p-1}} \\
	&\leq \meanwidth{\ssetgen} \lesssim \sqrt{s \cdot \log(2p/s)},
\end{align}
as claimed in \eqref{eq:unif:mwl1}. \qed

\paragraph{Proof of Proposition~\ref{prop:unif:mwloc}.}
We begin with writing out the definition of the local mean width as follows:
\begin{align}
	\meanwidth[t]{\sset - \rset} \leq \tfrac{1}{t} \cdot \meanwidth{(\sset - \rset) \intersec t \ball[2][p]} = \tfrac{1}{t} \cdot \mean\Big[\sup_{\substack{\pv \in \sset, \grtr \in \rset \\ \lnorm{\pv - \grtr} \leq t}} \sp{\gaussian}{\pv - \grtr}\Big],
\end{align}
where $\gaussian \distributed \Normdistr{\vnull}{\I{p}}$.
Now, let $\rset_t$ be a minimal $t$-net for $\rset$; in particular, we have that $\cardinality{\rset_t} = \covnumber[t]{\rset}$.
Hence, for every $\pv \in \sset$ and $\grtr \in \rset$ with $\lnorm{\pv - \grtr} \leq t$, there exists $\vdir_{\grtr} \in \rset_t$ such that $\lnorm{\vdir_{\grtr} - \grtr} \leq t$ and $\lnorm{\vdir_{\grtr} - \pv} \leq 2t$.
Therefore, we obtain
\begin{align}
	\mean\Big[\sup_{\substack{\pv \in \sset, \grtr \in \rset \\ \lnorm{\pv - \grtr} \leq t}} \sp{\gaussian}{\pv - \grtr}\Big] 
	&= \mean\Big[\sup_{\substack{\pv \in \sset, \grtr \in \rset \\ \lnorm{\pv - \grtr} \leq t}} \sp{\gaussian}{\pv - \vdir_{\grtr}} + \sp{\gaussian}{\vdir_{\grtr} - \grtr}\Big] \\
	&\leq \mean\Big[\sup_{\substack{\pv \in \sset, \vdir \in \rset_t \\ \lnorm{\pv - \vdir} \leq 2t}} \sp{\gaussian}{\pv - \vdir} + \sup_{\substack{\grtr \in \rset,\vdir \in \rset_t \\ \lnorm{\grtr - \vdir} \leq t}} \sp{\gaussian}{\vdir - \grtr}\Big] \\
	&= \meanwidth{(\sset - \rset_t) \intersec 2t \ball[2][p]} + \meanwidth{(\rset_t - \rset) \intersec t \ball[2][p]} \\
	&= \meanwidth{{\textstyle\bigunion_{\vdir \in \rset_t} (\sset - \vdir) \intersec 2t \ball[2][p]}} + \underbrace{\meanwidth{(\rset_t - \rset) \intersec t \ball[2][p]}}_{\leq \meanwidth{\rset_t - \rset} \leq 2 \cdot \meanwidth{\rset}}.
\end{align}
Next, we make use of the fact that $\rset_t$ is a finite set. Indeed, applying \cite[Lem.~6.1]{dgjs21}, this allows us to ``pull out'' the union over $\rset_t$ in the following way:
\begin{equation}
	\meanwidth{{\textstyle\bigunion_{\vdir \in \rset_t} (\sset - \vdir) \intersec 2t \ball[2][p]}} \lesssim \sup_{\vdir \in \rset_t} \ \meanwidth{(\sset - \vdir) \intersec 2t \ball[2][p]} + t \cdot \sqrt{\log\covnumber[t]{\rset}}.
\end{equation}
By Sudakov minoration (e.g., see \cite[Thm.~7.4.1]{ver18}), we have that $t \cdot \sqrt{\log\covnumber[t]{\rset}} \lesssim \meanwidth{\rset}$.
Hence, overall, we have that
\begin{equation}
	\meanwidth[t]{\sset - \rset} \lesssim \sup_{\vdir \in \rset_t} \ \underbrace{\meanwidth{\tfrac{1}{2t}(\sset - \vdir) \intersec \ball[2][p]}}_{= \meanwidthalt[t]{\sset - \vdir}} {}+{} t^{-1} \cdot \meanwidth{\rset} \lesssim \sup_{\grtr \in \rset} \ \meanwidthalt[t]{\sset - \grtr} + t^{-1} \cdot \meanwidth{\rset}.
\end{equation}
The `moreover'-part of Proposition~\ref{prop:unif:mwloc} follows directly from a stability bound derived in \cite[Prop.~2.6]{gen19}. \qed


\section*{Acknowledgments}
{\smaller
M.G.\ acknowledges support by the DFG Priority Programme DFG-SPP 1798 Grant DI 2120/1-1.
A.S.\ acknowledges support by the Fonds de la Recherche Scientifique -- FNRS under Grant n$^\circ$ T.0136.20 (Learn2Sense).
The authors thank Sjoerd~Dirksen and Maximilian~März for fruitful discussions. Moreover, the authors would like to thank the anonymous referees for their very useful comments and suggestions which have helped to improve the present article.

}

\renewcommand*{\bibfont}{\smaller}
\begin{refcontext}[sorting=nyt]
	\printbibliography[heading=bibintoc]
\end{refcontext}


\end{document}